\documentclass[12pt]{iopart}

\usepackage{titlesec}

\titleformat{\paragraph}
{\normalfont\normalsize\bfseries}{\theparagraph}{1em}{}
\titlespacing*{\paragraph}
{0pt}{3.25ex plus 1ex minus .2ex}{1.5ex plus .2ex}
\usepackage{iopams}
\expandafter\let\csname equation*\endcsname\relax
\expandafter\let\csname endequation*\endcsname\relax
\usepackage{amsmath}
\usepackage{tikz}
\usepackage{subcaption}
\pdfoutput=1
\usepackage{graphicx,psfrag,color}
\usepackage[utf8]{inputenc}
\usepackage[T1]{fontenc}
\usepackage{lmodern}
\usepackage{times}
\usepackage{mathptmx}
\usepackage{xcolor}
\usepackage[colorlinks,linkcolor=blue,urlcolor=blue,citecolor=blue]{hyperref}
\usepackage[normalem]{ulem}
\usepackage{xspace}
\usepackage{float}
\usepackage{bm}
\usepackage{cite}
\usepackage{lineno}
\hypersetup{
    colorlinks=true,
    linkcolor=blue,
    filecolor=magenta,
    urlcolor=cyan,
    citecolor=cyan
}

\DeclareSymbolFont{bbold}{U}{bbold}{m}{n}
\DeclareSymbolFontAlphabet{\mathbbold}{bbold}
\makeatletter
\newcommand*\bigcdot{\mathpalette\bigcdot@{.55}}
\newcommand*\bigcdot@[2]{\mathbin{\vcenter{\hbox{\scalebox{#2}{$\m@th#1\bullet$}}}}}
\makeatother
\def\bra#1{\mathinner{\langle{#1}|}}
\def\ket#1{\mathinner{|{#1}\rangle}}

\begin{document}
\newcommand{\gae}{\lower 2pt \hbox{$\, \buildrel {\scriptstyle >}\over {\scriptstyle
\sim}\,$}}
\newcommand{\lae}{\lower 2pt \hbox{$\, \buildrel {\scriptstyle <}\over {\scriptstyle
\sim}\,$}}

\title{Quantum synchronization: A tutorial guide}
\author{Sourabh Lahiri$^1$ and Shamik Gupta$^2$}
\address{$^1$Department of Physics, Birla Institute of Technology, Mesra, Ranchi, Jharkhand 835215, India
\\ $^2$Department of Theoretical Physics, Tata Institute of Fundamental Research, Homi Bhabha Road, Mumbai 400005, India}
\begin{abstract}
Synchronization is a collective phenomenon in which interacting nonlinear oscillators develop a persistent relation between their phases and frequencies. In quantum systems, its description requires connecting the classical dynamics of limit-cycle oscillators with quantum fluctuations, dissipation, and the notion of phase. In this tutorial review, we offer a systematic framework for analyzing synchronization in classical and quantum dynamical systems. Starting from amplitude–phase dynamics, we introduce successively topics of relevance, such as phase locking, the Adler equation, Arnold tongues, and the Kuramoto model, and use these concepts to analyze coupled oscillators and their stability. We then formulate the corresponding description for quantum limit-cycle oscillators using Lindblad master equations and phase-space methods, with explicit examples of two and many coupled oscillators. We further discuss noise, synchronization measures, and the extreme quantum regime, including synchronization in spin systems. The emphasis throughout is on analytical derivations and a practical methodology for identifying synchronized states, determining their stability, and characterizing synchronization in quantum systems. The review is designed as a self-contained guide from which we hope that students can learn the essential concepts and techniques, while researchers can readily find the analytical tools and methodological insights needed to approach new synchronization problems.
\end{abstract}

\maketitle
\tableofcontents

Synchronization is a ubiquitous phenomenon in nonlinear dynamical systems, whereby interacting nonlinear oscillators adjust their rhythms, developing a persistent relation between their phases or frequencies \cite{Strogatz2004sync}. It occurs in systems ranging from classical mechanical and biological oscillators to driven nonlinear systems \cite{Pikovsky2001synchronization, Pikovsky2015dynamics, Boccaletti2002,Gupta2018statistical} and, more recently, in quantum systems \cite{nadolny2026quantum,kehrer2024quantum,kehrer2025quantum,nadolny2025nonreciprocal,solanki2024exotic,nadolny2023macroscopic,daniel2023geometric,koppenhoefer2019optimal,koppenhoefer2020quantum,roulet2018quantum,roulet2018synchronizing,lorch2017quantum,lorch2016genuine,walter2015quantum,walter2014quantum,Vigneshwar2025,Vigneshwar2026,Vigneshwar2026b,Bandyopadhyay2023,Paul2024,Bandyopadhyay2021,Preethi2025,Thomas2026,solanki2026quantumsynchronization}. The study of synchronization in the quantum regime raises a number of questions that have no direct classical counterpart: how should the phase of a quantum oscillator be characterized, what constitutes a synchronized quantum state, and which observables or phase-space representations provide useful measures of synchronization? Addressing these questions requires combining the tools of nonlinear dynamics with those of open quantum systems.

Our objective here is not to provide a comprehensive review of the rapidly developing literature on classical or quantum synchronization. Rather, our aim is to review in a pedagogical manner a set of tools and techniques for studying synchronization, and to demonstrate how these tools can be applied systematically to concrete examples. We therefore deliberately choose a small number of representative models and work them out in full detail, starting from the underlying dynamical equations and proceeding step by step to the characterization of limit cycles, phase dynamics, synchronization conditions, stability, and synchronization measures. This approach is intended to make the underlying methodology transparent and reproducible, rather than to provide an exhaustive account of the many physical systems in which synchronization has been investigated.

We begin with the classical description, since it provides the simplest setting in which the basic ideas behind synchronization can be introduced without the additional complications associated with quantum mechanics. We first discuss stable limit cycles and the separation of amplitude and phase dynamics. This allows us to introduce the notion of a natural frequency and to identify the phase as the slow, neutrally stable variable associated with motion along a limit cycle. We then consider a periodically-driven oscillator and derive the phase equation leading to the celebrated Adler equation. The resulting phase-locking condition provides a simple route to the so-called Arnold tongue and illustrates the general mechanism by which detuning and coupling compete to determine whether and how synchronization occurs.

We next turn to interacting oscillators. For two coupled limit-cycle oscillators, we explicitly derive the amplitude and phase equations and show how in the weak-coupling regime the amplitude variables remain close to their stable values while the relative phase becomes the relevant dynamical variable. This reduction leads naturally to an Adler-type equation and provides a direct connection between synchronization and phase locking. We then extend the discussion to many oscillators and introduce the Kuramoto description and its complex order parameter to quantify synchronization \cite{kuramoto1984chemical,strogatz2000kuramoto,acebron2005kuramoto,gupta2014kuramoto,rodrigues2016kuramoto}. In this way, the familiar many-body synchronization transition emerges as a natural extension of the two-oscillator problem rather than as an independent phenomenological model. The classical examples thus establish a hierarchy of descriptions, from full amplitude-phase dynamics to reduced phase models and collective order parameters, that will be useful in the quantum setting as well.

The quantum part of the work follows the same philosophy. We first introduce quantum limit-cycle oscillators and discuss how the notion of a classical limit cycle is represented in an open quantum system. The density-matrix description \cite{cohen1977quantum1}, Lindblad dynamics \cite{Breuer,carmichael1999statistical,Manzano2020}, and phase-space representations such as the Wigner and Husimi \cite{gardiner2004quantum} functions provide complementary ways of characterizing the resulting dynamics. Rather than assuming a particular synchronization measure at the outset, we develop these descriptions explicitly and examine what information about amplitude, phase, and coherence can be extracted from them. This also makes clear where the classical intuition remains useful and where genuinely quantum features enter.

As representative examples, we consider coupled quantum limit-cycle oscillators and analyze both identical and detuned oscillators. For the latter, we explicitly derive the stationary synchronized state and its stability from the amplitude-phase equations. In particular, we emphasize that finite detuning generally produces unequal stationary amplitudes and a nontrivial phase difference, so that synchronization is more naturally understood as a frequency-locked state than as simple in-phase motion. Linearization around the synchronized solution and analysis of the corresponding Jacobian then provide a systematic criterion for its stability. Such calculations illustrate a general procedure that can be applied to a broad class of nonlinear open quantum systems.

We subsequently consider many-body and spin-system examples, together with the role of noise. These examples are chosen not to survey all known realizations of quantum synchronization, but to illustrate how the same basic analytical ideas can be adapted when the natural variables, phase-space geometry, or source of fluctuations changes. In particular, the comparison between oscillator and spin descriptions highlights the importance of choosing an appropriate phase-space representation and synchronization observable.

A central theme throughout the discussion is therefore the construction of a practical workflow for synchronization problems. Given a dynamical model, one first identifies the underlying oscillatory state and its stable amplitude dynamics; one then isolates the phase degree of freedom, determines the effective phase dynamics, identifies phase-locked solutions, and finally tests their stability and quantifies the resulting synchronization. In quantum systems, this procedure must be supplemented by an appropriate treatment of dissipation, quantum fluctuations, and the choice of a phase-space or operator-based synchronization measure. Working through these steps explicitly for a few representative models allows the reader to see not only the final synchronization criterion, but also where it comes from and which approximations are involved.

In \cite{solanki2026quantumsynchronization}, the authors survey the range of models, analytical tools, synchronization measures, and applications developed in the field. Our emphasis here is complementary: we focus on the derivation and use of the analytical tools themselves. The examples are consequently selected for their pedagogical and methodological value, and wherever possible the calculations are carried out in full detail. We hope that this makes the review useful not only as an introduction to quantum synchronization, but also as a practical guide for analyzing new synchronization problems beyond the particular examples considered here.

%In this respect, the objective of the present work is different from that of recent broad reviews of quantum synchronization, e.g., \cite{solanki2026quantumsynchronization}, whose primary goal is to survey the range of models, analytical tools, synchronization measures, and applications developed in the field. 

A broader aim of this review is to provide a unified entry point to the theoretical tools used to study synchronization across classical and quantum systems. The discussion brings together concepts from classical nonlinear dynamics and statistical mechanics, quantum mechanics and open quantum systems, and quantum optics and phase-space methods, highlighting how these approaches complement one another. In this sense, the review is intended to serve both as a pedagogical starting point for newcomers and as a convenient reference for researchers interested in the classical–quantum connection in synchronization.

The remainder of this review is organized as follows. Section~\ref{sec:sec1} introduces the basic concepts of classical limit-cycle oscillators, including the van der Pol oscillator, generic nonlinear oscillators, and the characterization of stable limit cycles. Section~\ref{sec:sec2} discusses driven classical limit cycles and introduces phase locking and the Arnold tongue. Section~\ref{sec:sec3} develops the theory of classical synchronization, first for two interacting limit cycles and then for many interacting oscillators, culminating in the Kuramoto model and its continuum and noisy extensions. Section~\ref{sec:sec4} turns to quantum limit cycles, beginning with the quantum harmonic oscillator and subsequently discussing their characterization using phase-space representations, including the Wigner function. Section~\ref{sec:sec5} considers driven quantum limit cycles, providing the quantum counterpart of the classical driven-oscillator problem. Section~\ref{sec:sec6} then develops the central theme of quantum synchronization, considering both two interacting quantum limit cycles and many interacting quantum limit cycles. Section~\ref{sec:sec7} extends the discussion to quantum synchronization in spin systems, with particular emphasis on spin-1 systems. Finally, Section~\ref{sec:sec9} summarizes the main results and discusses possible directions and perspectives for future work. Technical details and supplementary derivations are collected in Appendices A--F, covering coherent states, phase-space representations, Wigner-function dynamics, the Lindblad equation, vectorization techniques, and the Bloch-sphere representation.

%\chapter{Chapter on Classical Synchronization}

%%%%%%%%%%%%%%%%%%%%%%%%%%%%%%%%%%%%%%%%%%%%%%  
\section{Classical limit cycles}
\label{sec:sec1}
Oscillators in general are systems that exhibit  periodic motion or oscillations. Limit cycles are special cases of self-sustained oscillations exhibited by nonlinear dynamical systems. Limit cycles come up in a wide range of applications~\cite{strogatz-book}: the predator-prey problem, rhythmic beating of the heart, chemical oscillators (chemical components exhibiting periodic changes in concentration), vibrations in bridges, etc. Before distinguishing a limit cycle from a generic oscillatory motion, let us first graphically understand the concept of periodic motion.

%\paragraph{}
Consider a set of coupled nonlinear differential equations of the form
\begin{eqnarray}
    \dot x &=& f(x,y), \nonumber\\ \\
    \dot y &=& g(x,y), \nonumber
\end{eqnarray}
where either or both $f(x,y)$ and $g(x,y)$ are nonlinear functions of $x$ and $y$, and the overhead dot denotes  time derivative. It is easy to extend the entire formalism to higher dimensions. In compact notation, we can write the above set of equations as 
\begin{equation}
    \dot{\textbf{Q}} = \textbf{F}(\textbf{Q}),
    \label{eq:compact}
\end{equation}
where $\textbf{Q} = (x~~y)^T$ is a $2\times 1$ column vector and so is $\textbf{F}(\textbf{Q})$; here, $T$ denotes transpose operation.

\begin{figure}[!h]
    \centering
    \begin{subfigure}{0.4\linewidth}
    \includegraphics[width=\linewidth]{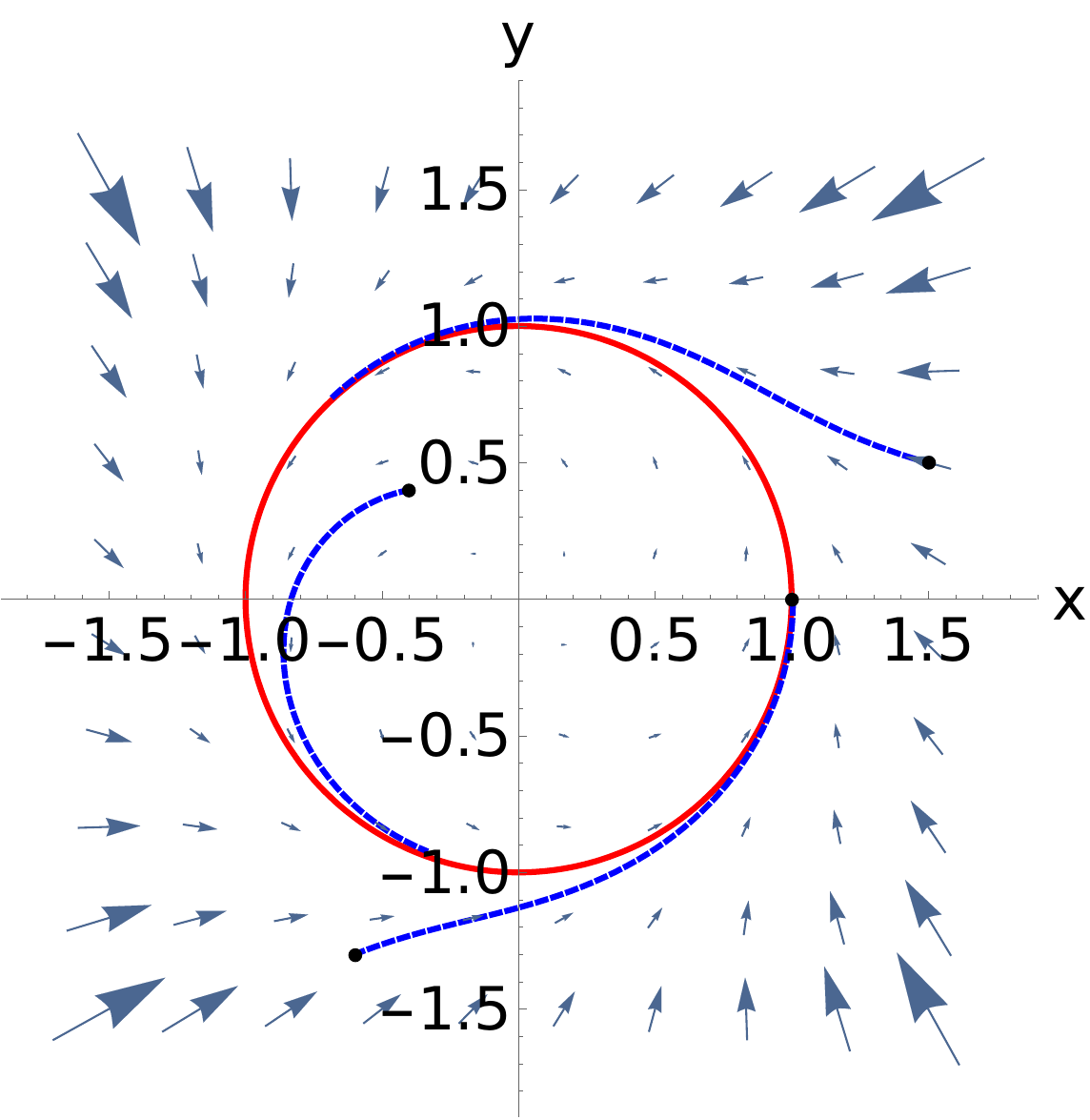}
    \caption{}
    \end{subfigure}
    \hfill
    \begin{subfigure}{0.4\linewidth}
    \includegraphics[width=\linewidth]{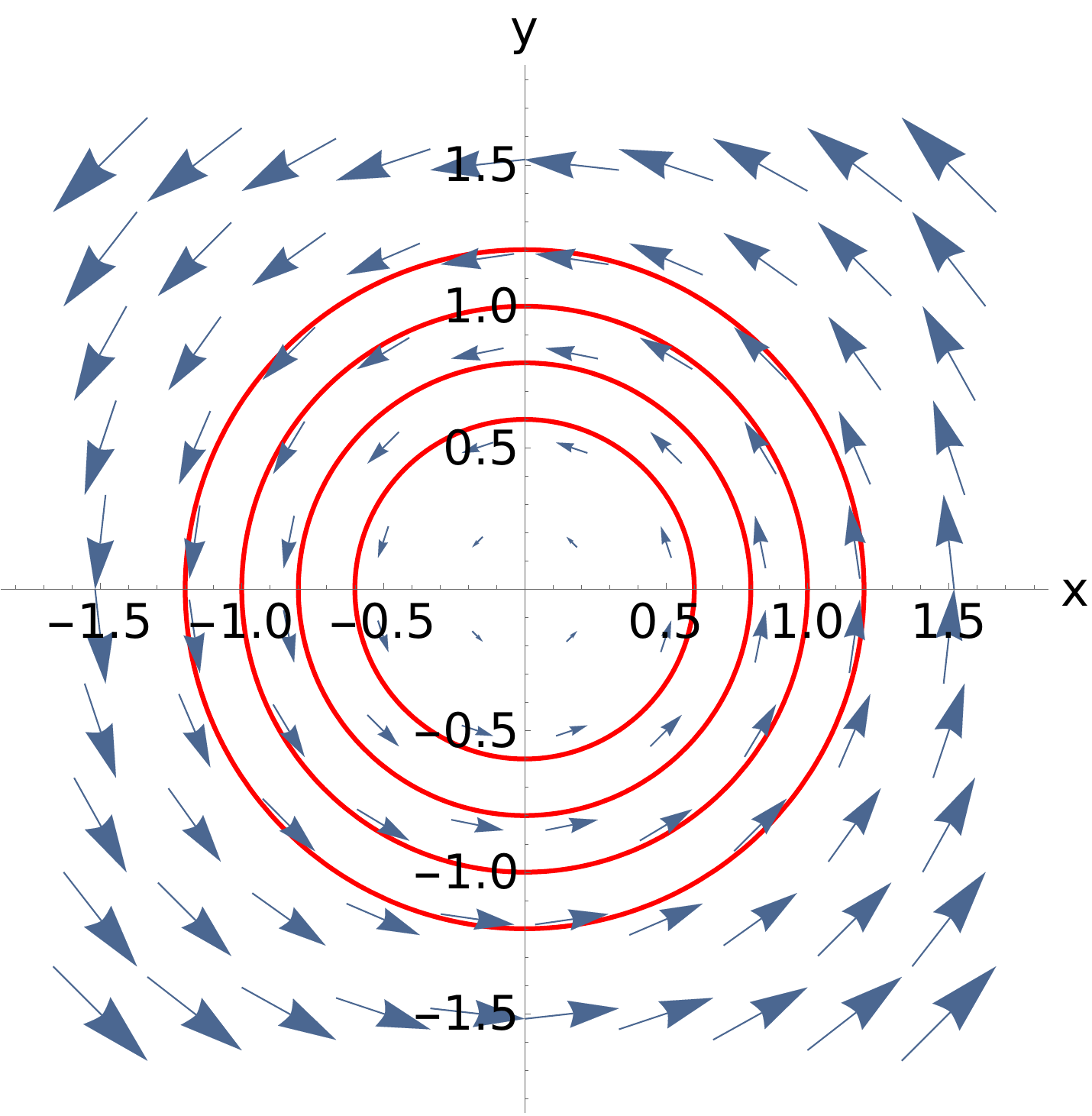}
    \caption{}
    \end{subfigure}
    \caption{(a) A stable limit cycle (red solid line) produced from Eq. \eqref{eq:limit cycle example}. Different initial conditions (blue dashed lines) originating close to the limit cycle asymptotically approach it in course of time. The underlying vector field $(\dot x,\dot y)$ is indicated by the arrows. (b) A family of concentric circles produced by the set of linear differential equations $\dot x = -ay$ and $\dot y = ax$, with $a>0$. None of these orbits is a limit cycle.}
    \label{fig:limit cycle}
\end{figure}

Suppose we find a set of initial conditions $(x_0,y_0)$ such that in the $xy$-plane, which is the phase space of the above system, the resulting trajectory or orbit (the terms ``trajectory'' and ``orbit'' would be used interchangeably)  is a closed one. It is clear that any set of initial conditions lying on this closed trajectory will trace the same orbit in phase space. Moreover, if we choose another set of initial conditions $(x'_0,y'_0)$ close to this orbit and find that the resulting trajectory asymptotically approaches the closed orbit in course of dynamical evolution, then we call the orbit a stable limit cycle. As an example, consider the set of differential equations $\dot r = r(1-r^2)$ and $\dot\theta=1$ in polar coordinates. The corresponding equations in $(x,y)$-coordinates are  
\begin{eqnarray}
    \dot x &=& x-y-x^3-xy^2,\nonumber\\ \label{eq:limit cycle example} \\
    \dot y &=& x+ y -x^2 y - y^3. \nonumber
\end{eqnarray}
In Fig. \ref{fig:limit cycle}(a), we have plotted the limit cycle (red solid line) corresponding to the above system. The vector field, given by the vector $(\dot x,\dot y)$, have also been shown in the same graph. The orientations of the arrows indicate the direction of the field, while their lengths provide the field magnitude. Clearly, the limit cycle is being traced out in the counter-clockwise direction in this vector field. The fact that the equation for $r$ has a stable fixed point at $r=1$ implies that the limit cycle is a unit circle in the $xy$-plane, which is as observed in the figure. The blue dashed lines in the figure denote trajectories originating from initial conditions close to the limit cycle, which are observed to approach it asymptotically in time, thereby making the latter a stable limit cycle.

\begin{figure}[!h]
    \centering
    \includegraphics[width=0.3\linewidth]{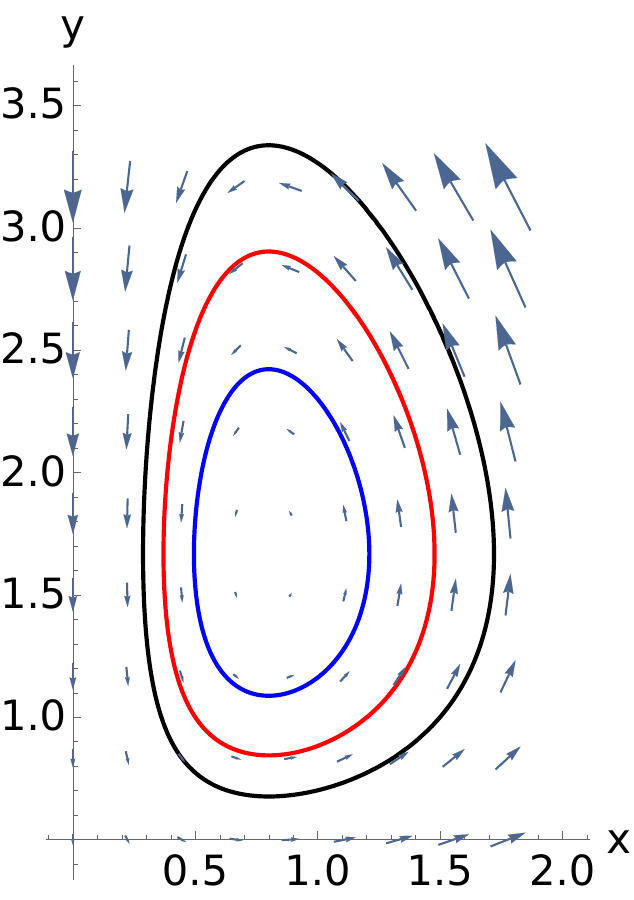}
    \caption{Orbits obtained from Eq.~\eqref{eq:Lotka-Volterra-basic1}, by varying the initial conditions $(x(0),y(0))$. A point originating close to one of the closed orbits forms another closed orbit. The underlying vector field $(\dot x,\dot y)$ is indicated by the arrows.}
    \label{fig:Lotka-Volterra basic}
\end{figure}

An orbit such as that corresponding to a stable limit cycle is said to be isolated. To understand the meaning of an isolated orbit, consider the set of differential equations $\dot x = -ay$ and $\dot y = ax$, with $a>0$. They can be trivially solved to obtain the trajectory as a circle of radius $a$ centered at the origin. However, if we choose two different initial conditions that are infinitesimally close to one another but not lying on the circle, then we will obtain two concentric circular orbits, differing infinitesimally in their radii. A family of such orbits has been shown in Fig. \ref{fig:limit cycle}(b). These orbits are not limit cycles, since a slightly different initial condition gives rise to another trajectory that also closes on itself. In general, a set of linear differential equations cannot generate a limit cycle, although they might produce closed trajectories in phase space~\cite{strogatz-book}. 
On the other hand, any nonlinear oscillator need not produce a limit cycle either. For instance, consider the basic Lotka-Volterra model that is used to study the population dynamics of predators and preys \cite{Berryman1992,Kingsland}:
\begin{eqnarray}
    \dot x &=& x(a-by),\nonumber\\ 
    \label{eq:Lotka-Volterra-basic1}\\
    \dot y &=& -y(c-dx).\nonumber
\end{eqnarray}
Here, $x(t)$ is the instantaneous prey population and $y(t)$ is the instantaneous predator population. The solution is as shown in Fig. \ref{fig:Lotka-Volterra basic}, where a slight change in the initial values of the two population produces another closed orbit in the neighborhood of the previous orbit. Thus, these orbits are not isolated, and do not satisfy the condition to become sacrosanct limit cycles. However, more realistic models for predator and prey populations, like the Rosenzweig–MacArthur model \cite{Rosenzweig1963}, have been shown to produce genuine limit cycles.

For completeness, let us mention that there also exist limit cycles that are unstable or half stable. An initial phase point close to the former keeps diverging from the orbit in time instead of approaching it asymptotically. In the case of a half-stable limit cycle, a trajectory originating from a phase point outside (inside) the orbit asymptotically diverges (converges).

\subsection{The van der Pol oscillator}
\begin{figure}
    \centering
    \begin{subfigure}{0.45\linewidth}
        \includegraphics[width=\linewidth]{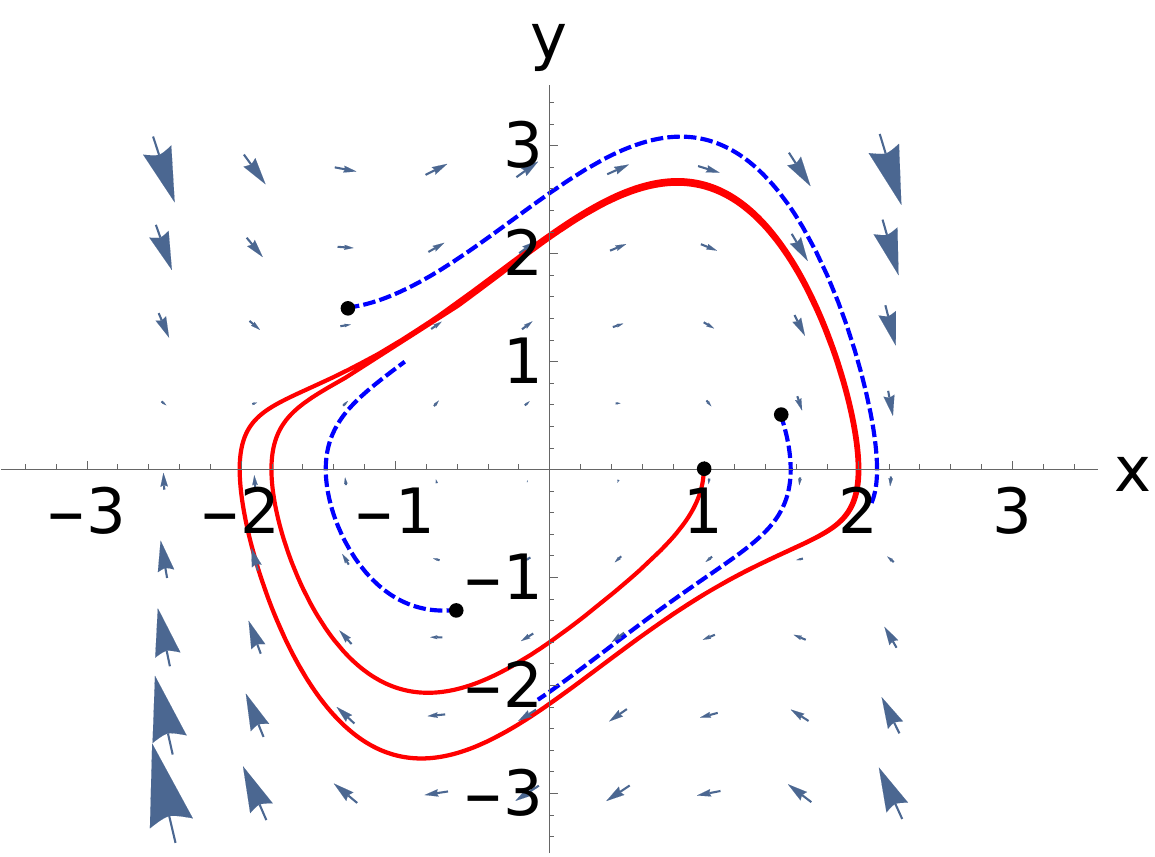}
        \caption{}
    \end{subfigure}
    \hfill
    \begin{subfigure}{0.5\linewidth}
        \includegraphics[width=\linewidth]{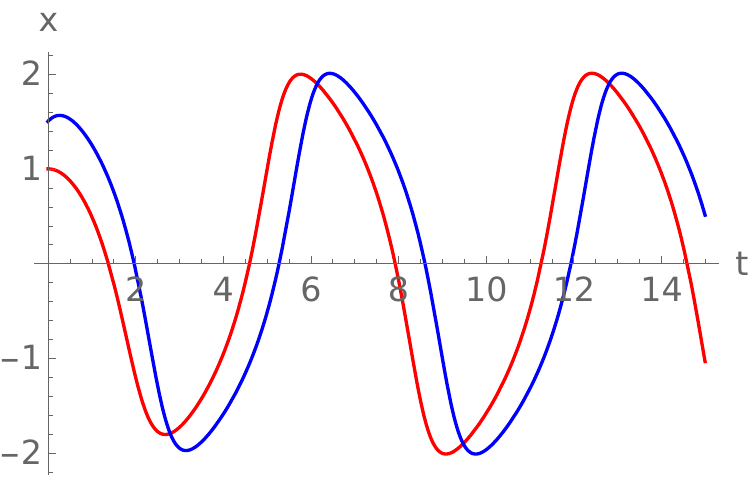}
        \caption{}
    \end{subfigure}
    
    \caption{For the van der Pol oscillator, Eq.~\eqref{eq:vdP split}, (a) shows trajectories traced out in the $(x,y)$-plane starting from different initial conditions, overlaid on the vector field $(\dot x, \dot y)$, with $\mu = 1$. (b) The variation of $x$ with time and its approach towards a periodic motion, for two different initial conditions.}
    \label{fig:vdP}
\end{figure}

Systems exhibiting limit cycles can be broadly classified as oscillators, since they show periodic motion. An important example is the van der Pol oscillator, which has played an important role in the evolution of the field of nonlinear dynamics. Balthasar van der Pol was working on the theory of triode oscillations, which led him to an equation of the form 
\begin{eqnarray}
    \ddot x + \mu(x^2 -1)\dot x + x = 0,
    \label{eq:van der Pol}
\end{eqnarray}
with $\mu\ge 0$. For $\mu=0$ or for the special case $|x|=1$, the system reduces to a simple harmonic oscillator and does not yield a genuine limit cycle. For $\mu > 0$, the second term gives rise to a nonlinear damping and generates a limit cycle. If $|x|>1$, the damping is positive (energy gets dissipated), while $|x|<1$ gives rise to negative damping (energy gets absorbed). The above equation can be split into two first-order equations as follows:
\begin{eqnarray}
    \dot x &=& y, \nonumber\\ 
    \label{eq:vdP split}\\
    \dot y &=& -\mu(x^2 -1)y - x.\nonumber
\end{eqnarray}
Let us graphically study the limit cycle of this oscillator. Figure~\ref{fig:vdP}(a) shows the trajectories traced out starting from different initial conditions (marked by black solid circles). The presence of the limit cycle is clear from the figure, with each of the trajectories approaching this closed orbit asymptotically in time. The red trajectory has been generated by running the dynamics for a longer time in order to reveal the approximate outline of the limit cycle. In Fig. \ref{fig:vdP}(b), we have shown the variation with time of the variable $x$, starting from two different initial conditions. It can be readily seen that the periodicity in both cases approach the same value, although a phase difference is present. 
\begin{figure}
    \centering
    \includegraphics[width=0.32\linewidth]{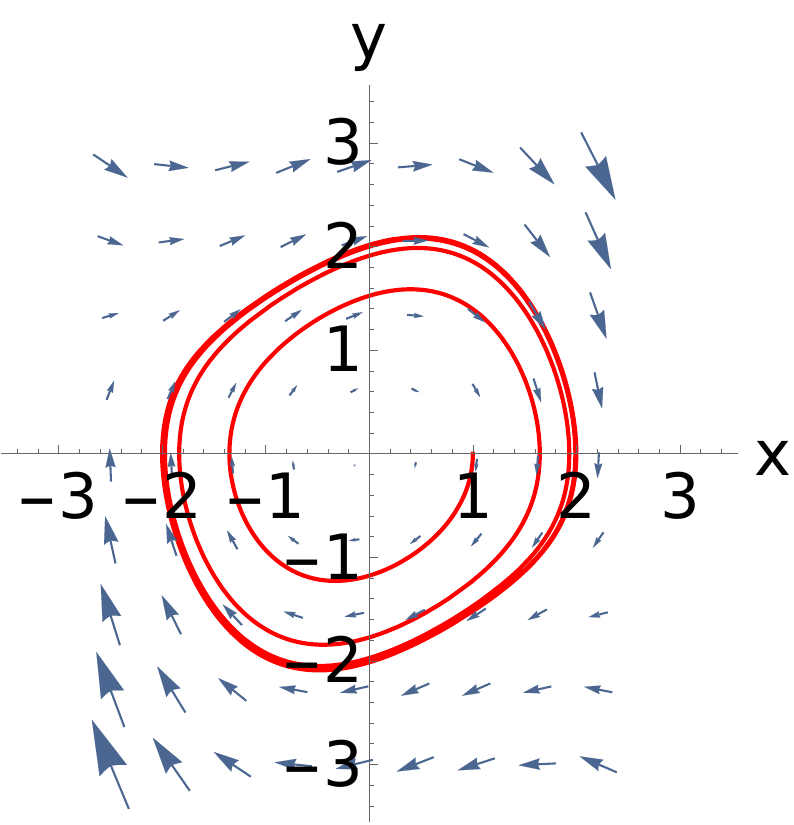}\hfill
    \includegraphics[width=0.32\linewidth]{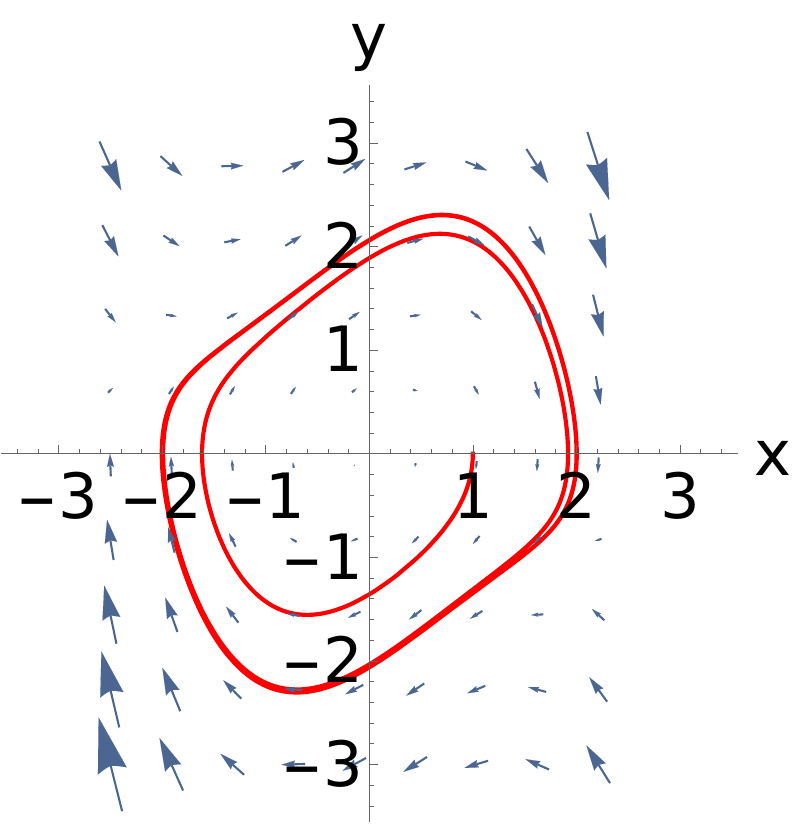}\hfill
    \includegraphics[width=0.32\linewidth]{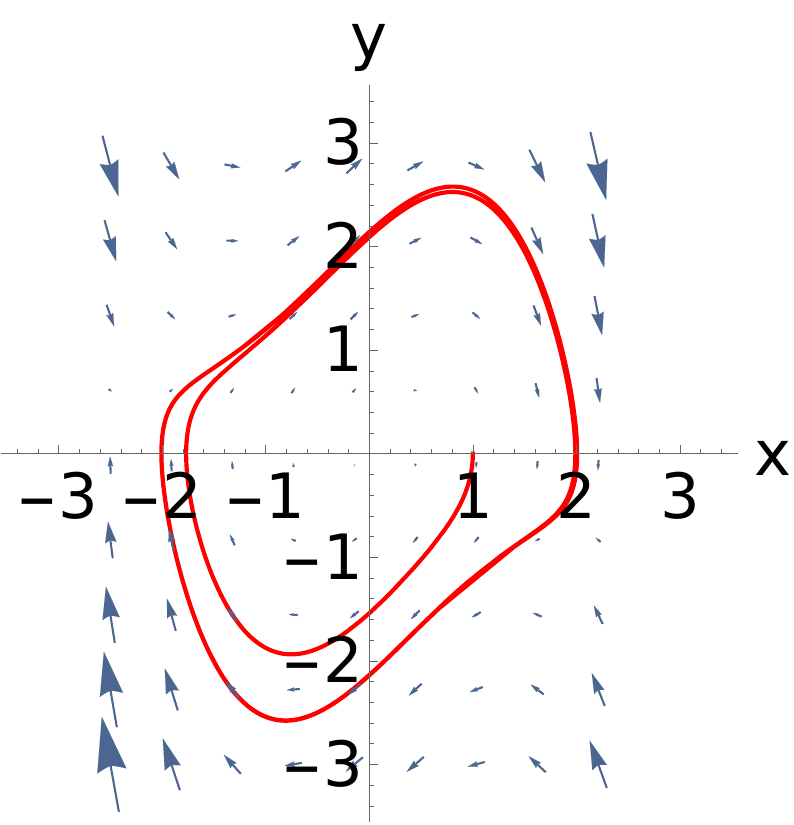}
    \caption{For the van der Pol oscillator, Eq.~\eqref{eq:vdP split}, the figure shows the change in the shape of the limit cycle for different values of the parameter $\mu$ (left to right: $\mu=0.3,~0.6,~0.9$). In each case, the dynamics has been run for a long-enough time in order to reveal the approximate outline of the limit cycle. As $\mu$ increases, the shape gets increasingly distorted from that of a circle, and there is an increase in the area under the cycle.}
    \label{fig:vdP different mu}
\end{figure}
Figure \ref{fig:vdP different mu} shows the change in the shape of the limit cycle with change in the parameter $\mu$ appearing in the damping term; in each case, the dynamics has been run for a long-enough time in order to reveal the approximate outline of the limit cycle. With increase in $\mu$, the orbit deviates more and more from a circular shape, and the area under the limit cycle expands. In particular, there is a sharp rise in the $y$-range covered by the cycles, in comparison to the $x$-range.

\subsection{Generic nonlinear oscillators}
We discuss here how limit cycles may arise in the dynamics of a single generic nonlinear oscillator. We consider a nonlinear oscillator subject to linear gain and nonlinear saturation, for which one can write down for the complex amplitude $\alpha$ the time evolution in the so-called normal form, as 
\begin{align}
\dot{\alpha}=-i\omega\alpha+
\alpha\left(\kappa_1-2\kappa_2|\alpha|^2\right).
\label{eq:amplitude}
\end{align}
Here, both the quantities $\kappa_1>0$ and $\kappa_2$ are real.  Let us now understand the various terms on the rhs. The first term, $-i\omega\alpha$, describes the free oscillation of the system at angular frequency $\omega$. If no other term is present in the dynamics, the solution is simply $\alpha(t)=\alpha(0)e^{-i\omega t}$, which corresponds to a constant-amplitude oscillation with a uniformly-rotating phase. Including next the linear gain term, $\alpha \kappa_1$, the amplitude grows exponentially in time, $|\alpha(t)|\propto e^{\kappa_1 t}$,
which is physically unrealistic, and which can be arrested by including nonlinear effects. The simplest nonlinear correction to the factor $\kappa_1$ that preserves the phase symmetry
$\alpha \rightarrow \alpha e^{i\phi}$ can 
depend only on the intensity $|\alpha|^2$. Consequently, the lowest-order nonlinear contribution is cubic, $-2\kappa_2|\alpha|^2\alpha$, where $\kappa_2$ characterizes the strength of the nonlinear damping. Incorporating this saturation term yields Eq.~\eqref{eq:amplitude}.

Writing $\alpha=re^{i\theta}$, where both $0\le r\le 1$ and $\theta \in [0,2\pi)$ are real, 
the dynamics~\eqref{eq:amplitude} separate into independent amplitude and phase equations,
\begin{align}
\dot{r}= r\left(\kappa_1-2\kappa_2r^2\right),\qquad \dot{\theta}= -\omega.
\end{align}
The phase rotates uniformly at the oscillator frequency, while the the steady-state amplitude satisfies
\begin{equation}
r\left(\kappa_1-2\kappa_2r^2\right)=0,
\end{equation}
giving the fixed points 
\begin{equation}
r_\mathrm{fp}=0,
\qquad
r_\mathrm{fp}=r_0=\sqrt{\frac{\kappa_1}{2\kappa_2}}.
\end{equation}
With both $\kappa_1>0$ and $\kappa_2>0$, the nonzero solution is stable and corresponds to limit-cycle oscillations with amplitude $\kappa_1/(2\kappa_2)$ and frequency $\omega$. In passing, we note that Eq.~\eqref{eq:amplitude} is mathematically equivalent to the Stuart--Landau equation, which is the universal amplitude equation describing oscillators near a supercritical Hopf bifurcation.

\subsection{Characterizing a stable limit cycle}
Having discussed the occurrence of a stable limit cycle, we now discuss how one may characterize such a limit cycle in the context of the dynamics of a generic autonomous dynamical system comprising many interacting
dynamical variables $\{x_\alpha\}_{1\le \alpha \le d};~d \gg 1$, with a
dynamics given by
\begin{align}
\frac{dx_\alpha}{dt}=F_\alpha(x_1,x_2,\ldots,x_d).
\label{eq:generic-dynamical-system}
\end{align}
By autonomous, it is meant that the functions $F_\alpha$ have no explicit time dependence, which implies that if $\{x_\alpha(t)\}$ is a solution, so is $\{x_\alpha(t+t_0)\}$ for
any $t_0$, i.e., the choice of the origin of
time is irrelevant. Limit cycles are a
particular class of solutions corresponding to one-dimensional periodic orbits in
the phase space. 

Consider a stable limit cycle with period $\mathcal{T}$. The length of the orbit that is traversed in the $d$-dimensional phase space in time $t$ equals $s(t)=\int_0^t dt'~\sqrt{ \sum_{\alpha=1}^d\left(dx_\alpha/dt\right)^2(t')}$, where the origin of time being irrelevant means that one has complete arbitrariness in choosing the origin from which
the orbit length is measured. Note that the time rate of variation of
$s$ is not required to be a constant along the limit cycle. Nevertheless, one may transform to a new variable $\theta=\theta(s)$, such that its 
time rate of variation along the limit cycle is a constant called the
 natural frequency $\omega \equiv 2\pi/\mathcal{T}$ of the cycle. The transformation reads as
\begin{align}
 \theta(s)\equiv \frac{2\pi}{\mathcal{T}}\int_0^s~\frac{ds'}{\left[\left(\frac{ds}{dt}\right)(s')\right]} = \frac{2\pi}{\mathcal{T}}\int_0^{t(s)} dt(s'),   
\end{align} implying $d\theta/dt=(d\theta/ds)(ds/dt)=2\pi/\mathcal{T}=\omega$. Moreover, defining $s_0\equiv s(\mathcal{T})$,which is equivalent to the condition $t(s_0) = \mathcal{T}$,
one has $\theta(s_0)=2\pi$. This implies that at the end of one time period $\mathcal{T}$, corresponding to one complete
traversal of the periodic orbit, the value of $\theta$ increases by $2\pi$. We conclude that a limit-cycle
oscillator is completely characterized by a phase $\theta \in [0,2\pi)$ that changes uniformly in time with period $\mathcal{T}$ and frequency
$\omega$, as
\begin{align}
\frac{d\theta}{dt}=\omega.
\label{eq:dtheta-dt}
\end{align}

Since we may associate a unique value of the phase with each point
on the limit cycle, we have $\theta=\theta(\{x_{\alpha,0}\})$, where the $\{x_{\alpha,0}\}$ denotes values of $\{x_\alpha\}$ on the limit cycle.
From Eq.~\eqref{eq:dtheta-dt}, we find that $\theta$ is a neutrally-stable variable: any (small)
perturbation neither grows nor decays in time. By contrast, the amplitude of oscillations has a definite
stable value on the limit cycle; any (small) perturbations in a direction transverse to
the phase will decay in time.

\begin{figure}[!htbp]
\centering
\includegraphics[width=0.5\linewidth]{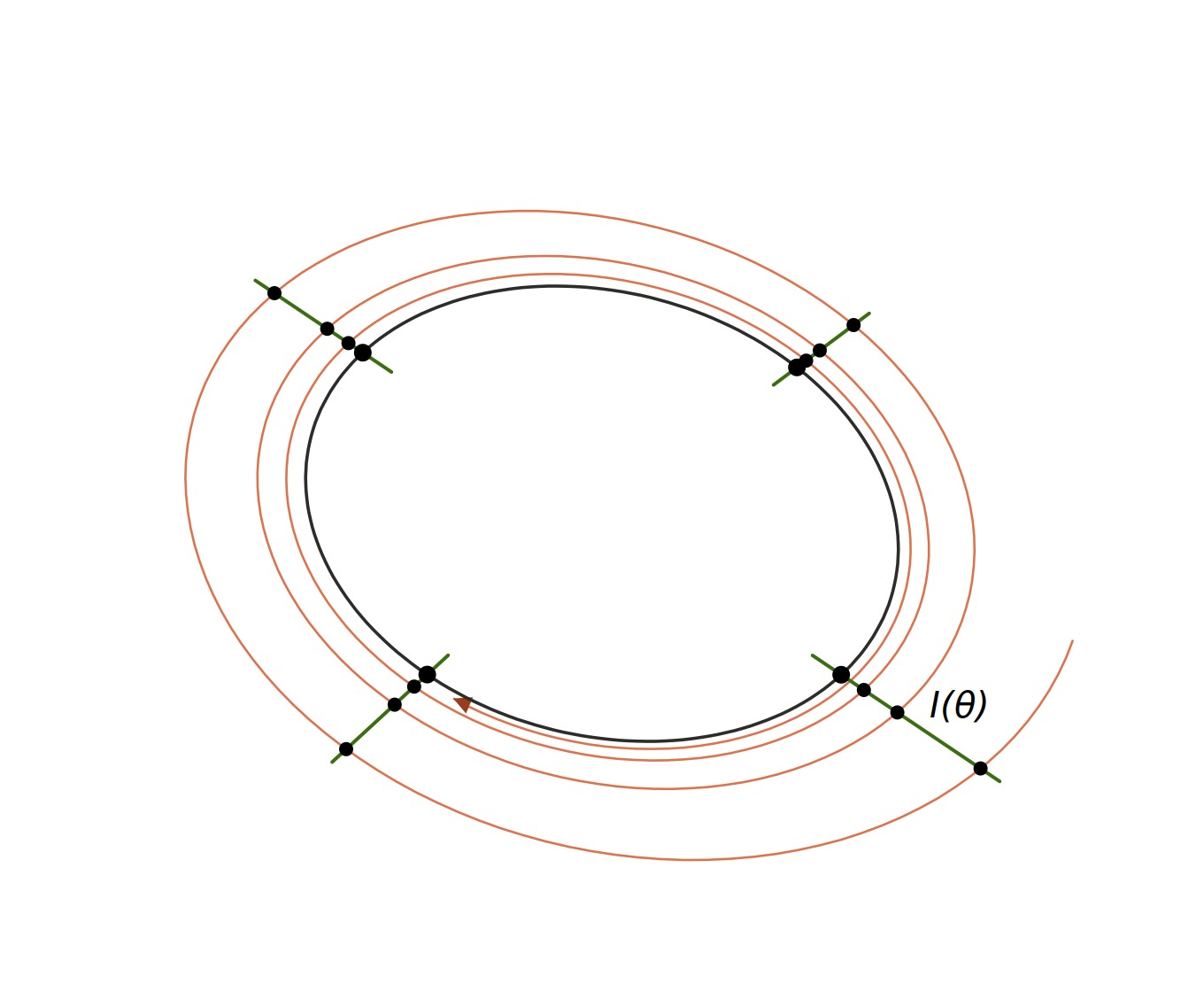}
\caption{Construction of isochrones $I(\theta)$. An initial
phase-space point sufficiently close to the limit cycle in course of time traces out an orbit that asymptotically approaches the limit cycle. As the point
traverses this orbit, we observe its
successive positions in time only stroboscopically, i.e., at times $t=k\mathcal{T};
~k=1,2,3,\ldots$. The limit of this sequence of points as $k \to \infty$ is a point on the
limit cycle. The aforementioned successive positions are marked by black filled circles. }
\label{fig:isochrone}
\end{figure}

A phase description such as above applies not only to the limit cycle but also to close by 
orbits. Consider an initial
phase-space point sufficiently close to the limit cycle. Since the limit cycle is stable, the point in course of time traces out an orbit that asymptotically approaches the limit cycle. As the point
traverses this orbit, let us observe its
successive positions in time only stroboscopically, i.e., at times $t=k\mathcal{T};
~k=1,2,3,\ldots$. Evidently, the limit of this sequence of points as $k \to \infty$ is a point on the
limit cycle, which has a particular value of the phase $\theta$. One can then associate this latter phase value with the sequence of points that are said to lie on a
$(d-1)$-dimensional hypersurface $I(\theta)$ called an isochrone. More precisely, isochrones are sets of initial conditions that asymptotically converge to the same phase point on the limit cycle. Figure~\ref{fig:isochrone} illustrates the construction of isochrones. Using the construction of isochrones, we may
associate a phase $\theta$ with each point of the phase space lying in
close proximity of the limit cycle, and hence, Eq.~\eqref{eq:dtheta-dt} remains valid also in close neighborhood of the limit cycle. 

\section{Driven classical limit cycles}
\label{sec:sec2}
Consider a limit-cycle oscillator, whose phase changes according to Eq.~\eqref{eq:dtheta-dt}, and which is subject to a periodic force with frequency $\Omega$ and amplitude $\epsilon \ll 1$. Up to linear order in $\varepsilon$, the dynamics reads as 
\begin{align}
\frac{d\theta}{dt} = \omega - \epsilon G(\theta,\psi),
\label{eq:dtheta-dt-2}
\end{align}
with $\psi = \Omega t$. The function $G$ depends on the form of the limit cycle and the periodic forcing, and which is $2\pi$-periodic in both its arguments. 
In general, the two frequencies $\omega$ and $\Omega$ are not integer multiples of one another. 

The negative sign in the coupling term implies that if the drive phase lags behind the oscillator phase, then $G(\theta,\psi)$ will be positive and the coupling will tend to decrease the instantaneous frequency of the oscillator below its natural frequency $\omega$.
%The drive, however, always obeys $\psi=\Omega$, since by definition it is an externally controlled perturbation that retains its exact time-dependence. 
%
Introducing 
\begin{align}
\Delta \theta \equiv \theta - \psi 
\end{align}
as the difference between the phases of the limit-cycle oscillator and the external forcing, and
\begin{equation}
    \Delta\omega = \omega-\Omega
\end{equation}
as the detuning frequency,  one gets 
\begin{equation}
    \frac{d\Delta\theta}{dt} = \Delta\omega - \epsilon G(\theta,\psi).
    \label{eq:Delta theta dot}
\end{equation}
Now, assume that we change the reference for measuring the phase angle by $\alpha$, such that $G(\theta,\psi)\to G(\theta_1,\psi_1) = G(\theta+\alpha,\psi+\alpha)$. This keeps the coupling invariant only if $G(\theta,\psi)$ is a function of the relative phase $(\theta-\psi) = (\theta_1-\psi_1) = \Delta\theta$. In short, $G(\theta,\psi) =G(\Delta\theta)$. Any other functional dependence will make the coupling show a spurious dependence on the arbitrary reference phase $\alpha$. 
 
\begin{figure}[!h]
    \centering
    \includegraphics[width=0.6\linewidth]{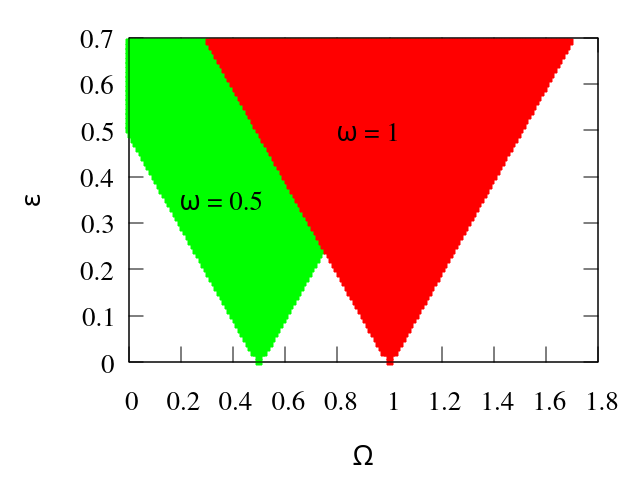}
    \caption{Arnold tongues (synchronization region in the $\varepsilon$ vs $\Omega$ plane) generated from the Adler equation, where the natural frequencies are: $\omega = 1$ (red region) and $\omega = 0.5$ (green region).}
    \label{fig:arnold}
\end{figure}

Next, we expand $G(\Delta\theta)$ in a Fourier series as
\begin{align}
    G(\Delta\theta) &= \frac{1}{2}a_0 + \sum_{n=1}^\infty \left[a_n\cos(n\Delta\theta) +  b_n\sin(n\Delta\theta)\right],
    \label{eq:Fourier}
\end{align}
where 
\begin{align}
    a_0 &= \frac{1}{\pi}\int_{-\pi}^\pi d(\Delta\theta) ~G(\Delta\theta), \nonumber\\
    a_n &= \frac{1}{\pi}\int_{-\pi}^\pi d(\Delta\theta) ~G(\Delta\theta) \cos(n\Delta\theta),\nonumber\\
    b_n &= \frac{1}{\pi}\int_{-\pi}^\pi d(\Delta\theta) ~G(\Delta\theta) \sin(n\Delta\theta).
\end{align}
Typically, the driven oscillator tends to synchronize with the external drive in the asymptotic time limit. In the resonant case $\Delta\omega=0$, the coupling function $G(\Delta\theta)$ must have the same sign as $\Delta\theta$ in the neighborhood of the synchronized state $\Delta\theta=0$, ensuring that the coupling acts to reduce the phase difference. Thus, the leading contribution to $G$ is expected to be odd in $\Delta\theta$, i.e., $G(-\Delta\theta) \simeq -G(\Delta\theta)$. Neglecting higher-order even components, we retain only the dominant odd part of the Fourier expansion and set $a_0=a_n=0$. The Fourier expansion of $G$ now reduces to 
\begin{eqnarray}
    G(\Delta\theta) &=& \sum_{n=1}^\infty b_n\sin(n\Delta\theta).
\end{eqnarray}
Next, we take the time-average of both sides of Eq.~\eqref{eq:Delta theta dot}, to obtain
 \begin{align}
     \left\langle \frac{d\Delta\theta}{dt}\right\rangle &= \frac{1}{\tau} \int_{0}^{\tau}dt~\frac{d\Delta \theta}{dt} 
     = \Delta\omega -  \frac{\varepsilon}{\tau} \sum_{n=1}^\infty b_n \int_0^\tau dt\sin(n\Delta\theta).
     \label{eq:time-average}
 \end{align}
 We choose the time $\tau$  such that $\tau \ll (n\Delta\omega)^{-1}$, which is long compared with fast oscillations, but short compared with slow phase evolution. Under this choice, only the lower frequency modes contribute to the time averaging, and the higher frequency modes can be safely ignored. A fairly good approximation is to retain only the $n=1$ term of the summation. After dropping the angular brackets from the left hand side for convenience, Eq.~\eqref{eq:time-average} then gives us 
 \begin{align}
     \frac{d\Delta\theta}{dt} &= \Delta\omega - \frac{b_1\varepsilon}{\tau} \int_0^\tau dt \sin(\Delta\theta).
\end{align}
We note that $\sin(\Delta\theta)$ does not change appreciably in the time interval $\tau$, so that it can be taken out of the integral. Now, on redefining $b_1\varepsilon \to \varepsilon$,  we readily obtain the so-called Adler equation:
\begin{equation}
    \frac{d\Delta\theta}{dt} = \Delta\omega - \varepsilon \sin(\Delta\theta).
    \label{eq:Adler}
\end{equation}
We conclude that provided the right hand side is negative, the phase lag $\Delta \theta$ will approach zero in course of time, and there will be synchrony between the phase of the limit-cycle oscillator and that of the external forcing. In this case, there will be phase locking, in the sense that the phase $\theta$ will just follow the phase of the forcing, $\theta=\Omega t+\mathrm{constant}$, regardless of $\omega$. One also has frequency entrainment, i.e., the observed frequency of the oscillator is the same as the forcing frequency $\Omega$. The synchronized (phase-locked) state corresponds to a fixed point $\Delta \theta^\ast$ of Eq.~\eqref{eq:Adler}, for which the phase difference remains constant in time. Thus, we require the left hand side of the equation to be zero, yielding
\begin{align}
\sin(\Delta\theta^{\ast})
=
\frac{\Delta\omega}{\varepsilon}.
\label{eq:locking_condition}
\end{align}
A real solution to Eq.~\eqref{eq:locking_condition} exists only when $\left|\Delta\omega/\varepsilon\right|\le 1$,
which yields the condition for phase locking,
\begin{align}
|\Delta\omega|
\le
\varepsilon.
\label{eq:locking_range}
\end{align}
In the $(\Omega,\epsilon)$-plane, the above equation defines a region in which synchronization is possible, and this region is the so-called $1{:}1$ Arnold tongue (the region has somewhat the shape of a tongue!), see Fig.~\ref{fig:arnold}, left panel. The dependence of the observed frequency $\omega_\mathrm{obs}$ on the external frequency displays a synchronization plateau, as shown in Fig.~\ref{fig:arnold}, right panel. The quantity $\omega_\mathrm{obs}$ is obtained by taking the time average of the quantity $d\theta/dt$ at long times. 

\subsection{The Arnold Tongue}

\paragraph{1:1 locking}
Consider the Adler equation, Eq. \eqref{eq:Adler}. The frequency  of the driven oscillator depends  on both $\varepsilon$ and $\Delta\omega$. Recall that $\omega$ is the natural frequency of the oscillator, $\dot\theta(t)$ is its instantaneous frequency, $\Omega$ is the frequency of the external drive, $\varepsilon$ is the coupling strength between the oscillator and the drive, and $\Delta\omega = \omega - \Omega$. 
We have a transient regime, where the instantaneous frequency of the oscillator changes with time. Let us assume that beyond a large enough time interval $\tau_\text{trans}$, this frequency becomes equal to the drive frequency $\Omega$ and thereby becomes independent of time. When $\dot\phi = \dot\theta - \Omega$ vanishes on time-averaging, i.e., 
\begin{equation}
    \lim_{\tau\to\infty} \frac{1}{\tau} \int_{\tau_\text{trans}}^{\tau_\text{trans}+\tau} \dot\phi(t)dt = 0,
    \label{eq:time average}
\end{equation}
the system's oscillation frequency is said to have got ``locked'' to that of the drive. The Adler's equation is an example of ``1:1 locking'', because under this condition, we have $\omega/\Omega = 1$. More general locking conditions are discussed in the next subsection.  This locking takes place only in a finite region on the $(\Omega,\varepsilon)$ plane. This region is often referred to as the Arnold Tongue. Figure \ref{fig:arnold} shows two Arnold tongues, when the natural frequency of the oscillator is $\omega=1$ (red region) and $\omega = 0.5$ (green region). When there is no coupling ($\epsilon=0$), the oscillator retains its natural frequency, so the tip of the tongues are found to appear at $\omega$. The total time of observation was up to $t=500$, out of which the range $t=300$ to $t=500$ has been chosen for the time averaging given by Eq. \eqref{eq:time average}. We find that as the coupling $\varepsilon$ with the external drive increases, so does the range of the drive frequency $\Omega$ over which a frequency locking can be sustained.

\paragraph{$\bm{p:q}$ locking}
Systems with 1:1 locking are special cases of ``$p:q$ locking''. The latter happens in systems where the locking of oscillator and drive takes place when $\omega/\Omega$ is a rational number $p/q$ (ratio of two integers).  As an example, let us get back to the driven van der Pol oscillator (see Eq. \eqref{eq:van der Pol}):
\begin{eqnarray}
    \ddot x + x + \mu f(x,\dot x) = 0,
    \label{eq:vdP1}
\end{eqnarray}
where we have defined $f(x,\dot x) = (x^2 -1)\dot x$.
It is often simpler to obtain approximate analytical expressions for the locking regions by means of Method of Averaging \cite{Bonnin2010}. We briefly describe the method below.

Using the trial solution $x(t) = A(t) \cos(t + \alpha(t))$ in Eq. \eqref{eq:vdP1}, and imposing the uniqueness condition on the solution $x(t)$, one arrives at the following nonlinear time dependences of $A(t)$ and $\alpha(t)$:
\begin{align}
    \dot A(t) &= \mu F(A(t)); \nonumber\\
    \dot\alpha(t) &= \mu G(A(t))/A(t),
\end{align}
where
\begin{align}
    F(A(t)) &= \frac{1}{2\pi} \int_0^{2\pi} g(A(t),\alpha(t)) \sin(t + \alpha(t)) d\alpha; \nonumber\\
    G(A(t)) &= \frac{1}{2\pi} \int_0^{2\pi} g(A(t),\alpha(t)) \cos(t + \alpha(t)) d\alpha,
\end{align}
where $g(A,\alpha) = f(x,\dot x)$.
Thus, we have the expressions for the amplitude and frequency in terms of quantities that are averaged over the phase $\alpha(t)$.
\begin{figure}
    \centering
    \includegraphics[width=0.7\linewidth]{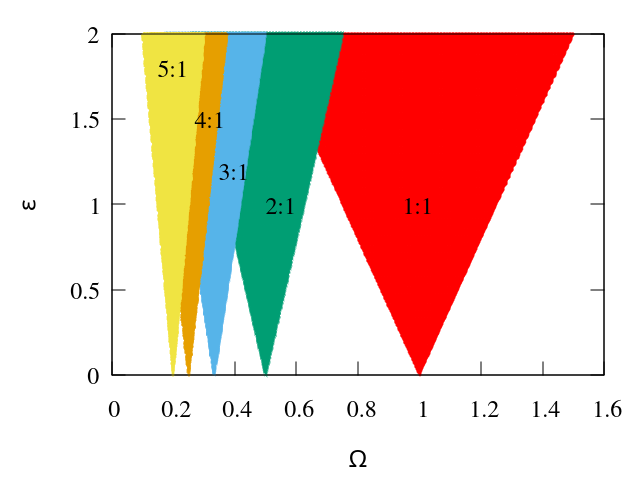}
    \caption{Approximate solutions for the Arnold tongue for the van der Pol oscillator, obtained by the Method of Averaging, for various values of $p:q$, with $q=1$ for convenience.}
    \label{fig:vdP_pq}
\end{figure}
For the initial conditions $(A_0, \alpha_0)$, one arrives at the solutions
\begin{align}
    A(t) &= \frac{2A_0}{\sqrt{A_0^2 + (4 - A_0^2) e^{-\mu t}}}; \nonumber\\
    \alpha(t) &= \alpha_0.
    \label{eq:A and Omega for vdP}
\end{align}

Now, let $\phi(t)$ be the free oscillation phase. The trajectory $\textbf{Q}(t) = (x, y)^\text{T}$ is mapped to the phase angle $\phi(\textbf{Q})$, such that $\dot\phi = 1$. If the equation $\dot{\textbf{Q}} = \textbf{F}(\textbf{Q})$ (see Eq. \eqref{eq:compact}) changes to $\dot{\textbf{Q}} = \textbf{F}(\textbf{Q}) + \varepsilon \textbf{G}(\textbf{Q})$ under a weak perturbation, then we obtain
\begin{equation}
    \dot\phi \simeq 1 + \varepsilon\nabla\phi(\textbf{Q}_0)\cdot \textbf{G}(t),
    \label{eq:phidot}
\end{equation}
where $\textbf{Q}_0(t) = \lim_{\varepsilon\to 0}\textbf{Q}(t)$. 
Let the value of $A(t)$ converge to $A_\infty$ in asymptotic time:  $A_{\infty} = \lim_{t\to\infty} A(t)$. The phase trajectory $\textbf{Q}(t)$ can be written in terms of this asymptotic amplitude by the mapping $\textbf{Q}_0(t) = \textbf{M}(\phi, A_\infty)$. Eq. \eqref{eq:phidot} can then be rewritten as
\begin{align}
    \dot\phi \simeq 1 + \varepsilon\nabla\phi(\textbf{M}(\phi,A_\infty))\cdot \textbf{G}(t).
    \label{eq:phidot approx}
\end{align}
Around the region $\dot\phi/\Omega \approx p/q$, let us define $\psi(t) = \phi(t) - (p/q)\Omega t$  to be the deviation of the free phase from the forcing phase. Taking time derivatives of both sides and using Eq. \eqref{eq:phidot approx}, we get
\begin{equation}
    \dot\psi = \nu + \varepsilon \nabla\phi(\textbf{M}(\phi,A_\infty))\cdot \textbf{G}(t),
    \label{eq:psidot for vdP}
\end{equation}
where $\nu = 1 - (p/q)\Omega$. 

The phase function $\phi(t)$ is related to the phase space coordinates $(x,y)$ where $y=\dot x$, via the relations
\begin{align}
    x(t) &= A(t)\cos\phi(t); \nonumber\\
    y(t) &= \dot A(t)\cos\phi(t) - A(t)\sin\phi(t) \dot\phi(t) \approx - A(t)\sin\phi(t),
\end{align}
assuming slow temporal variation of amplitude, $\dot A(t)\sim O(\mu)$, and $\dot\phi \approx 1$. Thus, we have $\phi = -\tan^{-1}(y/x)$. The gradient of phase is then given by
\begin{align}
    \nabla\phi &= \left( \frac{\partial\phi}{\partial x}, \frac{\partial\phi}{\partial y} \right)^\text{T}
    = \left( \frac{y}{x^2 + y^2},  \frac{-x}{x^2 + y^2}\right)^\text{T} \nonumber\\
    &= \left( y/A^2, -x/A^2\right)^\text{T} = \frac{1}{A}(\cos\phi, -\sin\phi)^\text{T}.
\end{align}
From Eq. \eqref{eq:A and Omega for vdP}, we note that $A_\infty = 2$. Substituting this for $A$ in the above equation gives
\begin{equation}
    \nabla\phi = \frac{1}{2}(\cos\phi, -\sin\phi)^\text{T}.
    \label{eq: grad phi for vdP}
\end{equation}
Also, we choose
\begin{equation}
    \textbf{G}(t) = (\cos(p\Omega t),0)^\text{T}.
    \label{eq:G for vdP}
\end{equation}
Considering $q=1$ for convenience, we find $\phi = \psi + p\Omega t$ and $\nu = 1 - p\Omega$. 
Next, we use Eq. \eqref{eq: grad phi for vdP} and \eqref{eq:G for vdP} in Eq. \eqref{eq:psidot for vdP} to obtain the equation \cite{Bonnin2010}:
\begin{equation}
    \dot\psi(t) = \nu - (\varepsilon/4) \cos\psi.
    \label{eq:vdP MoA}
\end{equation}
Here, we have used the fact that $\langle\cos^2(p\Omega t)\rangle = 1/2$ to arrive at the factor of $1/4$ in the second term on the right-hand side. The phase locking region lies in the range $-(\varepsilon/4) \le \nu \le (\varepsilon/4)$. 
These regions, for various values of $p$, have been plotted in Fig. \ref{fig:vdP_pq}. Setting $\dot\psi = 0$ in Eq. \eqref{eq:vdP MoA}, we obtain the relation 
\begin{equation}
    \varepsilon = 4\sec\psi (1 - p\Omega).
\end{equation}
Thus, for each value of $p$, the boundaries of the corresponding Arnold tongue have slopes of $\pm 4p\sec\psi$.

\section{Classical synchronization}
\label{sec:sec3}
As is clear from the above discussion, an external drive influences an oscillator to oscillate at a frequency that approaches the frequency of the drive, a phenomenon known as frequency locking. In the preceding example, the drive frequency was assumed to be fixed externally. Now consider three independent oscillators that are weakly coupled, but strongly enough for each oscillator to respond to the influence of the others. An example would be three simple pendulums suspended from a common support. Suppose two of the oscillators initially have the same phase, while the third is out of phase with them. Over time, it is generally observed that the oscillators tend toward a common phase. The force exerted on any oscillator by the others acts as an effective drive, although in this case the drive is not externally imposed; its phase and frequency evolve dynamically as the oscillators interact. Each oscillator therefore responds not to a fixed external signal but to the collective behavior of the system.

An oscillator whose phase differs substantially from the collective phase experiences a larger effective influence from the others and consequently tends to adjust its phase more rapidly. Oscillators that are already close to the collective phase experience a smaller correction and therefore evolve more gradually. As a result, an oscillator that is initially far from the synchronized state often approaches it more quickly than those already near it. For example, consider the initial phases $\theta_1=\theta_2=0$ and $\theta_3=\pi$. Oscillator 3 is maximally out of phase with oscillators 1 and 2 and therefore experiences the strongest tendency to adjust toward the collective state. Consequently, oscillator 3 tends to catch up with the other two, leading the system toward synchronization.

Similar behavior can occur in larger populations of oscillators whose natural frequencies differ slightly from one another. More generally, such synchronization phenomena arise in a wide variety of systems of interacting oscillators and, equivalently, in interacting limit cycles associated with those oscillators. When the coupling is sufficiently strong relative to the spread in natural frequencies, the oscillators may evolve toward a phase-locked state characterized by fixed phase differences between them. Such a phase-locked state is necessarily frequency-locked as well, since any persistent difference in frequency would cause the phase differences to drift with time and thereby destroy phase locking.

\subsection{Two interacting limit cycles}
\label{sec:2-interacting-classical}
Let us first discuss the case of synchronization in the dynamics of two interacting limit cycles. To this end, we consider the case of two coupled oscillators that are individually of the form of Eq.~\eqref{eq:amplitude}. The dynamics of the coupled system is given by
\begin{align}
&\dot\alpha_1= -i\omega_1\alpha_1 + \alpha_1(\kappa_1-2\kappa_2|\alpha_1|^2) + V(\alpha_2-\alpha_1), \label{eq:2osc-1}\\
&\dot\alpha_2= -i\omega_2\alpha_2 + \alpha_2(\kappa_1-2\kappa_2|\alpha_2|^2) + V(\alpha_1-\alpha_2). \label{eq:2osc-2}
\end{align}
Substituting $\alpha_n = r_n e^{i\theta_n}$ and defining $\theta \equiv \theta_2-\theta_1$, one gets
\begin{align}
&\dot r_1= r_1(\kappa_1-2\kappa_2 r_1^2) - V(r_1 - r_2\cos\theta), \\
&\dot\theta_1 = -\omega_1 + V\frac{r_2}{r_1}\sin\theta,\\
&\dot r_2= r_2(\kappa_1-2\kappa_2 r_2^2) - V(r_2 - r_1\cos\theta), \\
&\dot\theta_2= -\omega_2 - V\frac{r_1}{r_2}\sin\theta.
\end{align}
On subtracting, we get
\begin{align}
\dot\theta = -\Delta - V\left(\frac{r_1}{r_2} + \frac{r_2}{r_1}\right)\sin\theta,
\qquad \Delta \equiv \omega_2-\omega_1.
\label{eq:2classical}
\end{align}

A phase-locked state requires $\dot r_1=\dot r_2=\dot\theta=0$ simultaneously.
Setting $\dot\theta=0$ gives $\sin\theta = -\Delta/[V\left(r_1/r_2+r_2/r_1\right)]$, which has a real solution only if the right-hand side has magnitude $\le 1$.
Since $r_1/r_2+r_2/r_1 \ge 2$ (with equality at $r_1=r_2$), the
most favorable case is $r_1\approx r_2\approx r_0$. Let us now argue that such a situation is achieved when one has $V\ll\kappa_1$, which corresponds to the case when the coupling is much weaker than the intrinsic amplitude relaxation of each oscillator. To see why this limit implies $r_1\approx r_2\approx r_0$, consider first the uncoupled dynamics ($V=0$),
\begin{align}
\dot r = r(\kappa_1-2\kappa_2 r^2).
\end{align}
The stable fixed point is
\begin{align}
r_0=\sqrt{\frac{\kappa_1}{2\kappa_2}}.
\end{align}
The relaxation rate of amplitude fluctuations is obtained by linearizing about $r_0$. Defining $f(r) \equiv r(\kappa_1-2\kappa_2 r^2)$, one finds $f'(r)=\kappa_1-6\kappa_2 r^2$. Evaluating at the fixed point, one gets $f'(r_0)=-2\kappa_1$. Thus, a Taylor expansion of $f(r)$ around $r=r_0$, truncated at the linear order, shows that the deviations of the amplitude from $r_0$ decay at a rate of order $2\kappa_1$. Now include the coupling, if we have $V \ll \kappa_1$, the coupling acts as only a small perturbation compared with the intrinsic restoring force that keeps each oscillator amplitude at $r_0$. 
Therefore, when $V\ll\kappa_1$, the amplitudes remain nearly equal, $r_1\simeq r_2\simeq r_0$. Consequently, Eq.~\eqref{eq:2classical} reduces to the Adler equation,
\begin{align}
\dot\theta
=
-\Delta
-
2V\sin\theta,
\end{align}
yielding the phase-locking condition
\begin{align}
|\Delta|\le 2V.
\end{align}
In other words, in the $(\Delta,V)$-plane, one has phase locking inside the region $V \ge V_c(\Delta) = \frac{|\Delta|}{2}$ and no phase locking outside the region. For $V$ comparable to $\kappa_1$, the amplitudes $r_1,r_2$ are pulled
away from $r_0$, and the boundary between the phase locked and and the phase unlocked region must be found numerically.

\subsection{Many interacting limit cycles}
\label{sec:many-classical}
Consider a collection of limit-cycle oscillators, with the $i$-th oscillator characterized by a natural frequency $\omega_{i}$ and a phase $\theta_i$. The natural frequencies are generally distributed according to a probability density $P(\omega)$. The most general form of the phase dynamics for $N$ such oscillators interacting globally with one another may be written as~\cite{Kuramoto1975}
\begin{equation}
\dot{\theta}_i=\omega_{i}+\sum_{j=1}^{N}\Gamma_{ij}(\theta_j-\theta_i).
\end{equation}
In the absence of coupling, the second term vanishes, and each oscillator evolves independently according to $\dot{\theta}_i=\omega_{i}$. When the coupling is present, however, each oscillator experiences the collective influence of all the others through the interaction function $\Gamma_{ij}$.

Kuramoto proposed a particularly simple form of the coupling that favors the reduction of phase differences between oscillators: $\Gamma_{ij}(\theta_j-\theta_i)=(K/N)\sin(\theta_j-\theta_i)$, where $K$ is the coupling strength. Substituting this form into the general phase equation yields
\begin{equation}
\dot{\theta}_i
=
\omega_{i}
+
\frac{K}{N}
\sum_{j=1}^{N}
\sin(\theta_j-\theta_i),
\label{eq:Kuramoto}
\end{equation}
for $i=1,2,\ldots,N$. This system of equations constitutes the celebrated Kuramoto model~\cite{Kuramoto1975,Kuramoto1984}. An excellent review of the model and its numerous extensions is provided in Ref.~\cite{gupta2014kuramoto}.
The factor $(1/N)$ ensures that the effective coupling experienced by each oscillator remains finite as the system size increases. Without this normalization, the total interaction term would grow proportionally to $N$, leading to a size-dependent dynamics and an ill-defined thermodynamic limit. The sinusoidal coupling tends to reduce phase differences between oscillators. To see this, consider two oscillators with a small phase difference $\Delta\theta=\theta_j-\theta_i$. Since $\sin(\Delta\theta)\approx\Delta\theta$ for small $\Delta\theta$, a positive phase difference contributes positively to $\dot{\theta}_i$, causing the lagging oscillator to advance more rapidly, while a negative phase difference produces the opposite effect. In this way, the interaction acts to bring the oscillators closer in phase. When the coupling is sufficiently strong compared with the spread of natural frequencies, the population can undergo a transition from incoherent behavior to a partially or fully synchronized state. In the following sections, we examine this synchronization phenomenon in greater detail within the framework of the Kuramoto model.

It is interesting to note that one can also obtain the Kuramoto model by considering the setup in Eqs.~\eqref{eq:2osc-1} - \eqref{eq:2osc-2} generalized to a system of $N$ oscillators that are coupled all-to-all. The dynamics reads as 
\begin{align}
\dot\alpha_n = -i\omega_n\alpha_n + \alpha_n(\kappa_1-2\kappa_2|\alpha_n|^2)
+ \frac{V}{N}\sum_{m=1}^{N}(\alpha_m-\alpha_n).
\end{align}
In the interaction term, the self-term ($m=n$) vanishes, giving the interaction term to simplify to $V(\bar\alpha - \alpha_n)$, with $\bar\alpha \equiv \frac{1}{N}\sum_m \alpha_m$. Hence, we get
\begin{align}
\dot\alpha_n = -i\omega_n\alpha_n + \alpha_n(\kappa_1-2\kappa_2|\alpha_n|^2)
+ V(\bar\alpha - \alpha_n).
\end{align}
Writing $\bar\alpha = Re^{i\Theta}$ and $\alpha_n = r_n e^{i\theta_n}$ gives
\begin{align}
&\dot r_n= r_n(\kappa_1-2\kappa_2 r_n^2) + V\left(r_n - R\cos(\Theta-\theta_n)\right), \\
&\dot\theta_n= -\omega_n - \frac{VR}{r_n}\sin(\Theta-\theta_n).
\end{align}
In the small-$V$, small-$R$ limit, $r_n \to r_0$ and
\begin{align}
\dot\theta_n = -\omega_n - \frac{VR}{r_0}\sin(\Theta-\theta_n),
\end{align}
which under the rescaling $R \to R/r_0$, $\theta_n \to -\theta_n$, $\psi \to -\psi$ reads as
\begin{align}
\dot\theta_n =\omega_n - VR\sin(\Theta-\theta_n),
\end{align}
which is the Kuramoto model with effective coupling $K=-V$ and order parameter
$R=|\bar\alpha|$.

\subsection{Properties of the Kuramoto model}
\begin{figure}
    \centering
    \begin{subfigure}{0.3\linewidth}
        \includegraphics[width=\linewidth]{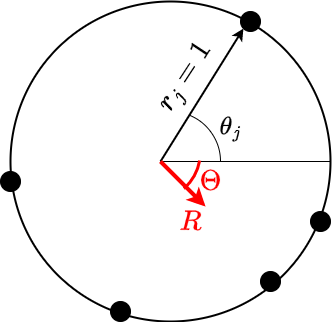}
        \caption{}
    \end{subfigure}
    \hspace{3cm}
    \begin{subfigure}{0.3\linewidth}
        \includegraphics[width=\linewidth]{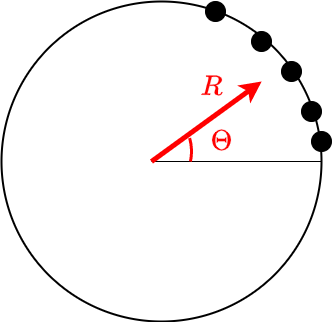}
        \caption{}
    \end{subfigure}
    \caption{(a) The collection of oscillators depicted by the solid circles on the circumference of the unit circle. (b) Change in the magnitude and angle of the order parameter as the oscillator phases close in on each other.}
    \label{fig:unit circle}
\end{figure}

The phases of different oscillators appearing in the Kuramoto model can be mapped to points on the circumference of a unit circle \cite{Kuramoto1984,Strogatz2000}, as shown in Fig. \ref{fig:unit circle}. The angle made by any such point with a reference line, say, the horizontal, can then be referred to as the phase of the oscillator. These points move about on the circumference with angular velocities dictated by the Kuramoto equation\footnote{A source of confusion might be the mapping of a simple harmonic motion to the projections of the position of a particle moving on a circle, on the diameter of the circle. The main difference in the case of a nonlinear oscillator is that, when the motion is not simple harmonic, the values of angular velocity of this particle will not be constant with time. Consequently, if we take projections of its motion, then the resulting plot will show a periodic variation that is different from a sinusoid.}. The position vector associated with the $j$-th point (which is the $j$-th oscillator) is given by $\textbf{r}_j\equiv (r_j,\theta_j)=(1,\theta_j)$. A compact form to denote this on the complex plane (consider the horizontal axis to be the real axis and the vertical one to be the imaginary axis) is $\textbf{r}_j = r_j e^{i\theta_j} = e^{i\theta_j}$. Let us perform a vector sum of all such position vectors $\textbf{r}_j$, with $j=1,2,\cdots,N$, and divide the resultant vector by $N$. We call this mean vector the complex order parameter:
\begin{equation}
    \textbf{R} = Re^{i\Theta} = \frac{1}{N}\sum_{j=1}^N \textbf{r}_j = \frac{1}{N}\sum_{j=1}^N e^{i\theta_j}.
    \label{eq:order parameter}
\end{equation}

\begin{figure}
    \centering
    \includegraphics[width=0.6\linewidth]{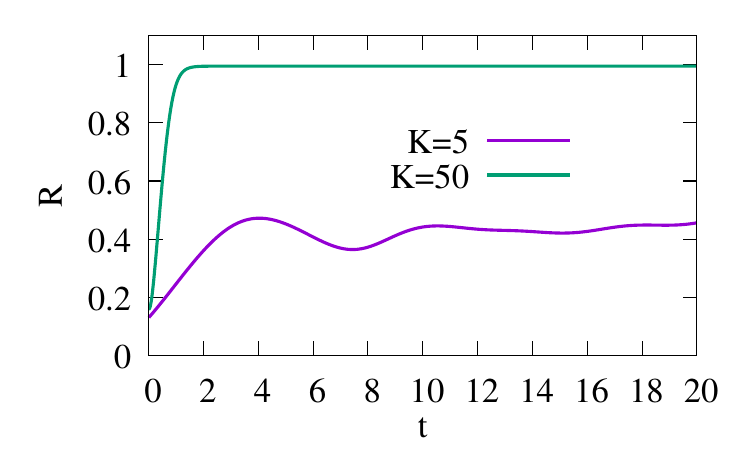}
    \caption{Variation of $R$ with time for 20 particles with $K=5$ (purple line) and $K=50$ (green line). The initial values of $\omega_{j}$ and $\theta_j$ are sampled from uniform distributions, with ranges $(-1/2,1/2)$ and $(0,2\pi)$, respectively.}
    \label{fig:R vs t}
\end{figure}

Note that Eq.~\eqref{eq:Kuramoto} is a system of $N$ first-order nonlinear and coupled ordinary differential equations. When this system of equations is solved numerically and the solution is mapped to the unit circle as discussed above, we find a set of $N$ points moving about on the circle, each point in general having a different instantaneous angular velocity than the other. This is similar to a set of runners running on a circular track, with the exception that the track is of zero width, and two runners are allowed to pass through each other without collision.  Each runner continuously adjusts his/her angular velocity by taking note of the average angular velocity of all the other runners. This adjustment will be very small if the coupling parameter $K$ is very small, and the tendency to maintain one's own angular velocity may overpower the urge to catch up with the others. However, if $K$ is large enough,  they are likely to get phase-locked in due time. In this case, the runners will huddle within a small segment of the track and continue running with approximately the same speed. Such a state will have a value of $R$ that is very close to unity, since the magnitude of the vector sum extends towards the periphery of the circle, as is clear from Fig. \ref{fig:unit circle}. The phase locking has been shown in Fig. \ref{fig:R vs t}, where $R(t)$ is observed to saturate at $R_\infty=1$ (when all the points almost converge to a single point on the track) for a large value of $K$ (green curve). The phase locking is absent for a small value of $K$ (purple curve). In fact, below a \textit{critical coupling parameter} $K=K_c$, the oscillators are unsynchronized, and the order parameter vanishes in the steady state: $R_\infty=0$. However, when $K>K_c$, partial synchronization takes place, and $R_\infty$ becomes non-zero. The value of $R_\infty$ grows with $K$, and becomes unity as $K\to\infty$. The variation of $R$ with time for two different values of $K > K_c$, have been shown in Fig. \ref{fig:R vs t}.

A compact way of writing Eq. \eqref{eq:Kuramoto} is by incorporating the definition \eqref{eq:order parameter} of the order parameter, and using the identity $\sin(x) = \mathrm{Im}(e^{ix})$ to get
\begin{eqnarray}
    \dot\theta_i &=& \omega_{i} + \frac{K}{N}~\mathrm{Im}\sum_j e^{i(\theta_j-\theta_i)} \nonumber\\
    &=& \omega_{i} + K~\mathrm{Im}\left[e^{-i\theta_i}R e^{i\Theta}\right] \nonumber\\
    &=& \omega_{i}  + KR\sin(\Theta-\theta_i).
    \label{eq:Kuramoto alternative}
\end{eqnarray}

\subsubsection{The co-moving frame} \label{sec:co-moving} 
\begin{figure}
     \centering
     \includegraphics[width=0.35\linewidth]{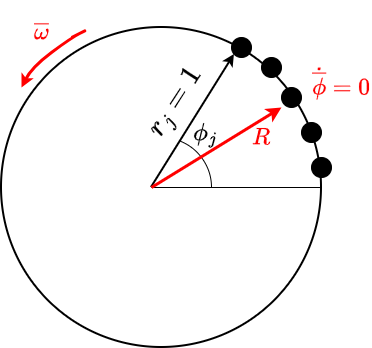}
     \caption{A co-moving frame: the unit circle itself rotates with an angular velocity of $\overline{\omega}$, and the phase of the $j^\text{th}$ particle is given by $\phi_j = \theta_j - \overline{\omega}t$. The order parameter \textbf{R} bisects the arc containing the particles.}
     \label{fig:comoving}
 \end{figure}

The depiction of the dynamics of the swarm of particles can be further simplified by subtracting the mean natural frequency, $\overline{\omega}=\int P(\omega)\omega d\omega$, from the values of $\omega$ of each particle. In other words, we move to a reference frame, called the \textit{co-moving frame}, that is also moving along the circular track with the velocity $\overline{\omega}$, as shown in Fig. \ref{fig:comoving}.
%Simply put, the observer in the co-moving frame is sitting on the $\textbf{R}$ vector itself. In this case, only a change in the magnitude $R$ is observed, whereas $\Omega\equiv \dot\Theta$ is found to remain constant with time. 
We further assume that $P(\omega)$ is a symmetric distribution about its mean, so that the number of particles ahead of $\overline{\omega}$ is same as that below $\overline{\omega}$ to begin with. 
With these simplifications, the observer in the new frame will find that the points no longer make complete laps of their tracks while simultaneously adjusting their phases. We only observe a fanning out or huddling of the points within a segment of the track, with the centre of the segment remaining immobile. This makes the effect of the coupling parameter on the particle phases more visualizable. 
Consider the variable $\phi_j=\theta_j-\overline{\omega} t$.
By construction, the average initial angular velocity in the co-moving frame vanishes, $\dot{\overline{\phi}}(0) = \frac{1}{N}\sum_{j=1}^{N}\dot\phi_j(0) =0$. Then the Kuramoto equation in this frame, \textit{as deduced by an observer in the original frame}, becomes
\begin{equation}
    \dot \phi_i = \omega_{i} - \overline{\omega} + \frac{K}{N}\sum_{j=1}^N \sin(\phi_j-\phi_i).
    \label{eq:Kuramoto co-moving}
\end{equation}
We may check the outcome of taking the mean of both sides of the equation. We get
\begin{eqnarray}
    \dot{\overline{\phi}} &=& \overline{\omega} - \overline{\omega} + \frac{1}{N}\sum_{i=1}^N \frac{K}{N}\sum_{j=1}^N \sin(\phi_j-\phi_i) \nonumber\\
    &=& \frac{K}{N^2} \sum_{i,j} \sin(\phi_j-\phi_i) = 0.
\end{eqnarray}
The last summation vanishes due to pairwise cancellations of the summand.  
This implies that the average angular velocity $\dot{\overline{\phi}}(t)$ remains zero for all values of $t$.

 Consider the arc over which the particles with angular velocities $\omega_{j}$ are uniformly distributed at initial time. The value of $\overline{\omega}$ must clearly be located at the mid-point of this arc \textit{in the steady state}, since all faster moving particles appear to its left and all slower moving particles appear to its right (this is strictly true either for a large number of particles, or considering an ensemble of experimental realizations).  
 Thus, if we move to the frame that rotates with this mean angular velocity, the mid-point of the arc remains fixed at a constant phase and has the angular velocity $\overline{\omega}$. However, the length of the arc might increase (decrease) if $K$ is small (large).
 The order parameter $\textbf{R}$ is the average of all the $\textbf{r}_j$ vectors, the  latter having angular orientations that are symmetrically distributed about the former. Accordingly, $\textbf{R}$ always bisects the angle subtended by the above arc to the centre of the circle.  As a result, the geometry of the setup   implies that $\textbf{R}$ always points towards $\overline{\omega}$. However, it may be noted that its magnitude $R$ may change over time, such that it increases when the arc length decreases and \textit{vice versa}.

% Since $\Omega=\overline{\omega}$ for a \textcolor{red}{synchronized state},
% Eq.~\eqref{eq:Kuramoto co-moving} may equivalently be written as 
% %
% \begin{equation}
%     \dot \phi_i = \omega_i - \Omega + \frac{K}{N}\sum_{j=1}^N \sin(\phi_j-\phi_i).
%     \label{eq:Kuramoto co-moving alternative}
% \end{equation}
% %

It is clear from the above discussion that the phase of the order parameter is stationary in this frame, i.e., $\dot{\Phi}=0$, where $\Phi=\Theta-\overline{\omega} t$. For an observer in the comoving frame, we have
\begin{equation}
    \dot \phi_i = \omega_i + \frac{K}{N}\sum_{j=1}^N \sin(\phi_j-\phi_i).
    \label{eq:Kuramoto co-moving alternative 1}
\end{equation}
The order parameter in this frame becomes
\begin{equation}
    Re^{i\Phi} = \frac{1}{N}\sum_{j=1}^N e^{i\phi_j}.
    \label{eq:order parameter co-moving}
\end{equation}

In the literature, it is common to retain the original notation after
transforming to the rotating frame, with the understanding that the
appropriate transformations have already been carried out. Thus one
typically writes
$\phi_j\rightarrow\theta_j$ and $\Phi\rightarrow\Theta$,
even while working in the rotating frame.

\subsubsection{A parallel with Adler Equation} Let us now  rewrite Eq. \eqref{eq:Kuramoto co-moving}  in a similar way as Eq. \eqref{eq:Kuramoto alternative}:
\begin{eqnarray}
    \dot\phi_i &=& \omega_i - \overline{\omega} + \frac{K}{N}~\mathrm{Im}\sum_j e^{i(\phi_j-\phi_i)} \nonumber\\
    &=& \omega_i - \overline{\omega} + KR\sin(\Phi-\phi_i).
    \label{eq:Kuramoto co-moving}
\end{eqnarray}
Now compare this with Eq. \eqref{eq:Adler}. Here, the frequency of the drive is replaced by $\overline{\omega}$, so that $\Delta\omega = \omega_i - \overline{\omega}$. Similarly, the phase of the drive is replaced by $\Phi$. If we make these replacements, the Kuramoto model in the co-moving frame takes the form of the Adler equation.

 \subsubsection{Continuum limit}

 Consider the limit of an infinite number of oscillators: $N\to \infty$. This limit lends itself to the assumption that the parameters $\omega$ and $\theta$ take continuous values. The system can then be described by means of a probability density $\rho(\theta;\omega,t)$, such that 
$\rho(\theta;\omega,t)d\theta$ is the probability that at time $t$, an oscillator with natural velocity $\omega$ has a phase in the range $(\theta,\theta+d\theta)$. Since the phase can take up any value from 0 to $2\pi$, we must impose the normalization condition \cite{Strogatz2000}
\begin{equation}
    \int_0^{2\pi} \rho(\theta;\omega,t) d\theta = 1.
    \label{eq:normalization}
\end{equation}
The order parameter is defined through the relation
\begin{eqnarray}
    R e^{i\Theta} = \langle e^{i\theta}\rangle = \int_0^{2\pi} d\theta\int_{-\infty}^{\infty} d\omega~ P(\omega) \rho(\theta;\omega,t) e^{i\theta}.
    \label{eq:order parameter continuum}
\end{eqnarray}
Conservation of the number of oscillators with a given natural frequency implies the following continuity equation \cite{Strogatz2000}:
\begin{eqnarray}
    \frac{\partial\rho}{\partial t} &=& - \frac{\partial}{\partial\theta} (\rho \dot\theta) \nonumber\\
    &=& - \frac{\partial}{\partial\theta}[\rho \{\omega + K\langle \sin(\theta'-\theta) \rangle \}]\nonumber\\
    &=& - \frac{\partial}{\partial\theta}[\rho \{\omega + K \int_0^{2\pi} d\theta'\int_{-\infty}^{\infty} d\omega' \rho(\theta';\omega',t) P(\omega') \sin(\theta'-\theta) \}].
\end{eqnarray}
In the second step, we have used the Kuramoto equation, Eq. \eqref{eq:Kuramoto}.

\subsubsection{The coupling threshold}

We now explore the critical coupling strength $K=K_c$, such that collective behaviour emerges as $K>K_c$, which was absent below this threshold. Recall Eq. \eqref{eq:Kuramoto co-moving}, where the Kuramoto equation has been cast in terms of the order parameter $\textbf{R} = (R,\Theta)$:
\[
\dot\phi_i = \omega_i - \overline{\omega} + KR\sin(\Phi-\phi_i).
\]
A reasonable assumption is that the distribution $P(\omega)$ of oscillator frequencies is an even function: $P(\omega) = P(-\omega)$. Then, the mean vanishes: $\overline{\omega} = 0$. We now consider the steady-state dynamics, when both $R$ and $\Phi$ become time independent. Without any loss of generality, we choose the reference axis for calculating the phase angle to be such that $\Phi = 0$. In other words, we make the transformation $\phi_i \to \phi_i + \Phi$. This simplifies the above equation to
\begin{equation}
    \dot\phi_i = \omega_i - KR\sin\phi_i.
    \label{eq:Kuramoto simplified co-moving}
\end{equation}
From the definition \eqref{eq:order parameter co-moving}, we can write $R e^{i\Phi} = \langle e^{i\phi}\rangle$. On substituting $\Phi = 0$, the left-hand side becomes real. Thus, we set the imaginary part of the right-hand side to zero to arrive at the relation
\begin{equation}
    R = \langle\cos\phi\rangle = \int_{-\infty}^{+\infty} d\omega~ P(\omega) \cos[\phi(\omega)],
    \label{eq:R threshold}
\end{equation}
where $\phi(\omega)$ is the solution of Eq. \eqref{eq:Kuramoto simplified co-moving}. Now, only the synchronized oscillators ($\dot\phi = 0$) contribute to this integral, while the contributions coming from the unsynchronized oscillators (denoted by the subscript ``unsync'' below) cancel out to zero:
\begin{equation}
    \langle \cos\phi\rangle_\text{unsync} = \frac{1}{2\pi}\int_0^{2\pi} \cos\phi = 0,
\end{equation}
which happens when $|\omega| > KR$ (see Eq. \eqref{eq:Kuramoto simplified co-moving}). Let us rewrite Eq. \eqref{eq:R threshold} as
\begin{eqnarray}
    R &=& \int_{-KR}^{+KR} d\omega~ P(\omega) \cos[\phi(\omega)] + \int_{|\omega|>KR} d\omega~ P(\omega) \cos[\phi(\omega)].
\end{eqnarray}
Since the second integral vanishes, only the oscillators with $-KR \le \omega \le KR$ contribute to the value of $R$:
\begin{eqnarray}
    R &=& \int_{-KR}^{+KR} d\omega~ P(\omega) \cos[\phi(\omega)]
    = \int_{-KR}^{+KR} d\omega~ P(\omega) \sqrt{1-\left(\frac{\omega}{KR}\right)^2}.
\end{eqnarray}
In the last step, we have made use of Eq. \eqref{eq:Kuramoto simplified co-moving}. The above equation may be rewritten as
\begin{eqnarray}
    R\left(1-K\int_{-1}^{+1} dx~ P(xKR) \sqrt{1-x^2}\right)=0.
\end{eqnarray}
We define the critical point as the point when $R=R_c=0^+$, i.e., $R$ assumes an infinitesimal positive value. Then, as $K \to K_c^+$, since $R \ne 0$, we must have  
\begin{eqnarray}
    1-K_cP(0)\int_{-1}^{+1} dx~\sqrt{1-x^2}=0.
\end{eqnarray}
The integral is the area $\pi/2$ of a unit semi-circle. The critical coupling constant is thus given by \cite{Strogatz2000}
\begin{eqnarray}
    K_c = \frac{2}{\pi P(0)}.
    \label{eq:critical coupling classical}
\end{eqnarray}

 \subsubsection{Dynamics in presence of noise}

\begin{figure}
    \centering
    \includegraphics[width=0.6\linewidth]{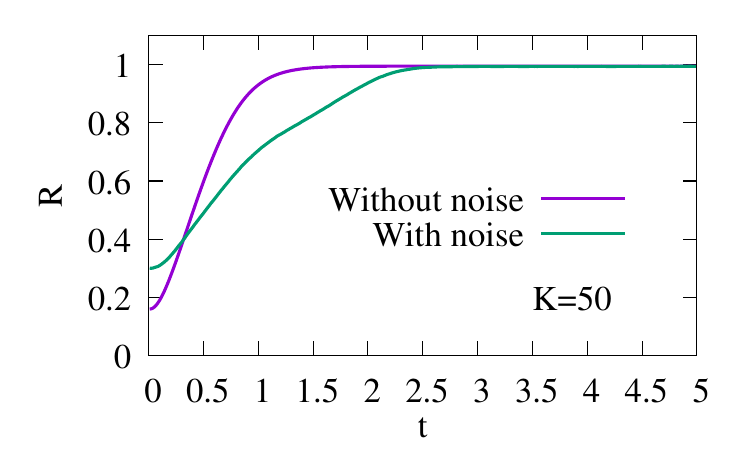}
    \caption{Plots showing the variation in $R$ as a function of time, in presence and in absence of noise, for $K=50$.}
    \label{fig:noise_R_t}
\end{figure}

 So far, the only randomness in the model has been due to the initial distributions of phase and angular velocity of the particles. The time evolution has been deterministic. One can add noise in the dynamics as well, so that Eq. \eqref{eq:Kuramoto} becomes
 \begin{equation}
     \dot\theta_i = \omega_i + \xi_i + \frac{K}{N}\sum_{j=1}^N \sin(\theta_j-\theta_i),
     \label{eq:noise}
 \end{equation}
 where $\langle \xi_i\rangle=0$, and $\langle \xi_i(t)\xi_j(t')\rangle = 2D\delta_{ij}\delta(t-t')$. Here, $D$ is the noise strength. The delta-correlation in time indicates white noise, while the presence of the Kr\"onecker delta $\delta_{ij}$ excludes the correlation between noise terms associated with  different oscillators. This model, as it turns out, is more tractable analytically \cite{Acebron2005}. If we now move to the co-moving frame, then the above equation becomes
 \begin{eqnarray}
     \dot\theta_i &=& \omega_i + KR\sin(\Theta-\theta_i) + \xi_i \nonumber\\
     &=& v_i+\xi_i,
     \label{eq:noise_comoving}
 \end{eqnarray}
 where $v_i = \omega_i + KR\sin(\Theta-\theta_i)$. In the continuum limit ($N\to\infty$), the above equation becomes
 \begin{equation}
     \dot\theta = v(t) + \xi(t),
 \end{equation}
 where $v(t)=\omega + KR\sin(\Theta-\theta)$, with $R$ and $\Theta$ as defined in Eq. \eqref{eq:order parameter continuum}.  
 This equation, being of the form of an overdamped Langevin dynamics, must yield the Smoluchowsky equation for the probability density $\rho(\theta;\omega,t)$:
\begin{equation}
     \frac{\partial\rho}{\partial t} = -\frac{\partial (\rho v)}{\partial\theta} + D\frac{\partial^2\rho}{\partial\theta^2}.
 \end{equation}
In Fig. \ref{fig:noise_R_t}, we plot the results of a simulation of 20 Kuramoto oscillators with a high coupling strength ($K=50$), both in the absence (purple curve) and presence (green curve) of noise. The longer time taken to synchronize in the latter case is clearly visible.

\section{Quantum limit cycles}
\label{sec:sec4}
\subsection{Quantum harmonic oscillator (QHO)}
\label{subsec:closed-QHO}

We begin our study with a discussion of a one-dimensional quantum harmonic oscillator (QHO) of mass $m$ and frequency $\omega$ that is interacting with a photon reservoir. The photons induce transitions between the energy eigenstates of the oscillator. In the following, we work within the Schr\"{o}dinger picture in which operators are time independent, while states depend on time. The Lindblad equation for the time evolution of the density operator or the density matrix $\hat{\rho}(t)$ of the system reads as (see Appendix~A):
\begin{align}
 \label{eq:vdpo_lee}
    \frac{d\hat{\rho}}{dt}=-\frac{i}{\hbar}[\hat{H},\hat{\rho}]+\gamma_{1}\mathcal{D}[\hat{a}^\dagger]\hat{\rho}+\gamma_{2}\mathcal{D}[\hat{a}^2]\hat{\rho}, 
\end{align}
where for any operator $\hat{O}$, we have 
\begin{align}
\mathcal{D}[\hat{O}]\hat{\rho} \equiv 2\hat{O}\hat{\rho}\hat{O}^\dagger-\hat{O}^\dagger\hat{O}\hat{\rho}-\hat{\rho}\hat{O}^\dagger\hat{O},
\end{align}
called the collapse or the jump operator. Also, $\hat{H}=\hbar\omega(\hat{a}^\dagger\hat{a}+1/2)$ is the usual Hamiltonian of the quantum harmonic oscillator, and where $\hat{a}$ and $\hat{a}^\dagger$ are the annihilation and the creation operator, respectively, with the commutator $[\hat{a},\hat{a}^\dagger]=1$.  
The interaction of the harmonic oscillator with the reservoir occurs via exchange of photons. The term
$\mathcal{D}[\hat{a}^\dagger]\hat{\rho}$ denotes one-photon gain, while the two-photon loss term is given by $\mathcal{D}[\hat{a}^2]\hat{\rho}$. The real parameter $\gamma_1$ in Eq.~\eqref{eq:vdpo_lee} represents the rate for one-photon gain, while the real parameter $\gamma_2$ is the rate for two-photon loss; both $\gamma_1$ and $\gamma_2$ being rates are positive quantities. The two above-mentioned terms lead to the quantum harmonic oscillator jumping between its energy eigenstates $\{|n\rangle\}$, where we have $\hat{H}|n\rangle=(n+1/2)\hbar \omega |n\rangle$, with $n$ being the number of photons. These eigenstates, called the Fock states, span the Hilbert space of the system. The mentioned photon gain-loss features may be understood by expressing the density operator  as $\hat{\rho}(t)=\sum_n p_n(t)|n\rangle \langle n|$, where $p_n(t)$ is the probability at time $t$ to have $n$ photons, with $\mathrm{Tr}[\hat{\rho}(t)]=1$ implying $\sum_n p_n(t)=1$, and Tr denoting trace operation.
From Eq.~\eqref{eq:vdpo_lee}, on using $p_n(t)=\langle n|\hat{\rho}(t)|n\rangle=\rho_{n,n}(t)$ and $\langle n|[\hat a^\dagger \hat a,\hat\rho]|n\rangle = 0$, we get
\begin{align}
&\dot{p}_n(t)=\underbrace{2\gamma_1 \left(np_{n-1}(t)-(n+1)p_n(t)\right)}_{\text{one-photon gain}}+\underbrace{2\gamma_2\left((n+1)(n+2)p_{n+2}(t)-n(n-1)p_n(t)\right)}_{\text{two-photon loss}}, \nonumber \\
\label{eq:Lindblad-2}
\end{align}
which makes it evident that the term with $\gamma_1$ in it models one-photon gain, while the one with $\gamma_2$ in it models two-photon loss. 
It may be noted that if the system is in a coherent state, the basis that diagonalizes the density matrix is generally not the number basis, because the coherent state is a superposition of number states (see Eq. \eqref{eq:coherent-n} below). The only exception is when $\alpha=0$ in Eq. \eqref{eq:coherent-n}, where the coherent state coincides with the vacuum number state $|0\rangle$.

\subsubsection{Closed QHO} 
\label{sec:closed-QHO}
Let us first discuss the case of the corresponding closed system, namely, a one-dimensional quantum harmonic oscillator undergoing unitary evolution. The corresponding time evolution is obtained from Eq.~\eqref{eq:vdpo_lee} by setting $\gamma_1=\gamma_2=0$. For such a system, it is known that when prepared in a coherent state $|\psi(0)\rangle=|\alpha\rangle$, defined as the eigenstate of the annihilation operator $\hat{a}$: $ \hat{a}|\alpha\rangle=\alpha|\alpha\rangle;~\alpha=|\alpha|e^{i\phi}$, the system continues to remain in the coherent state under the unitary evolution: the state at a later time instant is given by $|\psi(t)\rangle=e^{-i \omega t/2}|\alpha(t)=\alpha e^{-i\omega t}\rangle$~\cite{cohentannoudji}. The coherent state is given in terms of a linear superposition of the energy eigenstates, as 
\begin{align}
|\alpha\rangle=e^{-|\alpha|^2/2}\sum_{n=0}^\infty \frac{\alpha^n}{\sqrt{n!}}|n\rangle.
\label{eq:coherent-n}
\end{align}
Moreover, computing the expectation of the position and the momentum operator in a coherent state, one obtains an oscillatory motion identical to the one in the corresponding classical motion~\cite{cohentannoudji}:
$\langle \hat{x}\rangle_\alpha(t)=\langle \alpha(t)|\hat{x}|\alpha(t)\rangle=\sqrt{2\hbar/(m\omega)}|\alpha|\cos(\omega t-\phi)$ and $\langle \hat{p}\rangle_\alpha(t)=\langle \alpha(t)|\hat{p}|\alpha(t)\rangle=-\sqrt{2m \hbar\omega}|\alpha|\sin(\omega t-\phi)$. One may also obtain these results as $\langle \hat{x}\rangle_\alpha(t)=\mathrm{Tr}[\hat{x}\hat{\rho}(t)]$ and $\langle \hat{p}\rangle_\alpha(t)=\mathrm{Tr}[\hat{p}\hat{\rho}(t)]$, on using $\hat{\rho}(0)=|\alpha\rangle\langle \alpha|$ and $\hat{\rho}(t)=e^{-i\hat{H}t/\hbar}\hat{\rho}(0)e^{i\hat{H}t/\hbar}=|\alpha(t)\rangle \langle \alpha(t)|$. The position probability density in a coherent state at time $t$ is given by
\begin{align}
    |\psi_\alpha(x,t)|^2=|\langle x|\alpha(t)\rangle|^2=\frac{1}{\sqrt{2\pi} \Delta \hat{x}_\alpha}\exp\Big(-\frac{(x-\langle \hat{x}\rangle_\alpha(t))^2}{2(\Delta \hat{x}_\alpha)^2}\Big),
    \label{eq:coherent-psi}
\end{align}
where $\Delta \hat{x}_\alpha\equiv \sqrt{\langle \hat{x}^2\rangle_\alpha(t)-[\langle \hat{x}\rangle_\alpha(t)]^2}=\sqrt{\hbar/(2m\omega)}$ is time independent, while we have from above that $\langle \hat{x}\rangle_\alpha(t)=x_0\cos(\omega t-\phi)$, with $x_0=\sqrt{2\hbar/(m\omega)}|\alpha|$. Thus, the center of the coherent state position probability density oscillates back and forth in time with frequency $\omega$, exhibiting a motion identical to that of the position in a classical HO. On the other hand, the width of this probability density remains unchanged as a function of time. In this sense, coherent states move like classical particles in a harmonic potential~\cite{cohentannoudji}. Let us now discuss an important point. From above, we see that $\Delta \hat{x}_\alpha/x_0=1/(2|\alpha|)$, so that for $|\alpha|=\sqrt{\langle \hat{a}^\dagger \hat{a}\rangle_\alpha}\gg 1$, the position probability density appears to have a vanishing width with respect to the length scale over which it moves, and the motion appears classical. This is the semi-classical limit of the QHO. Using Eq.~\eqref{eq:coherent-n}, we see that the semi-classical limit corresponds to a linear superposition of a very large number of energy eigenstates~\cite{cohentannoudji}. 

Let us note that the number of photons in the system is given by the number operator $\hat{N}=\hat{a}^\dagger \hat{a}$, which commutes with the Hamiltonian $\hat{H}$. Consequently, the number of photons in a closed QHO system is conserved under its unitary evolution. One may check that when in a coherent state $|\alpha\rangle$, the expectation value of the number of photons in the system equals $\langle \hat{N}\rangle_\alpha=|\alpha|^2$.

\subsubsection{Open QHO}

Having discussed the closed QHO, we now move on to discuss the case of open QHO. Unlike the closed QHO, now when the system is prepared in a coherent state, so that $\hat{\rho}(0)=|\alpha\rangle \langle \alpha|$, one no longer has unitary evolution leading to $\hat{\rho}(t)=e^{-i\hat{H}t/\hbar}\hat{\rho}(0)e^{i\hat{H}t/\hbar}=|\alpha(t)\rangle \langle \alpha(t)|$. Hence, for times $t>0$, the system does not remain in a coherent state. The time evolution of the expectation value of any operator $\hat{O}$, given by $\langle \hat{O}\rangle(t)=\mathrm{Tr}[\hat{O}\hat{\rho}(t)]$, reads as 
\begin{align}
    \frac{d}{dt}\langle \hat{O}\rangle(t)=\mathrm{Tr}[\hat{O}\dot{\hat{\rho}}(t)],
\end{align}
where the dot denotes as always the derivative with respect to time. Using Eq.~\eqref{eq:vdpo_lee} then leads to the following time evolution equation for the expectation of the annihilation operator: 
\begin{align}
    \frac{d}{dt}\langle \hat{a}\rangle=-i\omega\langle \hat{a}\rangle+\gamma_1 \langle \hat{a}\rangle-2\gamma_2\langle \hat{a}^\dagger \hat{a}^2\rangle.
\end{align}
When prepared in a coherent state $|\alpha\rangle$, we have $\langle \hat{a}\rangle=\alpha$, and the above equation gives
\begin{align}
    \frac{d\alpha}{dt}=\left(\gamma_1-i\omega\right)\alpha-2\gamma_2\langle \hat{a}^\dagger \hat{a}^2\rangle.
    \label{eq:a-evolution}
\end{align}
Note that setting $\gamma_1=\gamma_2=0$, one recovers the time evolution appropriate to the closed QHO, yielding $\alpha(t)=\alpha(0)e^{-i\omega t}$, in conformity with our discussion in Section~\ref{sec:closed-QHO}. For the open QHO, the time evolution~\eqref{eq:a-evolution} is evidently non-linear due to the presence of the last term on the right hand side. Further simplification can be made by considering the semi-classical limit $|\alpha|=\sqrt{\langle \hat{a}^\dagger \hat{a}\rangle_\alpha}\gg 1$. Decomposing $\hat{a}$ as $\hat{a}=\alpha+\delta \hat{a}$, with $\langle \delta \hat{a}\rangle_\alpha=0$, by definition, the semi-classical limit corresponds to having negligible quantum fluctuations, yielding $\langle \hat{a}^\dagger \hat{a}^2\rangle_\alpha=\langle (\alpha^*+\delta \hat{a}^\dagger)(\alpha+\delta \hat{a})^2\rangle_\alpha=|\alpha|^2\alpha+\langle \delta \hat{a}^\dagger\rangle_\alpha \alpha^2+\langle \delta \hat{a}^\dagger \delta \hat{a}^2\rangle_\alpha+\alpha^*\langle \delta \hat{a}^2\rangle +2|\alpha|^2 \langle \delta \hat{a}^\dagger \delta \hat{a}\rangle_\alpha \approx |\alpha|^2\alpha$. Under such an approximation, Eq.~\eqref{eq:a-evolution} gives the time evolution 
\begin{align}
    \frac{d\alpha}{dt}=-i\omega \alpha+\left(\gamma_1-2\gamma_2|\alpha|^2\right)\alpha.
    \label{eq:a-evolution-1}
\end{align} 

Equation~\eqref{eq:a-evolution-1} may be contrasted with the dynamics of a classical nonlinear system, the celebrated Landau-Stuart oscillator dynamics for a complex variable $Q$ given by 
\begin{align}
\frac{dQ}{dt}=i\Omega Q +(\alpha-\beta |Q|^2)Q,
\label{eq:LS}
\end{align}
with $\alpha,\beta,\Omega$ all real quantities, and additionally, one has $\alpha,\beta>0$. Let us write $Q=\rho
e^{i\theta}$ with $\rho,\theta$ being real quantities and $\theta \in [-\pi,\pi]$. It then follows from Eq.~\eqref{eq:LS} that $\theta$ rotates
uniformly in time with angular frequency equal to the parameter $\Omega$, while $\rho$ has a stable
value $\rho_\mathrm{stable}=\sqrt{\alpha/\beta}$, such that
$d\rho/d t|_{\rho=\rho_\mathrm{stable}}=0,d^2\rho/dt^2|_{\rho<\rho_\mathrm{stable}}
>0$, and $d^2\rho/dt^2|_{\rho>\rho_\mathrm{stable}}<0$. Then, setting $\rho=\rho_\mathrm{stable}$ leads to self-sustained limit-cycle oscillations at frequency
$\Omega$ with amplitude $\rho_\mathrm{stable}$; the phase $\theta$ varies
as a function of time as
\begin{align}
\frac{d\theta}{dt}=\Omega.
\label{eq:angular-velocity-definition}
\end{align}
On the basis of the above discussion, we conclude that the system~\eqref{eq:vdpo_lee} in the semi-classical limit $|\alpha|\gg 1$ has the dynamics of the quantum expectation $\langle \hat{a}\rangle_\alpha$ in a coherent state that maps on to the dynamics of the Stuart-Landau oscillator. That the latter exhibits a limit-cycle behavior suggests that the system~\eqref{eq:vdpo_lee} when viewed appropriately, namely, in terms of the so-called Wigner function $W(\alpha,\alpha^*)$, may exhibit a limit cycle in the $\mathrm{Re}(\alpha)-\mathrm{Im}(\alpha)$-plane, whose amplitude will be given by $\sqrt{\gamma_1/(2\gamma_2)}$. We take up this task in the following.

\subsection{Characterization using Wigner function}
\label{sec:wigner}

A convenient way to characterize the dynamics of either the closed or the open QHO is by invoking the well-known Wigner function, see~\ref{app:PQW}. The latter is a quasiprobability distribution function that plays a role in quantum mechanics similar to that of the classical phase space probability distribution function; in particular, quantum expectations may be computed as classical averages with respect to the Wigner function. For our case of a quantum harmonic oscillator, the Wigner function corresponding to a density operator $\hat{\rho}$ is defined in terms of the
characteristic function (see Appendix~B and in particular, Eq.~\eqref{eq:Wigner-app-0}):
\begin{align}
W(\alpha,\alpha^*)= \frac{1}{\pi^2}\int d^2\eta~\chi(\eta,\eta^*)e^{-\eta\alpha^*+\eta^*\alpha}, ~~\text{with}~~\chi(\eta,\eta^*) = \Tr[\hat{\rho} \hat{D}(\eta)],
\label{eq:Wigner-chi}
\end{align}
where
$\hat{D}(\eta) = e^{\eta \hat{a}^\dagger - \eta^* \hat{a}}$ is the displacement operator, and we have $d^2\eta = d(\mathrm{Re}(\eta))\,d(\mathrm{Im}(\eta))$. In the above equation, the integration is over the entire complex-$\eta$ plane. Any time dependence in $W(\alpha,\alpha^*)$ arises from the time dependence of $\hat{\rho}$; nevertheless, we will drop this explicit time dependence and continue to denote the Wigner function by $W(\alpha,\alpha^*)$. Equation~\eqref{eq:Wigner-chi} gives the prescription to define the Wigner transform of any operator $\hat{O}$, as 
\begin{align}
W_O(\alpha,\alpha^*)\equiv \frac{1}{\pi^2}\int d^2\eta~\Tr[\hat{O} \hat{D}(\eta)]e^{-\eta\alpha^*+\eta^*\alpha}.
\label{eq:Wigner-transform}
\end{align}
Obviously, with $\hat{O}=\hat{\rho}$, the above equation reduces to Eq.~\eqref{eq:Wigner-chi}: $W_\rho(\alpha,\alpha^*)=W(\alpha,\alpha^*)$.

The quantum expectation of any operator $\hat{O}$ is obtained as 
\begin{align}
    \langle \hat{O}\rangle \equiv \int d^2\alpha~O(\alpha)W(\alpha,\alpha^*),
    \label{eq:Oexpectation}
\end{align}
where $O(\alpha)$ is obtained by expressing the operator $\hat{O}$ in terms of the creation and the annihilation operator and then computing the expectation of $\hat{O}$ in a coherent state $|\alpha\rangle$. For example, we have $x(\alpha)=\langle \hat{x}\rangle_\alpha=\sqrt{\hbar/(2m\omega)}\,\langle \hat{a}+\hat{a}^\dagger\rangle_\alpha=\sqrt{\hbar/(2m\omega)}\,(\alpha+\alpha^*)$, where we have used $\langle \hat{a}\rangle_\alpha=\alpha$ and $\langle \hat{a}^\dagger\rangle_\alpha=\alpha^*$. Note that in Eq.~\eqref{eq:Oexpectation}, any time dependence in $\langle \hat{O}\rangle$ arises from the time dependence of the Wigner function appearing on the right hand side of the equation.

Next, with $W$ being real, we have
\begin{align}
W(\alpha,\alpha^*)= \frac{1}{\pi^2}\int d^2\eta~\chi^*(\eta,\eta^*)e^{-\eta^*\alpha+\eta\alpha^*},~~ \text{with}~~ \chi^*(\eta,\eta^*) = \Tr[ \hat{D}^\dagger(\eta)\hat{\rho}]=\Tr[\hat{\rho} \hat{D}^\dagger(\eta)],
\label{eq:Wigner-chi-again}
\end{align}
where we have used $(\Tr[\hat{A}])^*=\Tr[\hat{A}^\dagger]$ and the invariance of trace operation under cyclic permutation. Again, as done in Eq.~\eqref{eq:Wigner-transform}, we may write
\begin{align}
W_{O}(\alpha,\alpha^*)= \frac{1}{\pi^2}\int d^2\eta~\Tr[\hat{O} \hat{D}^\dagger(\eta)]e^{-\eta^*\alpha+\eta\alpha^*}.
\label{eq:Wigner-transform-1}
\end{align}

Our objective in this section is to use Eq.~\eqref{eq:vdpo_lee} in Eq.~\eqref{eq:Wigner-chi}, or, equivalently, Eq.~\eqref{eq:Wigner-chi-again}, and derive an evolution equation for $W(\alpha,\alpha^*)$. Taking time derivative of both sides of the first equality in Eq.~\eqref{eq:Wigner-chi-again}, the left hand side reads as $\partial W(\alpha,\alpha^*)/\partial t$, while the right hand side involves the term $\partial \hat{\rho}/\partial t$, which on using Eq.~\eqref{eq:vdpo_lee} leads to terms of the form $\mathcal{W}_{[\hat{H},\hat{\rho}]}=\Tr[[\hat{H},\hat{\rho}]\hat{D}^\dagger(\eta)]$,  $\mathcal{W}_{\mathcal{L}_1[\rho]}=\Tr[\mathcal{L}_1[\hat{\rho}]\hat{D}^\dagger(\eta)]$, and $\mathcal{W}_{\mathcal{L}_2[\rho]}=\Tr[\mathcal{L}_2[\hat{\rho}]\hat{D}^\dagger(\eta)]$, with
\begin{align}
&\mathcal{L}_1[\hat{\rho}] \equiv \gamma_1( 2\hat{a}^\dagger\hat{\rho}\hat{a}-\hat{a}\hat{a}^\dagger\hat{\rho}-\hat{\rho}\hat{a}\hat{a}^\dagger), \\
&\mathcal{L}_2[\hat{\rho}] \equiv \gamma_1( 2\hat{a}^2\hat{\rho}(\hat{a}^\dagger)^2-(\hat{a}^\dagger)^2\hat{a}^2\hat{\rho}-\hat{\rho}(\hat{a}^\dagger)^2\hat{a}^2.
\end{align}
To proceed requires evaluation of the terms $\mathcal{W}_{[\hat{H},\hat{\rho}]}$, $\mathcal{W}_{\mathcal{L}_2[\rho]}$ and $\mathcal{W}_{\mathcal{L}_2[\rho]}$. This will be done by first deriving a set of useful correspondences.

\section*{Deriving a set of useful correspondences}

We start with the defining equation for the displacement operator: $\hat{D}(\eta) = e^{\eta \hat{a}^\dagger - \eta^* \hat{a}}$. Using the Baker--Campbell--Hausdorff formula gives 
$\hat{D}(\eta) = e^{-|\eta|^2/2} e^{\eta \hat{a}^\dagger} e^{-\eta^* \hat{a}}$, giving
\begin{align}
\frac{\partial \hat{D}(\eta)}{\partial \eta}
  = \left(\hat{a}^\dagger - \frac{1}{2}\eta^*\right) \hat{D}(\eta);~~\frac{\partial \hat{D}(\eta)}{\partial \eta^*}
  = -\left(\hat{a} - \frac{1}{2}\eta\right) \hat{D}(\eta),
  \label{eq:Dderivs}
\end{align}
and hence,
\begin{align}
\frac{\partial \hat{D}^\dagger(\eta)}{\partial \eta^*}
  =  \hat{D}^\dagger(\eta)\left(\hat{a} - \frac{1}{2}\eta\right);~~\frac{\partial \hat{D}^\dagger(\eta)}{\partial \eta}
  = -\hat{D}^\dagger(\eta)\left(\hat{a}^\dagger - \frac{1}{2}\eta^*\right).
  \label{eq:Dderivs-1}
\end{align}
Using $\hat{D}^\dagger(\eta) = e^{-|\eta|^2/2} e^{-\eta \hat{a}^\dagger} e^{\eta^* \hat{a}}$ and $[(\hat{a}^\dagger)^n,\hat{a}]=-n(\hat{a}^\dagger)^{n-1}$ implying $[e^{-\eta \hat{a}^\dagger},\hat{a}]=\eta e^{-\eta \hat{a}^\dagger}$ gives 
\begin{align}
\frac{\partial \hat{D}^\dagger(\eta)}{\partial \eta^*}
  = \left(\hat{a} +  \frac{1}{2}\eta\right) \hat{D}^\dagger(\eta);~~\frac{\partial \hat{D}^\dagger(\eta)}{\partial \eta}
  = \left(-\hat{a}^\dagger - \frac{1}{2}\eta^*\right) \hat{D}^\dagger(\eta),
  \label{eq:Dderivs-2}
\end{align}
which indeed implies Eq.~\eqref{eq:Dderivs-1} on using $[\hat{D}^\dagger(\eta),\hat{a}]=\eta \hat{D}^\dagger(\eta)$.

Using Eq.~\eqref{eq:Dderivs}, we have
\begin{align}
\Tr[\hat{\rho}\hat{a} \hat{D}(\eta)]=-\Tr\left[\hat{\rho}\frac{\partial \hat{D}(\eta)}{\partial \eta^*}\right]+\frac{\eta}{2}\Tr[\hat{\rho} \hat{D}(\eta)]=\left(-\frac{\partial}{\partial \eta^*}+ \frac{\eta}{2}\right)\chi(\eta,\eta^*),
\label{eq:atrhochi}
\end{align}
which implies on using Eq.~\eqref{eq:Wigner-transform} that
\begin{align}
W_{\rho a}(\alpha,\alpha^*)&=\frac{1}{\pi^2}\int d^2\eta\,\Tr[\hat{\rho}\hat{a}\hat{D}(\eta)]
e^{-\eta\alpha^*+\eta^*\alpha}\nonumber \\
&=\frac{1}{\pi^2}\int d^2\eta\,\left(\left(-\frac{\partial}{\partial\eta^*}+ \frac{\eta}{2}\right)\chi(\eta,\eta^*)\right)e^{-\eta\alpha^*+\eta^*\alpha}.
\end{align}
Integrating the first term by parts and assuming that boundary contributions of $\chi(\eta,\eta^*)$ vanish transfer the derivative to the exponential factor: $(\partial/\partial \eta^*)e^{-\eta\alpha^*+\eta^*\alpha}= \alpha e^{-\eta\alpha^*+\eta^*\alpha}$.
Similarly, the factor of \(\eta\) can be replaced by a derivative with
respect to \(\alpha^*\) acting on the exponential: $\eta e^{-\eta\alpha^*+\eta^*\alpha}
  = -(\partial/\partial \alpha^*)
    e^{-\eta\alpha^*+\eta^*\alpha}$.
We finally get
\begin{align}
W_{\rho a}(\alpha,\alpha^*)
  = \left(\alpha - \frac{1}{2}\frac{\partial}{\partial\alpha^*}\right)
    W(\alpha,\alpha^*),
\end{align}
yielding the first Wigner correspondence:
\begin{align}
  \rho a\longleftrightarrow\
  \left(\alpha - \frac{1}{2}\frac{\partial}{\partial\alpha^*}\right)W,
  \label{eq:correspondence-1}
\end{align}
where by $\rho a$ is meant the quantity $W_{\rho a}(\alpha,\alpha^*)$.

Next, using Eq.~\eqref{eq:Wigner-transform-1}, we get 
\begin{align}
W_{a\rho}(\alpha,\alpha^*)&=\frac{1}{\pi^2}\int d^2\eta\,\Tr[\hat{a}\hat{\rho}\hat{D}^\dagger(\eta)]
e^{-\eta^*\alpha+\eta\alpha^*}\nonumber \\
&= \frac{1}{\pi^2}\int d^2\eta\,\Tr[\hat{D}^\dagger(\eta)\hat{a}\hat{\rho}]
e^{-\eta^*\alpha+\eta\alpha^*}\nonumber \\&=\frac{1}{\pi^2}\int d^2\eta\,\left[\left(\frac{\partial}{\partial\eta^*}+ \frac{\eta}{2}\right)\chi^*(\eta,\eta^*)\right]e^{-\eta^*\alpha+\eta\alpha^*}.
\end{align}
Integrating the first term by parts and assuming that boundary contributions of $\chi^*(\eta,\eta^*)$ vanish transfer the derivative to the exponential factor: $(\partial/\partial \eta^*)e^{-\eta^*\alpha+\eta\alpha^*}= -\alpha e^{-\eta^*\alpha+\eta\alpha^*}$.
Similarly, the factor of \(\eta\) can be replaced by a derivative with
respect to \(\alpha^*\) acting on the exponential: $\eta e^{-\eta^*\alpha+\eta\alpha^*}
  = (\partial/\partial \alpha^*)
    e^{-\eta^*\alpha+\eta\alpha^*}$.
We finally get
\begin{align}
  W_{a\rho}(\alpha,\alpha^*)
  = \left(\alpha + \frac{1}{2}\frac{\partial}{\partial\alpha^*}\right)
    W(\alpha,\alpha^*),
\end{align}
yielding the second Wigner correspondence:
\begin{align}
  a\rho \longleftrightarrow\
  \left(\alpha + \frac{1}{2}\frac{\partial}{\partial\alpha^*}\right)W.
  \label{eq:correspondence-2}
\end{align}
In a similar manner, one may obtain the remaining two correspondences:
\begin{align}
&a^\dagger\rho\longleftrightarrow\left(\alpha^* - \frac{1}{2}\frac{\partial}{\partial \alpha}\right)W,\label{eq:correspondence-3}\\
&\rho a^\dagger\longleftrightarrow\left(\alpha^* + \frac{1}{2}\frac{\partial}{\partial \alpha}\right)W.
\label{eq:correspondence-4}
\end{align}
These correspondences will prove essential for transforming the Lindblad equation~\eqref{eq:vdpo_lee} into a differential equation for the Wigner function. 

With the above background, let us do the following evaluation:
\begin{align}
W_{a\rho a^\dagger}(\alpha,\alpha^*)=&\frac{1}{\pi^2}\int d^2\eta\,\Tr[\hat{a}\hat{\rho}\hat{a}^\dagger\hat{D}^\dagger(\eta)]
e^{-\eta^*\alpha+\eta\alpha^*}\nonumber \\
&= \frac{1}{\pi^2}\int d^2\eta\,\Tr[\hat{a}^\dagger\hat{D}^\dagger(\eta)\hat{a}\hat{\rho}]
e^{-\eta^*\alpha+\eta\alpha^*}\nonumber \\&=\frac{1}{\pi^2}\int d^2\eta\,\Tr\left[\hat{a}^\dagger\left(\frac{\partial}{\partial\eta^*}+ \frac{\eta}{2}\right)\hat{D}^\dagger(\eta)\hat{\rho}\right]e^{-\eta^*\alpha+\eta\alpha^*}\nonumber \\
&=\frac{1}{\pi^2}\int d^2\eta\,\Tr\left[\left(\frac{\partial}{\partial\eta^*}+ \frac{\eta}{2}\right)\hat{a}^\dagger\hat{D}^\dagger(\eta)\hat{\rho}\right]e^{-\eta^*\alpha+\eta\alpha^*}\nonumber \\
&=\frac{1}{\pi^2}\int d^2\eta\,\left[\left(\frac{\partial}{\partial\eta^*}+ \frac{\eta}{2}\right)\left(-\frac{\partial}{\partial\eta}-\frac{\eta^*}{2}\right)\chi^*(\eta,\eta^*)\right]e^{-\eta^*\alpha+\eta\alpha^*}\nonumber \\
&=\frac{1}{\pi^2}\int d^2\eta\,\left[\left(-\frac{\partial^2}{\partial\eta^* \partial \eta}-\frac{1}{2}-\frac{\eta^*}{2}\frac{\partial }{\partial \eta^*}-\frac{\eta}{2}\frac{\partial }{\partial \eta}- \frac{|\eta|^2}{4}\right)\chi^*(\eta,\eta^*)\right]e^{-\eta^*\alpha+\eta\alpha^*}\nonumber \\
&=\left(|\alpha|^2+\frac{1}{2}\frac{\partial }{\partial \alpha}\alpha+\frac{1}{2}\frac{\partial }{\partial \alpha^*}\alpha^*+\frac{1}{4}\frac{\partial^2}{\partial \alpha \partial \alpha^*}-\frac{1}{2}\right)W\nonumber \\
&=\left(\alpha+\frac{1}{2}\frac{\partial}{\partial \alpha^*}\right)\left(\alpha^* + \frac{1}{2}\frac{\partial}{\partial \alpha}\right)W.
\label{eq:DDerivs-3}
\end{align}
Here, in obtaining the third step, we have used the first equality in Eq.~\eqref{eq:Dderivs-1}, while in obtaining the fifth step, we have used the second equality in Eq.~\eqref{eq:Dderivs-2}. We have also performed integration by parts as appropriate in different steps of the above computation. 

Let us evaluate one more term:
\begin{align}
W_{\rho a^\dagger a}(\alpha,\alpha^*)=&\frac{1}{\pi^2}\int d^2\eta\,\Tr[\hat{\rho}\hat{a}^\dagger \hat{a}\hat{D}^\dagger(\eta)]
e^{-\eta^*\alpha+\eta\alpha^*}\nonumber \\
&=\frac{1}{\pi^2}\int d^2\eta\,\Tr\left[\hat{\rho}\hat{a}^\dagger\left(\frac{\partial}{\partial\eta^*}- \frac{\eta}{2}\right)\hat{D}^\dagger(\eta)\right]e^{-\eta^*\alpha+\eta\alpha^*}\nonumber \\
&=\frac{1}{\pi^2}\int d^2\eta\,\Tr\left[\left(\frac{\partial}{\partial\eta^*} -\frac{\eta}{2}\right)\hat{a}^\dagger\hat{D}^\dagger(\eta)\hat{\rho}\right]e^{-\eta^*\alpha+\eta\alpha^*}\nonumber \\
&=\frac{1}{\pi^2}\int d^2\eta\,\left[\left(\frac{\partial}{\partial\eta^*}- \frac{\eta}{2}\right)\left(-\frac{\partial}{\partial\eta}-\frac{\eta^*}{2}\right)\chi^*(\eta,\eta^*)\right]e^{-\eta^*\alpha+\eta\alpha^*}\nonumber \\
&=\frac{1}{\pi^2}\int d^2\eta\,\left[\left(-\frac{\partial^2}{\partial\eta^* \partial \eta}-\frac{1}{2}-\frac{\eta^*}{2}\frac{\partial }{\partial \eta^*}+\frac{\eta}{2}\frac{\partial }{\partial \eta}+ \frac{|\eta|^2}{4}\right)\chi^*(\eta,\eta^*)\right]e^{-\eta^*\alpha+\eta\alpha^*}\nonumber \\
&=\left(|\alpha|^2+\frac{1}{2}\frac{\partial }{\partial \alpha}\alpha-\frac{1}{2}\frac{\partial }{\partial \alpha^*}\alpha^*-\frac{1}{4}\frac{\partial^2}{\partial \alpha \partial \alpha^*}-\frac{1}{2}\right)W\nonumber \\
&=\left(\alpha-\frac{1}{2}\frac{\partial}{\partial \alpha^*}\right)\left(\alpha^* + \frac{1}{2}\frac{\partial}{\partial \alpha}\right)W.
\label{eq:DDerivs-4}
\end{align}
Here, in obtaining the second step, we have used the first equality in Eq.~\eqref{eq:Dderivs-2}, while in obtaining the fourth step, we have used the second equality in Eq.~\eqref{eq:Dderivs-2}. We have also performed integration by parts as appropriate in different steps of the above computation.

On the basis of the above, we now lay down simple rules for obtaining the correspondence for any term $\ldots BA\rho CD\ldots$, by first applying the right action on $\rho$ by $C$, the term appearing just to the right of $\rho$, and placing the resulting term at the far right. One must continue performing all operations that act from the right, adding each new term successively to the left of the previous one. After completing the right actions, apply the left actions on $\rho$ one at a time, starting with $A$, again placing each resulting term successively from right to left. For example, the correspondence for the term $BA\rho CD$ will be $B'A'D'C'W$, where, e.g., $A'$ denotes the term obtained by left-acting on $\rho$ by $A$. By appling these rules, we get 
\begin{align}
    W_{a\rho a^\dagger}(\alpha,\alpha^*) \longleftrightarrow\ \left(\alpha + \frac{1}{2}\frac{\partial}{\partial\alpha^*}\right)\left(\alpha^* + \frac{1}{2}\frac{\partial}{\partial \alpha}\right)W,
    \label{eq:rule1}
\end{align}
which matches with Eq.~\eqref{eq:DDerivs-3}. Similarly, one gets 
\begin{align}
    W_{\rho a^\dagger a}(\alpha,\alpha^*) \longleftrightarrow\ \left(\alpha -\frac{1}{2}\frac{\partial}{\partial\alpha^*}\right)\left(\alpha^* + \frac{1}{2}\frac{\partial}{\partial \alpha}\right)W,
    \label{eq:rule2}
\end{align}
which matches with Eq.~\eqref{eq:DDerivs-4}.

In passing, we derive a handy expression for the Wigner function for the harmonic oscillator eigenstates $|n\rangle$, which will prove useful later in the discussion. We start with Eq.~\eqref{eq:Wigner-chi}, which we rewrite as
\begin{align}
W(\alpha,\alpha^*)=\mathrm{Tr}\!\left[\hat{\rho}\,\hat{A}(\alpha)\right],
\label{eq:W-A-def}
\end{align}
where we have introduced the operator
\begin{align}
\hat{A}(\alpha)\equiv \frac{1}{\pi^{2}}\int d^{2}\eta\;
\hat{D}(\eta)\,e^{\,\alpha\eta^{*}-\alpha^{*}\eta}.
\label{eq:A-def}
\end{align}
Using the well-known result for the displacement operator, see Eq.~\eqref{eq:DaDb},
\begin{align}
\hat{D}(\xi)\hat{D}(\zeta)
= e^{(\xi\zeta^{*}-\xi^{*}\zeta)/2}\,\hat{D}(\xi+\zeta),
\end{align}
and shifting the integration variable in Eq.~\eqref{eq:A-def} as $\eta=\xi+2\alpha$, we get
\begin{align}
\hat{A}(\alpha)=\frac{1}{\pi^{2}}\,\hat{D}(2\alpha)
\int d^{2}\xi\; \hat{D}(\xi).
\end{align}
Thus, $\hat{A}(\alpha)$ is just a displaced version of $\hat{A}(0)$:
\begin{align}
\hat{A}(\alpha)=\hat{D}(2\alpha)\,\hat{A}(0).
\label{eq:Aalpha}
\end{align}

The next step is to determine $\hat{A}(0)$, for which we compute its matrix elements between coherent states, as
\begin{align}
\langle\beta|\hat{A}(0)|\beta'\rangle
= \frac{1}{\pi^{2}}\int d^{2}\eta\;
\langle\beta|\hat{D}(\eta)|\beta'\rangle.
\end{align}
Using $
\langle\beta|\hat{D}(\eta)|\beta'\rangle
= \exp\!\left(-|\eta|^{2}/2
+\eta\beta^{*}-\eta^{*}\beta'\right)
\langle\beta|\beta'\rangle$ and the Gaussian-integration result $\int d^{2}\eta\;
\exp\!\left(-|\eta|^{2}/2
+u\,\eta^{*}+v\,\eta\right)
=2\pi\,e^{\,2uv}$, we obtain
\begin{align}
\langle\beta|\hat{A}(0)|\beta'\rangle
= \frac{2}{\pi}\,
\langle\beta|\beta'\rangle\,
e^{-2\beta^{*}\beta'}.
\end{align}

Now, the parity operator of the eigenstates is defined as
\begin{align}
\hat{P} \equiv e^{i\pi \hat{a}^\dagger \hat{a}}
= \sum_{n=0}^\infty (-1)^n |n\rangle\langle n|,
\end{align}
whose coherent-state matrix element, using $|\alpha\rangle=e^{-|\alpha|^2/2}\sum_{n=0}^\infty \alpha^n/\sqrt{n!}\,|n\rangle$, is given by
\begin{align}
\langle\beta|\hat{P}|\beta'\rangle
=\langle\beta|-\beta'\rangle
=\exp\!\left(
-\tfrac{1}{2}(|\beta|^{2}+|\beta'|^{2})
-\beta^{*}\beta'
\right).
\end{align}
Using $\langle\beta|\beta'\rangle=\exp[-\frac12(|\beta|^{2}+|\beta'|^{2})+\beta^*\beta']$, see Eq.~\eqref{eq:coherent-state-orthonormal-app}, 
one verifies that
\begin{align}
\langle\beta|\hat{A}(0)|\beta'\rangle
=\frac{2}{\pi}\;\langle\beta|\hat{P}|\beta'\rangle,
\end{align}
yielding
\begin{align}
\hat{A}(0)=\frac{2}{\pi}\,\hat{P}.
\label{eq:A0}
\end{align}

Combining Eqs.~\eqref{eq:Aalpha} and~\eqref{eq:A0}, we obtain
\begin{align}
\hat{A}(\alpha)=\frac{2}{\pi}\,\hat{D}(2\alpha)\,\hat{P},
\end{align}
and therefore, we obtain from Eq.~\eqref{eq:W-A-def} that
\begin{align}
W(\alpha,\alpha^*)=\frac{2}{\pi}\,
\operatorname{Tr}\!\big[\,\hat{\rho}\,\hat{D}(2\alpha)\,\hat{P}\,\big].
\label{eq:W-displaced-parity}
\end{align}
The above is the desired displaced--parity representation of the Wigner function.

\section*{Computation of $\mathcal{W}_{[\hat{H},\hat{\rho}]}$}
This computation will involve addition of two terms $-i\Tr[\hat{a}^\dagger \hat{a}\hat{\rho}\hat{D}^\dagger(\eta)]$ and $i\Tr[\hat{\rho}\hat{a}^\dagger \hat{a}\hat{D}^\dagger(\eta)]$. Using the short-hand notations
$D\equiv \partial/\partial\alpha$, $\overline{D}\equiv \partial/\partial \alpha^*$, $c=1/2$ and the rules explained around Eqs.~\eqref{eq:rule1} and~\eqref{eq:rule2}, we have the correspondences
\begin{align}
a^\dagger a \rho
&\longleftrightarrow
\left(\alpha^* - c D \right)\!\left(\alpha + c \overline{D} \right)W=\alpha \alpha^* W-cD(\alpha W)+\alpha^* c\overline{D}W-c^2D\overline{D}W, 
\label{eq:piece1-1}\\
\rho a^\dagger a
&\longleftrightarrow
\,\!\left(\alpha - c\overline{D} \right)\!\left(\alpha^* + c D \right)W=\alpha^* \alpha W-c\overline{D}(\alpha^* W)+\alpha cDW-c^2D\overline{D}W.
\label{eq:piece3-2}
\end{align}
Combining everything, we get 
\begin{align}
\mathcal{W}_{[\hat{H},\hat{\rho}]}
&=-i\omega(-2c\alpha DW+2\alpha^* c\overline{D}W)=i\omega\left(\alpha \frac{\partial }{\partial \alpha}-\alpha^* \frac{\partial }{\partial \alpha^*}\right)W.
\label{eq:zero-term}
\end{align}

\section*{Computation of $\mathcal{W}_{\mathcal{L}_1[\rho]}$}
Here, we have the correspondences
\begin{align}
2a^\dagger \rho a
&\longleftrightarrow
2\!
\left(\alpha^* - c D \right)\!\left(\alpha - c \overline{D} \right)W=2\alpha^*\alpha W - 2c\alpha^*\overline{D}W - 2cD(\alpha W) + 2c^2D\overline{D}W, 
\label{eq:piece1}\\
-\,a a^\dagger\rho
&\longleftrightarrow
-\,\!\left(\alpha + c\overline{D} \right)\!\left(\alpha^* - cD \right)W=-\alpha\alpha^* W +c\alpha DW  - c\overline{D} (\alpha^* W) + c^2D\overline{D}W,
\label{eq:piece2}\\
-\,\rho a a^\dagger
&\longleftrightarrow
-\,\left(\alpha^* +cD \right)\!\left(\alpha - c\overline{D}\right)W=-\alpha\alpha^* W + c\alpha^*\overline{D}W- cD(\alpha W)  + c^2D\overline{D}W.
\label{eq:piece3}
\end{align}
Combining everything, we get 
\begin{align}
\mathcal{W}_{\mathcal{L}_1[\rho]}=\gamma_1\left[-\left(\frac{\partial }{\partial \alpha}\alpha + \frac{\partial }{\partial \alpha^*}\alpha^*\right) \,W
+\frac{\partial^2W}{\partial \alpha \partial \alpha^*} \right].
\label{eq:first-term}
\end{align}

\section*{Computation of $\mathcal{W}_{\mathcal{L}_2[\rho]}$}
Here, we have the correspondences
\begin{align}
2a^2\rho a^{\dagger 2}
\longleftrightarrow
2\Big(\alpha + c\overline{D}\Big)^2
\Big(\alpha^* + cD\Big)^2 W&= 2\Big\{
\alpha^2\alpha^{*2}W
+ 2c\,\alpha^2\alpha^* D W
+ c^2\,\alpha^2 D^2 W\nonumber \\
&+ 2c\,\alpha\overline{D}\,\alpha^{*2} W
+ 4c^2\,\alpha\overline{D}\,\alpha^* D W
+ 2c^3\,\alpha\overline{D}\,D^2 W\nonumber \\
&+ c^2\,\overline{D}^2\alpha^{*2} W
+ 2c^3\,\overline{D}^2\alpha^* D W
+ c^4\,\overline{D}^2 D^2 W\Big\},
\end{align}
\begin{align}
-a^{\dagger2}a^2\rho\longleftrightarrow
- \Big(\alpha^* - cD\Big)^2 \Big(\alpha + c\overline{D}\Big)^2 W&=-\Big\{
\alpha^{*2}\alpha^2 W
+ 2c\,\alpha^{*2}\alpha\overline{D}W
+ c^2\,\alpha^{*2}\overline{D}^2W \nonumber \\
&- 2c\,\alpha^*D\,\alpha^2 W
- 4c^2\,\alpha^*D\,\alpha\overline{D}W
- 2c^3\,\alpha^*D\,\overline{D}^2W \nonumber \\
&+ c^2 D^2\alpha^2 W
+ 2c^3 D^2\alpha\overline{D}W
+ c^4 D^2\overline{D}^2 W
\Big\},
\end{align}
\begin{align}
-\rho a^{\dagger 2} a^2\longleftrightarrow
- \Big(\alpha - c\overline{D}\Big)^2 \Big(\alpha^* + cD\Big)^2 W&= -\Big\{
\alpha^2\alpha^{*2}W
+ 2c\,\alpha^2\alpha^* D W
+ c^2\,\alpha^2 D^2 W \nonumber \\
&
- 2c\,\alpha\overline{D}\,\alpha^{*2} W
- 4c^2\,\alpha\overline{D}\,\alpha^* D W
- 2c^3\,\alpha\overline{D}\,D^2 W\nonumber\\
&
+ c^2\,\overline{D}^2\alpha^{*2} W
+ 2c^3\,\overline{D}^2\alpha^* D W
+ c^4\,\overline{D}^2 D^2 W
\Big\}.
\end{align}
Combining the above expressions, we get
\begin{align}
\mathcal{W}_{\mathcal L_2[\rho]}
&= 2\gamma_2\Big[ \Big(\frac{\partial}{\partial \alpha}\alpha+\frac{\partial}{\partial \alpha^*}\alpha^*\Big)(|\alpha|^2-1)
+ \frac{\partial^2}{\partial \alpha\partial \alpha^*}(2|\alpha|^2-1)\Big]W\nonumber \\
&+\frac{\gamma_2}{2}\Big(\frac{\partial^2}{\partial \alpha^2}\frac{\partial}{\partial \alpha^*}\alpha
+ \frac{\partial }{\partial \alpha}\frac{\partial^2}{\partial (\alpha^*)^2} \alpha^*\Big)W.
\label{eq:second-term}
\end{align}

Using Eqs.~\eqref{eq:zero-term},~\eqref{eq:first-term}, and~\eqref{eq:second-term}, we finally arrive at the evolution equation for the Wigner function corresponding to the Lindblad dynamics given by Eq.~\eqref{eq:vdpo_lee}:
\begin{align}
\frac{\partial W}{\partial t}&=i\omega\left(\alpha \frac{\partial }{\partial \alpha}-\alpha^* \frac{\partial }{\partial \alpha^*}\right)W\nonumber \\
&+\Big\{\left(\frac{\partial }{\partial \alpha}\alpha + \frac{\partial }{\partial \alpha^*}\alpha^*\right)\left[-\gamma_1+2\gamma_2(|\alpha|^2-1)\right]
+\frac{\partial^2}{\partial \alpha \partial \alpha^*}\left[\gamma_1+2\gamma_2(2|\alpha|^2-1)\right] \nonumber \\
&+\frac{\gamma_2}{2}\Big(\frac{\partial^2}{\partial \alpha^2}\frac{\partial}{\partial \alpha^*}\alpha
+ \frac{\partial }{\partial \alpha}\frac{\partial^2}{\partial (\alpha^*)^2} \alpha^*\Big)\Big\}W.
\label{eq:Wigner-evolution}
\end{align}
Recall that $W$ is a function of both $\alpha$ and $\alpha^*$, and hence, the partial derivatives on the right hand side of the above equation also act on $W$. The mappings between the density operator and the Wigner function formalisms have been provided in Table~ \ref{tab:mapping}. Note that as per the rules discussed before Eq. \eqref{eq:rule1} and \eqref{eq:rule2}, the first four relations can be used to deduce the remaining ones.

\begin{table}[!t]
    \centering
    \begin{tabular}{|c|c|}
    \hline
        \textbf{Density operator formalism} & \textbf{Wigner function formalism} \\
        \hline
        $\rho a$ & $(\alpha - c \overline{D}) W$ \\
        \hline
        $a\rho$ & $(\alpha + c \overline{D}) W$ \\
        \hline
        $a^\dagger \rho$ & $(\alpha^* - cD) W$ \\
        \hline
        $\rho a^\dagger$ & $(\alpha^* + cD) W$ \\
        \hline
        $a\rho a^\dagger$ & $(\alpha + c\overline{D})(\alpha^* + cD) W$ \\
        \hline
        $\rho a^\dagger a$ & $(\alpha - c\overline{D})(\alpha^* + cD) W$ \\
        \hline
        $a^\dagger a \rho$ & $(\alpha^* - cD)(\alpha + c\overline{D}) W$ \\
        \hline
        $\rho a^\dagger a$ & $(\alpha - c \overline{D})(\alpha^* + cD) W$\\
        \hline
        $a^\dagger \rho a$ & $(\alpha^* - cD)(\alpha - c \overline{D}) W$ \\
        \hline
        $aa^\dagger \rho$ & $(\alpha + c \overline{D})(\alpha^* - cD) W$ \\
        \hline
        $\rho a a^\dagger$ & $(\alpha^* + cD)(\alpha - c \overline{D}) W$ \\
        \hline
        $a^2 \rho (a^\dagger)^2$ & $(\alpha + c \overline{D})^2(\alpha^* + cD)^2 W$ \\
        \hline
        $(a^\dagger)^2 a^2 \rho$ & $(\alpha^* - cD)^2(\alpha + c \overline{D})^2 W$ \\
        \hline
        $\rho (a^\dagger)^2 a^2$ & $(\alpha - c\overline{D})^2(\alpha^* + cD)^2 W$ \\
        \hline
    \end{tabular}
    \caption{Table showing the mapping between the density operator and Wigner formalisms. Here, $c=1/2$, $D = \partial/\partial\alpha$ and $\overline D = \partial/\partial\alpha^*$.}
    \label{tab:mapping}
\end{table}

\subsubsection{Closed QHO}

\begin{figure*}
    \centering
    \includegraphics[height=8cm]{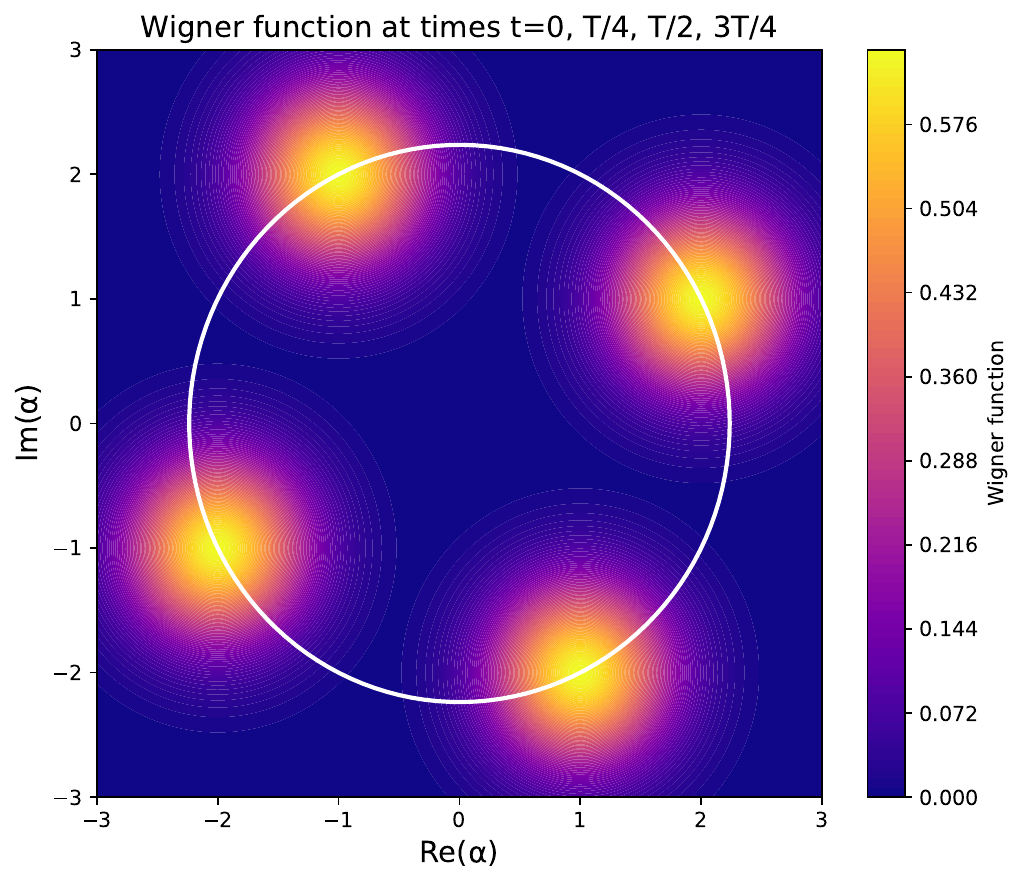}
    \caption{The figure shows for a closed quantum harmonic oscillator of frequency $\omega=1.0$ the time development of the Wigner function $W(\alpha,\alpha^*)$ at times $t=0$, $t=T/4$, $t=T/2$, and $t=3T/4$ (clockwise), starting from a coherent state $|\alpha_0\rangle$, with $\alpha_0=2.0+i$. The time evolution of the Wigner function is given by Eq.~\eqref{eq:Wigner-closed-QHO}. Here, $T=2\pi/\omega$ is the time period of the oscillator. The Wigner function performs a periodic motion on a circle of radius $|\alpha_0|=\sqrt{5}$ with frequency $\omega$ (the circle is depicted in white). The data are obtained by numerically integrating Eq.~\eqref{eq:vdpo_lee} with $\gamma_1=\gamma_2=0$ by using the software \textit{QuTip}.}
    \label{fig:closed-SHO}
\end{figure*}

For the closed QHO, Eq.~\eqref{eq:Wigner-evolution} gives
\begin{align}
\frac{\partial W}{\partial t}&=i\omega\left(\alpha \frac{\partial }{\partial \alpha}-\alpha^* \frac{\partial }{\partial \alpha^*}\right)W.
\label{eq:Wigner-closed-QHO}
\end{align}
Let us say that we prepare the system in a coherent state at the initial time $t=0$: $|\psi(0)\rangle=|\alpha_0\rangle$; using $\hat{\rho}=|\alpha_0\rangle \langle \alpha_0|$ gives $\chi(\eta,\eta^*) = \Tr[|\alpha_0\rangle \langle \alpha_0|e^{-|\eta|^2/2} e^{\eta \hat{a}^\dagger} e^{-\eta^* \hat{a}}]$, where we have used $\hat{D}(\eta)=e^{-|\eta|^2/2} e^{\eta \hat{a}^\dagger} e^{-\eta^* \hat{a}}$.
To proceed, we have 
\begin{align}
\chi(\eta,\eta^*)&=\Tr[e^{-|\eta|^2/2} e^{\eta \hat{a}^\dagger} e^{-\eta^* \alpha_0}|\alpha_0\rangle \langle \alpha_0|]\nonumber \\&= \Tr[e^{-|\eta|^2/2} e^{-\eta^* \alpha_0}|\alpha_0\rangle \langle \alpha_0|e^{\eta \hat{a}^\dagger} ]\nonumber \\
&=e^{-|\eta|^2/2+\eta \alpha_0^* - \eta^* \alpha_0}\Tr[\hat{\rho}].
\end{align}
Using $\Tr[\hat{\rho}]=1$, we get 
\begin{align}
\chi(\eta,\eta^*)=e^{-|\eta|^2/2+\eta \alpha_0^* - \eta^* \alpha_0}=\langle \alpha_0|e^{\eta \hat a^\dagger - \eta^{*}\hat a}|\alpha_0\rangle=\langle\alpha_{0}|\hat{D}(\eta)|\alpha_{0}\rangle.
\end{align}
Next, we use $\hat{D}(\eta)\,|\alpha_{0}\rangle
= e^{(1/2)(\eta \alpha_{0}^{*}-\eta^{*}\alpha_{0})}
\,|\alpha_{0}+\eta\rangle$, see Eq.~\eqref{eq:Dbetaalpha-1}, to get 
\begin{align}
\langle\alpha_{0}|\hat{D}(\eta)|\alpha_{0}\rangle
= e^{(1/2)(\eta \alpha_{0}^{*}-\eta^{*}\alpha_{0})}
\langle\alpha_{0}|\alpha_{0}+\eta\rangle.
\end{align}
Using 
$\langle\alpha_{0}|\alpha_{0}+\eta\rangle
= e^{-(1/2)|\eta|^{2}
+ (1/2)(\alpha_{0}^{*}\eta - \alpha_{0}\eta^{*})}$, see Eq.~\eqref{eq:coherent-state-orthonormal-app}, we get 
$\chi(\eta,\eta^*)
= e^{-(1/2)|\eta|^{2}
+ \eta\,\alpha_{0}^{*}
- \eta^{*}\alpha_{0}}$.
This yields the Wigner function as 
\begin{align}
    W(\alpha,\alpha^*)= \frac{1}{\pi^2}\int d^2\eta~e^{-|\eta|^2/2-\eta(\alpha^*-\alpha_0^*)+\eta^*(\alpha-\alpha_0)}=\frac{2}{\pi}\exp(-2|\alpha-\alpha_0|^2).
    \label{eq:Wigner t=0}
\end{align}
At a later instant $t>0$, using $|\psi(t)\rangle=e^{-i \omega t/2}|\alpha(t)=\alpha e^{-i\omega t}\rangle$ (see the discussion preceding Eq.~\eqref{eq:coherent-n}), one then obtains 
\begin{align}
    W(\alpha,\alpha^*)=\frac{2}{\pi}\exp(-2|\alpha-\alpha_0 e^{-i\omega t}|^2),
    \label{eq:Wigner t>0}
\end{align}
which exactly satisfies Eq.~\eqref{eq:Wigner-closed-QHO}. We thus obtain the Wigner function at all times to be given by  a Gaussian with a peak that drifts in time; in the $\mathrm{Re}(\alpha)-\mathrm{Im}(\alpha)$-plane, the Wigner function performs a periodic motion on a circle of radius $|\alpha_0|$ with frequency $\omega$. This periodic motion is depicted in Fig.~\ref{fig:closed-SHO}. 

\subsubsection{Open QHO}
\label{sec:open-QHO}

Here, our object of study is Eq.~\eqref{eq:Wigner-evolution}, which we may recall has been derived from Eq.~\eqref{eq:vdpo_lee}. The equation can be substantially simplified by resorting to the interaction picture, defined with respect to the QHO Hamiltonian $\hat{H}$. In terms of the unitary time-evolution operator due to $\hat{H}$ given by $\hat{U}(t) = e^{-\frac{i}{\hbar}\hat{H}t}$, we want to rewrite a given Lindblad equation
\begin{align}
 \frac{d\hat{\rho}(t)}{dt}=-\frac{i}{\hbar}[\hat{H},\hat{\rho}(t)]+\mathcal{D}[\hat{O}]\hat{\rho}(t)
 \label{eq:Lindblad-example}
\end{align}
in the interaction picture. In this picture, defining respectively the density operator and the Lindblad operator as 
\begin{align}
\hat{\widetilde\rho}(t) \equiv \hat{U}^\dagger(t)\,\hat{\rho}(t)\,\hat{U}(t),~~\hat{\widetilde O}(t) \equiv \hat{U}^\dagger(t)\,\hat{O}\,\hat{U}(t),
\end{align}
we obtain 
\begin{align}
 \frac{d\hat{\widetilde \rho}(t)}{dt}=\hat{U}^\dagger(t)\,\frac{d\hat{\rho}(t)}{dt}\,\hat{U}(t)+\frac{i}{\hbar}[\hat{H},\hat{\widetilde \rho}(t)].
\end{align}
Using Eq.~\eqref{eq:Lindblad-example}, we see that the Hamiltonian commutator part cancels out, and we get
\begin{align}
    \frac{d\hat{\widetilde \rho}(t)}{dt}=\mathcal{D}[\hat{\widetilde O}]\hat{\widetilde \rho}(t).
\end{align}
We thus see that in the interaction picture, we have in the time evolution equation for the density operator only the Lindblad operator that now has an explicit time dependence. 

In the context of the Lindblad equation~\eqref{eq:vdpo_lee}, using 
\begin{align}
&\hat{U}^\dagger(t)\hat{a}\hat{U}(t)
= \hat{a} e^{-i\omega t},~~\hat{U}^\dagger(t)\hat{a}^\dagger\hat{U}(t)
= \hat{a}^\dagger e^{+i\omega t},
\end{align}
we get
\begin{align}
\hat{\widetilde a}^2(t) = \hat{a}^2 e^{-2i\omega t},~~
\hat{\widetilde a}^\dagger(t) = \hat{a}^\dagger e^{+i\omega t},
\end{align}
leading to the Lindblad equation in the interaction picture as 
\begin{align}
\frac{d\hat{\widetilde \rho}(t)}{dt}
= \gamma_1\,\mathcal{D}\!\big[\hat{a}^\dagger e^{i\omega t}\big]\hat{\widetilde \rho}(t)
+ \gamma_2\,\mathcal{D}\!\big[\hat{a}^2 e^{-2i\omega t}\big]\hat{\widetilde \rho}(t).
\end{align}
Next, noting that for any phase factor $\phi(t)$, one has $\mathcal{D}[e^{i\phi(t)}\hat{O}]\hat{\rho}=2e^{i\phi(t)}\hat{O}\hat{\rho}e^{-i\phi(t)}\hat{O}^\dagger--\hat{O}^\dagger\hat{O}\hat{\rho}-\hat{\rho}\hat{O}^\dagger\hat{O}=\mathcal{D}[\hat{O}]\hat{\rho}$, we get 
\begin{align}
\frac{d\hat{\widetilde \rho}(t)}{dt}
= \gamma_1\,\mathcal{D}\!\big[\hat{a}^\dagger \big]\hat{\widetilde \rho}(t)
+ \gamma_2\,\mathcal{D}\!\big[\hat{a}^2 \big]\hat{\widetilde \rho}(t),
\label{eq:wigner-evolution-interaction-picture}
\end{align}
i.e., the same form as the starting equation~\eqref{eq:vdpo_lee}, but without the Hamiltonian commutator term. From now on, while discussing the open QHO, we will work in the interaction picture, unless stated otherwise. We will drop the tilde for brevity of notation. The equation for the time evolution of the Wigner function consequently reads as 
\begin{align}
\frac{\partial W}{\partial t}&=\Big\{\left(\frac{\partial }{\partial \alpha}\alpha + \frac{\partial }{\partial \alpha^*}\alpha^*\right)\left[-\gamma_1+2\gamma_2(|\alpha|^2-1)\right]
+\frac{\partial^2}{\partial \alpha \partial \alpha^*}\left[\gamma_1+2\gamma_2(2|\alpha|^2-1)\right] \nonumber \\
&+\frac{\gamma_2}{2}\Big(\frac{\partial^2}{\partial \alpha^2}\frac{\partial}{\partial \alpha^*}\alpha
+ \frac{\partial }{\partial \alpha}\frac{\partial^2}{\partial (\alpha^*)^2} \alpha^*\Big)\Big\}W.
\label{eq:Wigner-evolution-1}
\end{align}

\begin{figure*}
    \centering
    \includegraphics[height=5cm]{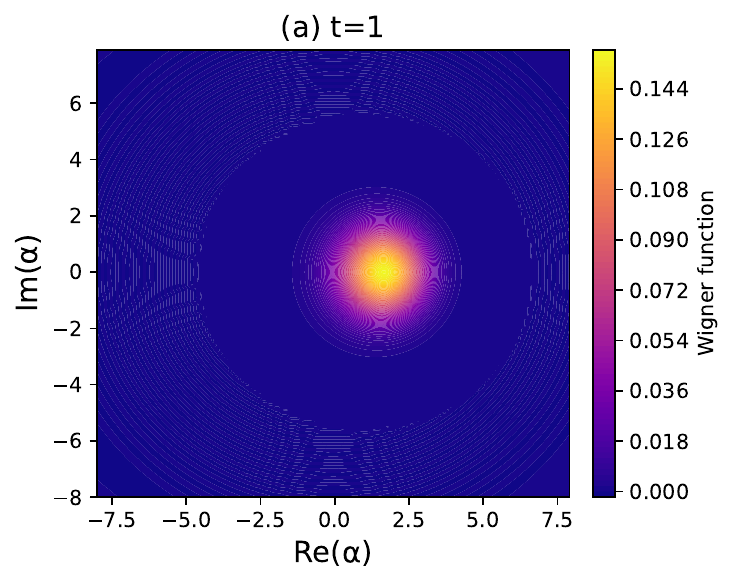}
    \includegraphics[height=5cm]{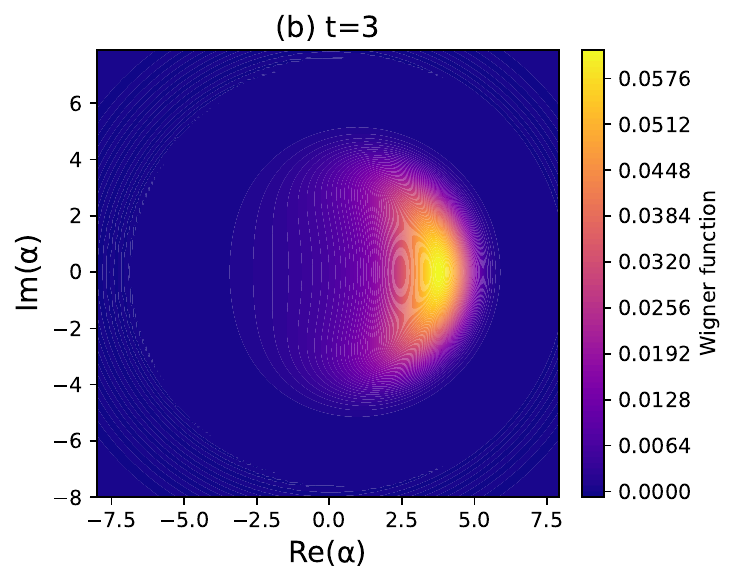}
    \includegraphics[height=5cm]{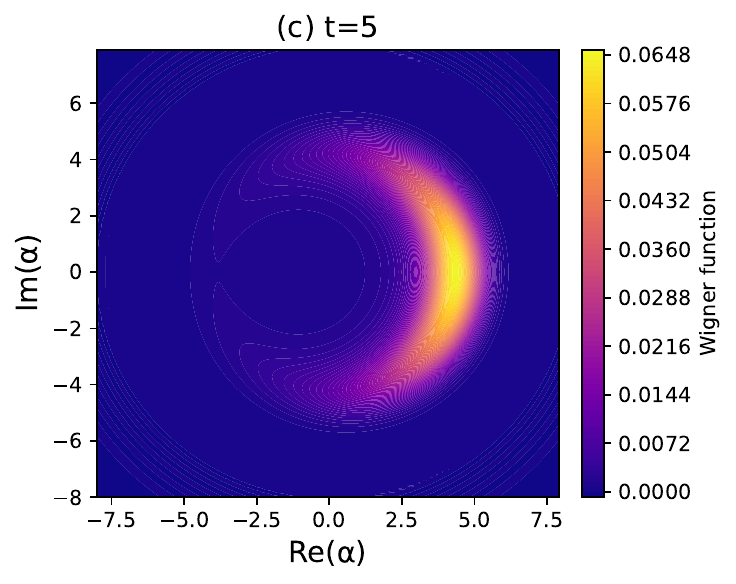}
    \includegraphics[height=5cm]{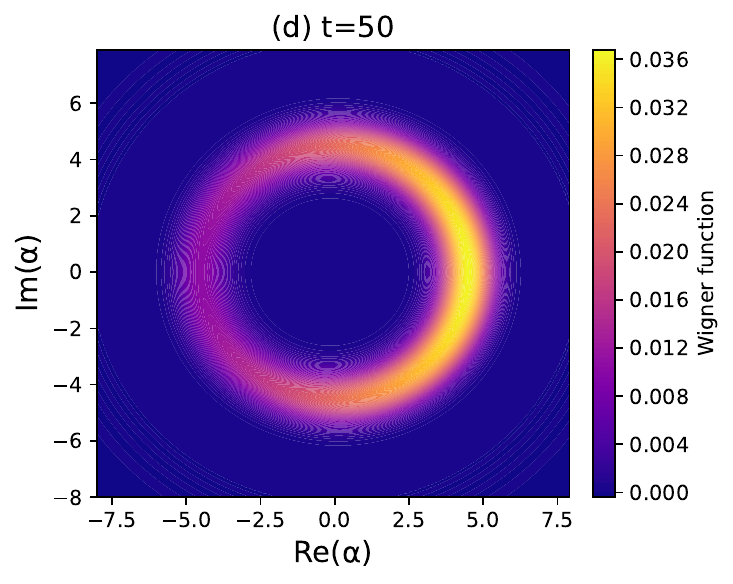}
    \includegraphics[height=5cm]{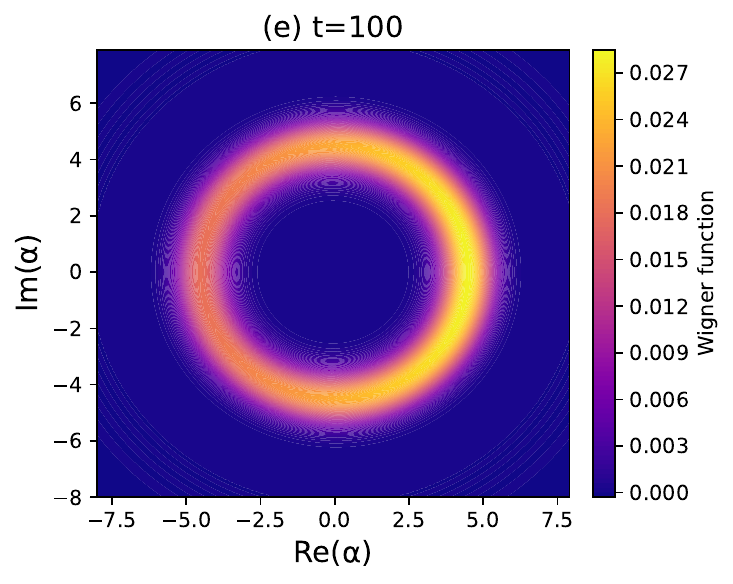}
    \includegraphics[height=5cm]{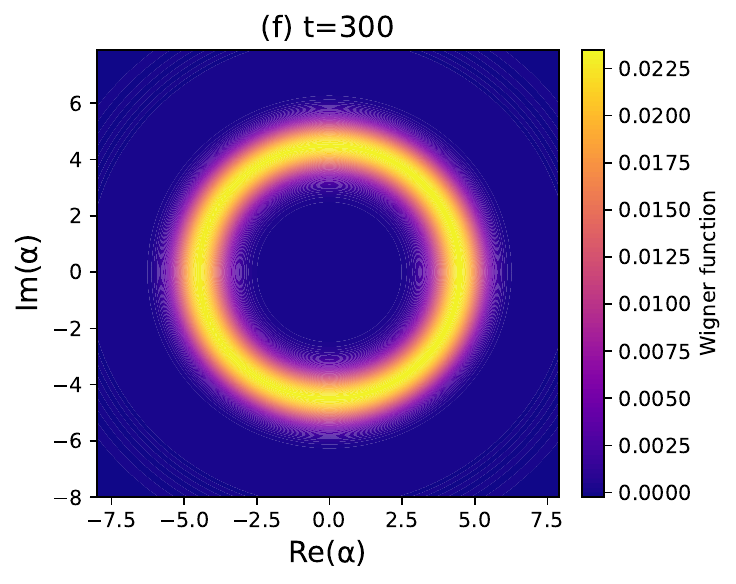}
    \includegraphics[height=5cm]{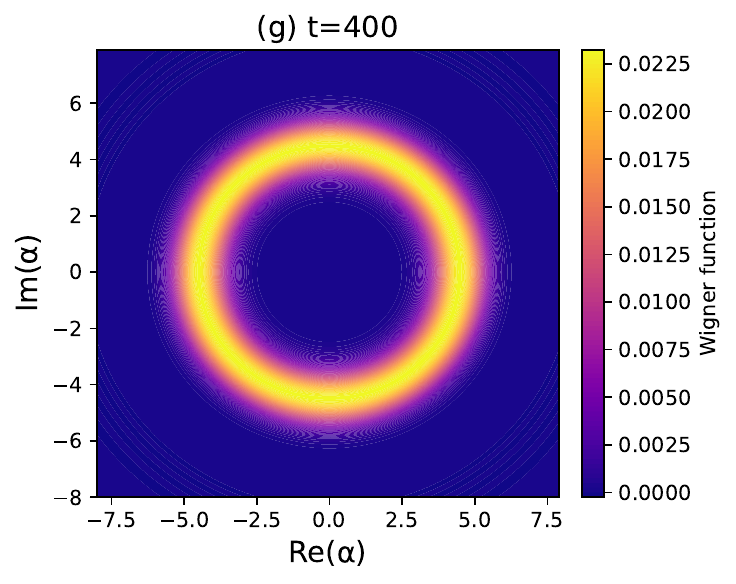}
    \includegraphics[height=5cm]{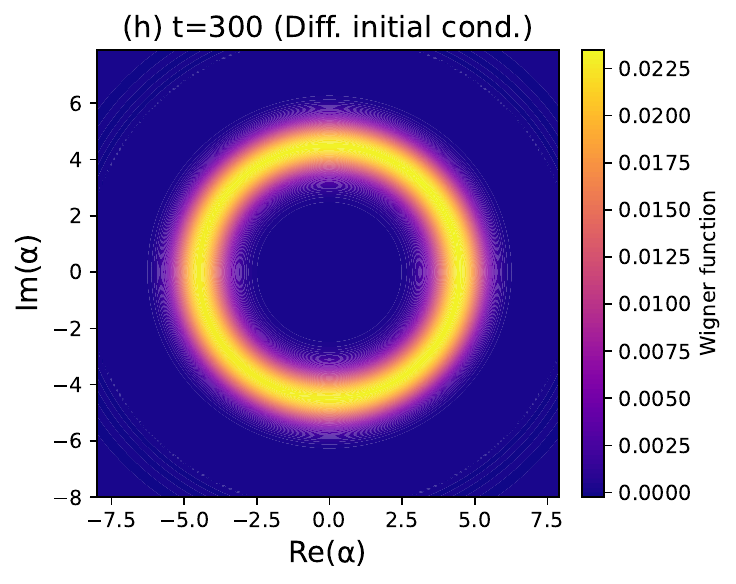}
    \caption{Panels (a) -- (g) show for an open quantum harmonic oscillator and described by the Lindblad equation~\eqref{eq:wigner-evolution-interaction-picture} the time development of  Wigner function $W(\alpha,\alpha^*)$ in the semi-classical ($\gamma_1 \gg \gamma_2$) limit and starting from a coherent state $|\alpha_0\rangle$, with $\alpha_0=1.0$. The time evolution of the Wigner function is given by Eq.~\eqref{eq:Wigner-evolution-1}. Panels (f) and (g) imply that stationary state has been reached. Panel (h) shows that the same stationary state is reached starting from another initial condition, namely, a coherent state $|\alpha_0\rangle$, with $\alpha_0=2+i$. The data are obtained by numerically integrating Eq.~\eqref{eq:wigner-evolution-interaction-picture} using the software \textit{QuTip}.  Here, we have taken $\gamma_1=1.0$, $\gamma_2=0.025$. }
    \label{fig:open-SHO}
\end{figure*}

\begin{figure*}
    \centering
    \includegraphics[height=5cm]{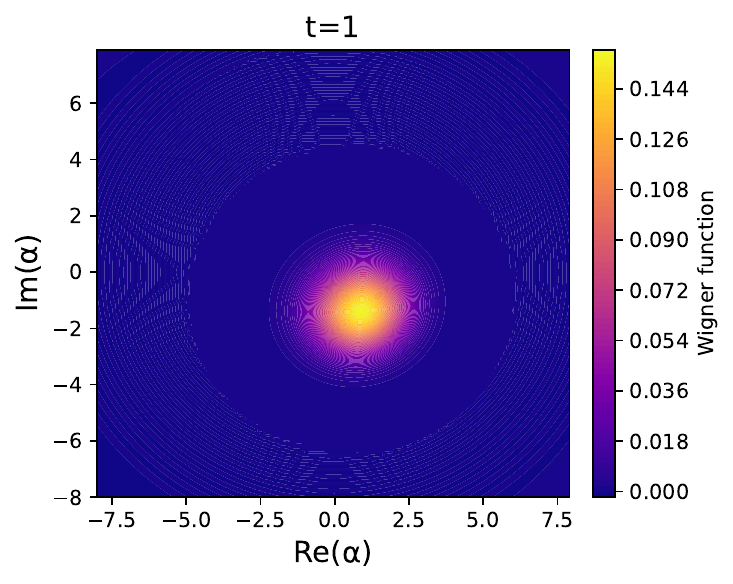}
    \includegraphics[height=5cm]{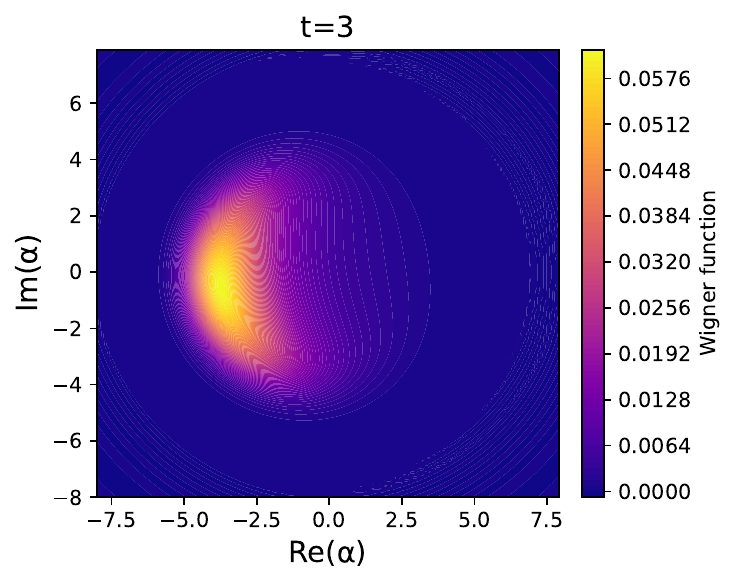}
    \includegraphics[height=5cm]{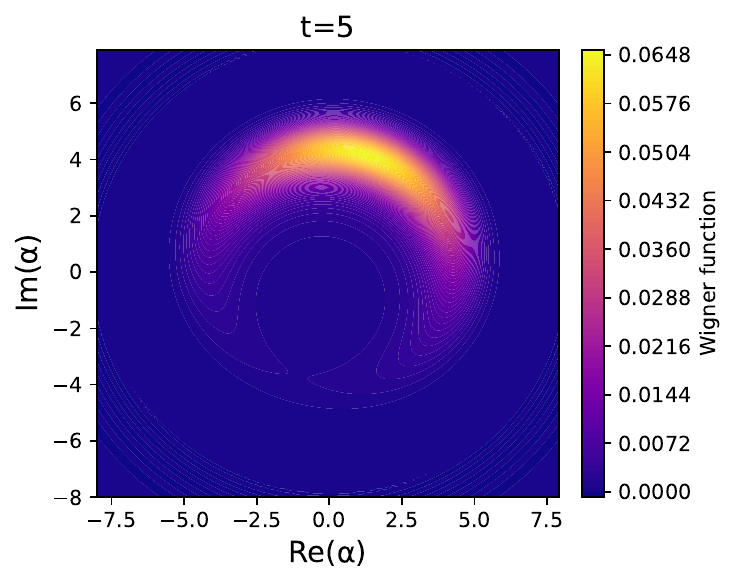}
    \includegraphics[height=5cm]{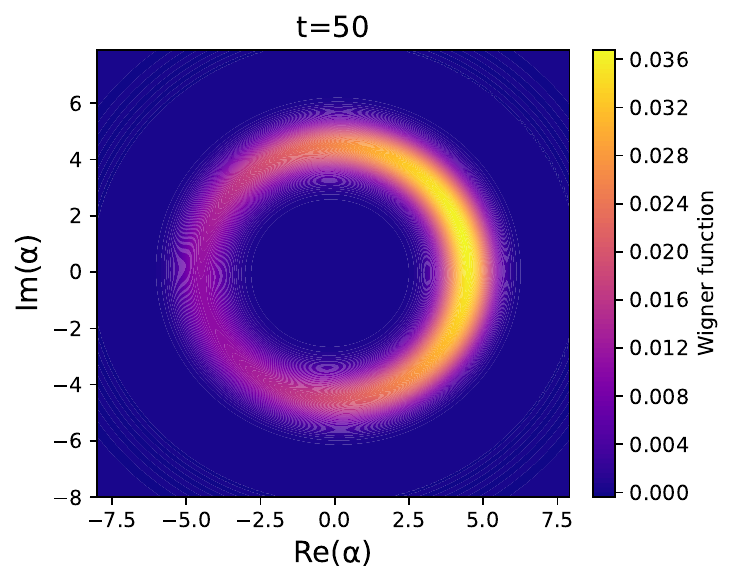}
    \includegraphics[height=5cm]{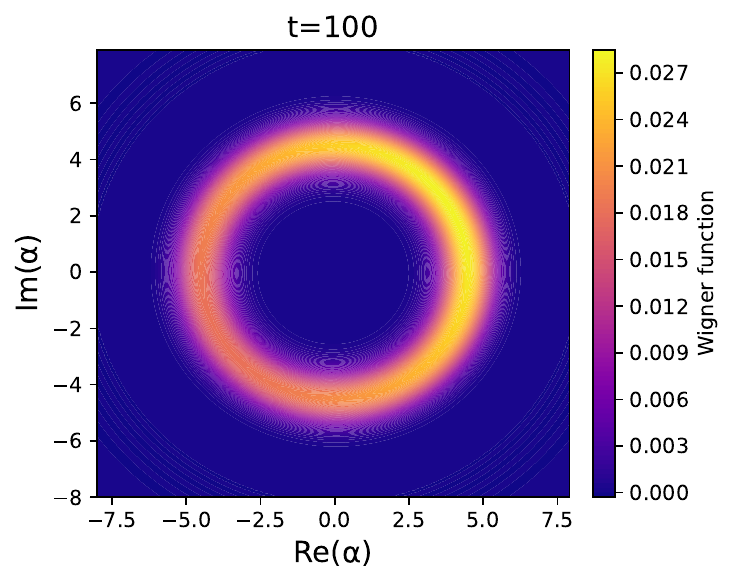}
    \includegraphics[height=5cm]{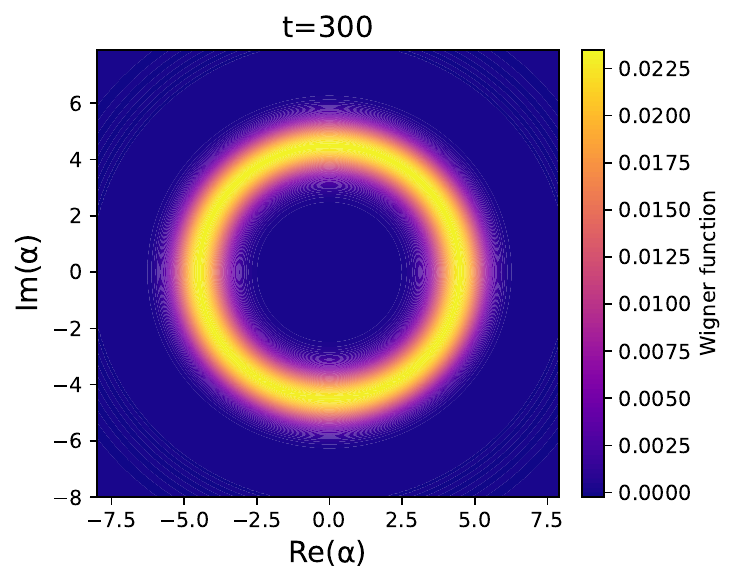}
    \includegraphics[height=5cm]{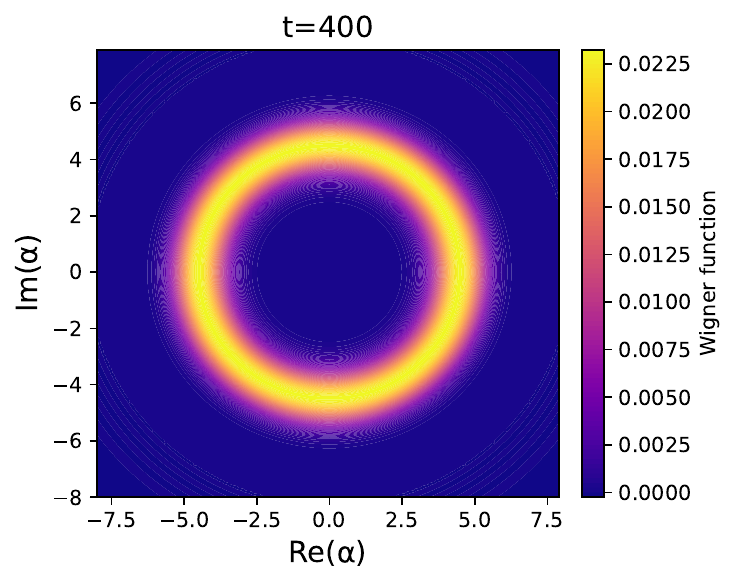}
    \caption{The data in the panels correspond to the same condition as in panels (a) -- (g) of Fig.~\ref{fig:open-SHO}, excepting that here, we are not working in the interaction picture, and hence, the time evolution is described by the Lindblad equation~\eqref{eq:vdpo_lee}, which we may recall describes time evolution in the Schr\"{o}dinger picture.}
    \label{fig:open-SHO-1}
\end{figure*}

For the open QHO under consideration in the interaction picture, the results depicted in Fig.~\ref{fig:open-SHO} and based on Eq.~\eqref{eq:wigner-evolution-interaction-picture} show that in contrast to Fig.~\ref{fig:closed-SHO}, the Wigner function in course of evolution attains a stationary state in which it is spread out over a time-independent closed path in the complex-$\alpha$ plane. The closed path is what is referred to in the literature as a quantum limit cycle. The closed path obtained in the stationary state is independent of the choice of the initial coherent state, as may be seen by comparing the results plotted in panels (g) and (h) in Fig.~\ref{fig:open-SHO}. 
The stationary-state Wigner function corresponds to a time-independent density operator obtained from Eq.~\eqref{eq:wigner-evolution-interaction-picture} by setting its left hand side to zero. In the stationary state, evidently, the photon loss and gain processes balance each other. Figure~\ref{fig:open-SHO} corresponds to the interaction picture; Fig.~\ref{fig:open-SHO-1} shows that not working in the interaction picture, the time evolution is so as to spread out the initial Wigner function but also to rotate it, an implication of the first term on the right hand side of the corresponding evolution, Eq.~\eqref{eq:vdpo_lee}. In this case, the system settles into a time-periodic stationary state. Both Figs.~\ref{fig:open-SHO} and~\ref{fig:open-SHO-1} correspond to the case $\gamma_1 \gg \gamma_2$, which stands for the semi-classical limit of the dynamics~\eqref{eq:wigner-evolution-interaction-picture}, as we discuss below.

\section*{The semi-classical limit and the Fokker-Planck equation}

\begin{figure*}
    \centering
    \includegraphics[height=4cm]{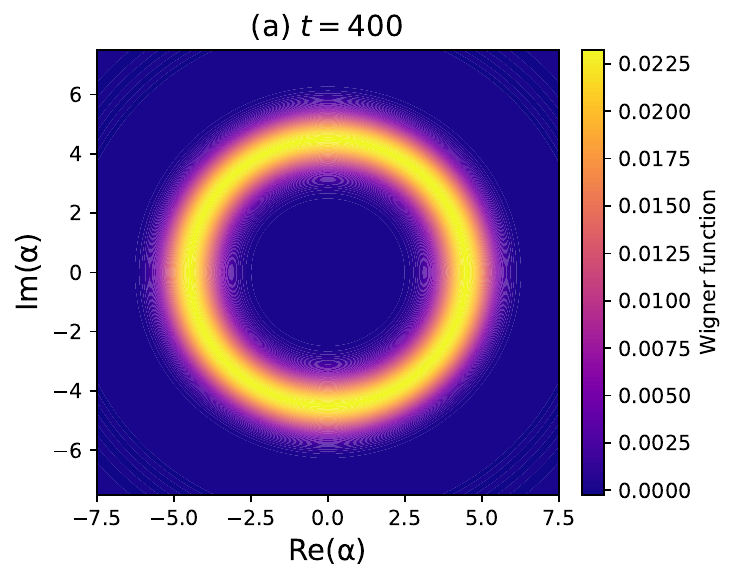}
    \includegraphics[height=4cm]{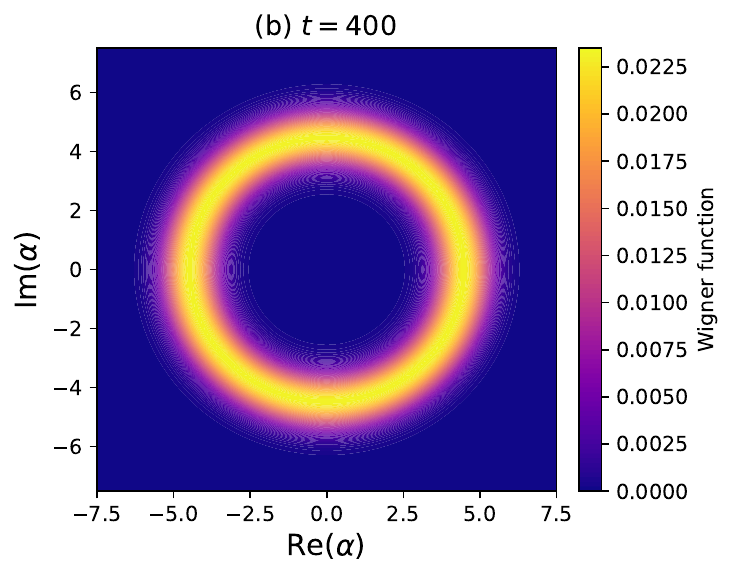}
    \includegraphics[height=4cm]{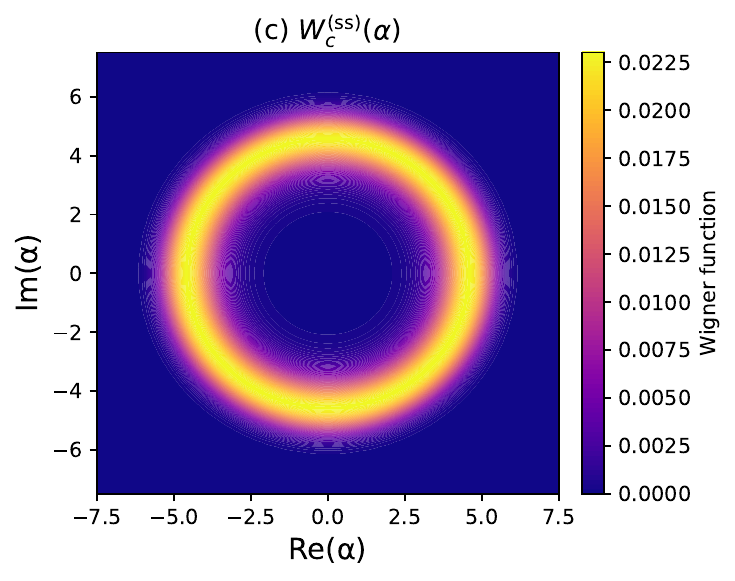}
       \caption{For an open quantum harmonic oscillator and described by the Lindblad equation~\eqref{eq:wigner-evolution-interaction-picture}, the plots show the stationary-state  Wigner function $W(\alpha,\alpha^*)$ in the semi-classical ($\gamma_1 \gg \gamma_2$) limit; here, $\gamma_1=1.0$ and $\gamma_2=0.025$. The time evolution of the Wigner function is given by Eq.~\eqref{eq:Wigner-evolution-1} for panel (a), and by Eq.~\eqref{eq:FP-1} for panel (b). For (a), the data are obtained by numerically integrating Eq.~\eqref{eq:wigner-evolution-interaction-picture} using the software \textit{QuTip} and starting from a coherent state $|\alpha_0\rangle$, with $\alpha_0=1.0$, and corresponds to time $t=400$. For (b), the data for the same initial condition are obtained by numerically integrating Eq.~\eqref{eq:FP-1}, and corresponds to time $t=400$. Panel (c) shows the analytical result~\eqref{eq:openSHO-analytical}; one may observe a very good agreement with panels (a) and (b).}
    \label{fig:open-SHO-semiclassical}
\end{figure*}

We begin from Eq.~\eqref{eq:Wigner-evolution-1}, which in terms of two functions
\begin{align}
&B(\alpha,\alpha^*)\equiv -\gamma_1 + 2\gamma_2(|\alpha|^2-1),\\
&A(\alpha,\alpha^*)\equiv \gamma_1 + 2\gamma_2(2|\alpha|^2-1), \label{eq:A-defn}
\end{align}
writes as
\begin{align}
\frac{\partial W}{\partial t}&=\Big\{\left(\frac{\partial }{\partial \alpha}\alpha + \frac{\partial }{\partial \alpha^*}\alpha^*\right)B(\alpha,\alpha^*)
+\frac{\partial^2}{\partial \alpha \partial \alpha^*}A(\alpha,\alpha^*) \nonumber \\
&+\frac{\gamma_2}{2}\Big(\frac{\partial^2}{\partial \alpha^2}\frac{\partial}{\partial \alpha^*}\alpha
+ \frac{\partial }{\partial \alpha}\frac{\partial^2}{\partial (\alpha^*)^2} \alpha^*\Big)\Big\}W.
\label{eq:Wigner-evolution-2}
\end{align}

The semi-classical limit of~\eqref{eq:Wigner-evolution-2} involves converting it into a Fokker-Planck equation with a diffusion coefficient that is a constant. To this end, we first ignore the third--order--derivative term on the right hand side, resulting in the equation
\begin{align}
    \frac{\partial W}{\partial t}=-\left(\frac{\partial }{\partial \alpha}[-\alpha BW]+\frac{\partial }{\partial \alpha^*}[-\alpha^* BW]\right)+\left(\frac{\partial^2}{\partial \alpha \partial \alpha^*}\right)[AW].
    \label{eq:FP-1}
\end{align}
The above has the form of a Fokker-Planck equation (see also~\ref{sec:Separation of FP}), albeit with a diffusion coefficient that is not a constant. To do the identification, let us note that the Fokker-Planck equation for a probability distribution $P(\alpha,\alpha^*,t)$ in two variables $\alpha,\alpha^*$ has the general form
\begin{align}
    \frac{\partial P}{\partial t}&=-\left(\frac{\partial}{\partial \alpha}F_\alpha+\frac{\partial}{\partial \alpha^*}F_{\alpha^*}\right)P\nonumber \\
    &+\left(\frac{\partial^2}{\partial \alpha^2}D_{\alpha,\alpha}+\frac{\partial^2}{\partial \alpha^{*2}}D_{\alpha^*\alpha^*}+\frac{\partial^2}{\partial \alpha \partial \alpha^*}D_{\alpha \alpha^*}+\frac{\partial^2}{\partial \alpha^* \partial \alpha}D_{\alpha^*\alpha}\right)P,
\end{align}
with $F_\alpha$, $F_{\alpha^*}$, $D_{\alpha \alpha}$, $D_{\alpha^*\alpha^*}$, $D_{\alpha \alpha^*}$, $D_{\alpha^* \alpha}$ all being functions of $\alpha$ and $\alpha^*$. Here, $F_\alpha$'s and the $D_{\alpha\alpha}$'s are the drift and the diffusion coefficients, respectively. When one has $D_{\alpha \alpha^*}=D_{\alpha^* \alpha}$, the equation has the form
\begin{align}
    \frac{\partial P}{\partial t}&=-\left(\frac{\partial}{\partial \alpha}F_\alpha+\frac{\partial}{\partial \alpha^*}F_{\alpha^*}\right)P+\left(\frac{\partial^2}{\partial \alpha^2}D_{\alpha,\alpha}+\frac{\partial^2}{\partial \alpha^{*2}}D_{\alpha^*\alpha^*}+2\frac{\partial^2}{\partial \alpha \partial \alpha^*}D_{\alpha \alpha^*}\right)P.
\end{align}
On comparing the above equation with Eq.~\eqref{eq:FP-1}, we obtain 
\begin{align}
    F_\alpha=-\alpha B,~F_{\alpha^*}=-\alpha^*B,~D_{\alpha \alpha^*}=A/2,~D_{\alpha \alpha}=D_{\alpha^*\alpha^*}=0.
\end{align}

We now move on to convert Eq.~\eqref{eq:FP-1} into one that involves a constant diffusion coefficient. To proceed, we note that the deterministic ($A=0$) dynamics implies a nontrivial stationary state, a stable fixed point, defined by the condition $B(\alpha_0,\alpha_0^*) = 0$, which yields
\begin{align}
|\alpha_0|^2 = 1 + \frac{\gamma_1}{2\gamma_2}.
\label{eq:steady-ring}
\end{align}
Correspondingly, one has 
\begin{align}
    W_0(\alpha,\alpha^*)=\frac{\delta(|\alpha|^2-|\alpha_0|^2)}{2\pi |\alpha_0|},
\end{align}
which describes a ring in the $\mathrm{Re}(\alpha)-\mathrm{Im}(\alpha)$-plane. Choosing $\gamma_1 \gg \gamma_2$ ensures that one has $|\alpha_0|^2 \gg 1$, and the aforementioned ring coincides with the limit cycle obtained in the semi-classical limit of the dynamics~\eqref{eq:vdpo_lee} and discussed in the paragraph in and around Eq.~\eqref{eq:angular-velocity-definition}. Indeed, the radius of the limit cycle is $\sqrt{\gamma_1/(2\gamma_2)}$, as anticipated there.

For $A\ne 0$, the ring will be not as sharp as for $A=0$. Let us explore the conditions under which $W$ will not be a $\delta$-function as for $A=0$, but will have a little spread. From the form of the function $A$, see Eq.~\eqref{eq:A-defn},  we know that it has the smallest possible value on the ring, and its value increases as one moves away from the ring. Indeed, on the ring, $A$ has the value 
\begin{align}
A_0 \equiv A(\alpha_0,\alpha_0^*)=\gamma_1 + 2\gamma_2(2|\alpha_0|^2-1) = 3\gamma_1 + 2\gamma_2.
\end{align}
As one moves away from the ring, we will have  $2|\alpha_0|^2-1 > \gamma_1/(2\gamma_2)$, so that one will have $A>A_0$. Then, to ensure the narrowness condition stated above, we must assume that $A$ remains pinned to the value it has on the ring. Under such an approximation, we obtain from Eq.~\eqref{eq:FP-1} the corresponding Fokker-Planck equation as 
\begin{align}
    \frac{\partial W}{\partial t}=-\left(\frac{\partial }{\partial \alpha}[-\alpha BW]+\frac{\partial }{\partial \alpha^*}[-\alpha^* BW]\right)+A_0\left(\frac{\partial^2}{\partial \alpha \partial \alpha^*}\right)W,
    \label{eq:FP-2}
\end{align}
which involves a constant diffusion coefficient. The above equation yields a Wigner function that describes small
fluctuations about the noiseless stationary ring 
$|\alpha|=|\alpha_0|$ given by Eq.~\eqref{eq:steady-ring}. In particular, in the semi-classical limit $\gamma_1 \gg \gamma_2$, when $|\alpha_0|^2 \gg 1$, the Wigner function peaked around the stationary ring will be even narrower, and the dynamics~\eqref{eq:FP-2} will be a faithful representation of the dynamics~\eqref{eq:FP-1}. Corresponding to the equivalent semi-classical model~\eqref{eq:FP-1}, one may write down a Langevin equation
\begin{align}
    \dot{\alpha}=\alpha(\gamma_1+2\gamma_2-2\gamma_2|\alpha|^2)+\xi^R(t)+i\xi^I(t),
    \label{eq:Langevin-1}
\end{align}
where both $\xi^R(t)$ and $\xi^I(t)$ are Gaussian, white noise with properties
\begin{align}
    &\langle \xi^R(t)\rangle=\langle \xi^I(t)\rangle=0,\\
    &\langle \xi^R(t)\xi^R(t')\rangle=\langle \xi^I(t)\xi^I(t')\rangle=\left(\frac{3\gamma_1}{2}+\gamma_2\right)\delta(t-t'),~~\langle \xi^R(t)\xi^I(t)\rangle=0.
    \label{eq:noise correlations}
\end{align}
Here, angular brackets denote averaging over noise realizations. Based on the above discussion, we conclude that the semi-classical limit of the dynamics~\eqref{eq:vdpo_lee} corresponds to the case $\gamma_2 \ll 1$ for a fixed $\gamma_1$, while the quantum limit corresponds to the case $\gamma_2 \gg 1$. In the latter scenario, the oscillator is near the ground state and quantum fluctuations
play an important role in the dynamics. By contrast, in the semi-classical limit, the system can access dynamically a very large number of energy eigenstates. The stationary state of Eq.~\eqref{eq:FP-2} can be checked to be given by 
\begin{align}
    W^{(\mathrm{ss})}(\alpha) \propto e^{\left\{(2/(3\gamma_1+2\gamma_2))[(\gamma_1+2\gamma_2)|\alpha|^2-\gamma_2|\alpha|^4]\right\}}. 
    \label{eq:openSHO-analytical}
\end{align}
In Fig.~\eqref{fig:open-SHO-semiclassical}, we show that in the semi-classical limit $\gamma_1 \gg \gamma_2$, the successive approximations, Eqs.~\eqref{eq:Wigner-evolution-1} and~\eqref{eq:FP-1}, describe quite faithfully the original system, Eq.~\eqref{eq:wigner-evolution-interaction-picture}. 

Equation~\eqref{eq:Wigner-evolution-1}, on the other hand, cannot be solved exactly. The way around is to consider the parent equation, namely, the Lindblad equation~\eqref{eq:vdpo_lee}, which can solved in the extreme quantum limit $\gamma_2 \to \infty$, when the system may be taken to be essentially accessing the ground state $|0\rangle$ with energy $E_0$, the first excited state $|1\rangle$ with energy $E_1$ and the second excited state $|2 \rangle$ with energy $E_2$. To this end, we first transform the equation into a linear one using the vectorization technique detailed in Appendix~E. The vectorized Lindblad equation reads as
\begin{align}
\dot{\vec{\rho}} = (-iH_\times + \gamma_1 \mathcal{W}_1+\gamma_2 \mathcal{W}_2)\vec{\rho},
\label{eq:Lindblad-vector}
\end{align}
with
\begin{align}
H_\times = I \otimes H - H^T \otimes I,
\end{align}
where $I$ is the $3\times3$ identity matrix. The jump operator $\mathcal{D}[\hat{a}^\dagger]\hat{\rho}$ (respectively, $\mathcal{D}[\hat{a}^2]\hat{\rho}$) when vectorized yields the $9\times 9$ matrix $\mathcal{W}_1$ (respectively, $\mathcal{W}_2$). Using the Fock states $|n\rangle$ as the basis, we have
\begin{align}
H_\times =
\begin{pmatrix}
0 & 0 & 0 & 0 & 0 & 0 & 0 & 0 & 0\\
0 & E_{10} & 0 & 0 & 0 & 0 & 0 & 0 & 0\\
0 & 0 & E_{20} & 0 & 0 & 0 & 0 & 0 & 0\\
0 & 0 & 0 & E_{01} & 0 & 0 & 0 & 0 & 0\\
0 & 0 & 0 & 0 & 0 & 0 & 0 & 0 & 0\\
0 & 0 & 0 & 0 & 0 & E_{21} & 0 & 0 & 0\\
0 & 0 & 0 & 0 & 0 & 0 & E_{02} & 0 & 0\\
0 & 0 & 0 & 0 & 0 & 0 & 0 & E_{12} & 0\\
0 & 0 & 0 & 0 & 0 & 0 & 0 & 0 & 0
\end{pmatrix},
\end{align}
where we have defined $E_{ij} = E_i - E_j$ for $i,j = 0,1,2$. Also, we have 
\begin{align}
\mathcal{W}_1 = 2L_1^\ast \otimes L_1 - I \otimes L_1^\dagger L_1 - (L_1^\dagger L_1)^T \otimes I,
\end{align}
with
\begin{align}
L_1 =
\begin{pmatrix}
0 & 0 & 0\\
1 & 0 & 0\\
0 & \sqrt{2} & 0
\end{pmatrix},
\quad
L_1^\dagger L_1 =
\begin{pmatrix}
1 & 0 & 0\\
0 & 2 & 0\\
0 & 0 & 0
\end{pmatrix},
\end{align}
yielding 
\begin{align}
\mathcal{W}_1 =
\begin{pmatrix}
-2 & 0 & 0 & 0 & 0 & 0 & 0 & 0 & 0\\
0 & -3 & 0 & 0 & 0 & 0 & 0 & 0 & 0\\
0 & 0 & -1 & 0 & 0 & 0 & 0 & 0 & 0\\
0 & 0 & 0 & -3 & 0 & 0 & 0 & 0 & 0\\
2 & 0 & 0 & 0 & -4 & 0 & 0 & 0 & 0\\
0 & 2\sqrt{2} & 0 & 0 & 0 & -2 & 0 & 0 & 0\\
0 & 0 & 0 & 0 & 0 & 0 & -1 & 0 & 0\\
0 & 0 & 0 & 2\sqrt{2} & 0 & 0 & 0 & -2 & 0\\
0 & 0 & 0 & 0 & 4 & 0 & 0 & 0 & 0
\end{pmatrix}.
\label{eq:W1}
\end{align}
Similarly, for
\begin{align}
L_2 =
\begin{pmatrix}
0 & 0 & \sqrt{2}\\
0 & 0 & 0\\
0 & 0 & 0
\end{pmatrix},
\quad
L_2^\dagger L_2 =
\begin{pmatrix}
0 & 0 & 0\\
0 & 0 & 0\\
0 & 0 & 2
\end{pmatrix},
\end{align}
we find
\begin{align}
\mathcal{W}_2 =
\begin{pmatrix}
0 & 0 & 0 & 0 & 0 & 0 & 0 & 0 & 4\\
0 & 0 & 0 & 0 & 0 & 0 & 0 & 0 & 0\\
0 & 0 & -2 & 0 & 0 & 0 & 0 & 0 & 0\\
0 & 0 & 0 & 0 & 0 & 0 & 0 & 0 & 0\\
0 & 0 & 0 & 0 & 0 & 0 & 0 & 0 & 0\\
0 & 0 & 0 & 0 & 0 & -2 & 0 & 0 & 0\\
0 & 0 & 0 & 0 & 0 & 0 & -2 & 0 & 0\\
0 & 0 & 0 & 0 & 0 & 0 & 0 & -2 & 0\\
0 & 0 & 0 & 0 & 0 & 0 & 0 & 0 & -4
\end{pmatrix}.
\label{eq:W2}
\end{align}

On the basis of the above, we get
\begin{align}
&-iH_\times + \gamma_1 \mathcal{W}_1 + \gamma_2 \mathcal{W}_2 \nonumber \\
&=
\scriptsize
\begin{pmatrix}
-2\gamma_1 & 0 & 0 & 0 & 0 & 0 & 0 & 0 & 4\gamma_2\\
0 & -iE_{10}-3\gamma_1 & 0 & 0 & 0 & 0 & 0 & 0 & 0\\
0 & 0 & -iE_{20}-\gamma_1-2\gamma_2 & 0 & 0 & 0 & 0 & 0 & 0\\
0 & 0 & 0 & -iE_{01}-3\gamma_1 & 0 & 0 & 0 & 0 & 0\\
2\gamma_1 & 0 & 0 & 0 & -4\gamma_1 & 0 & 0 & 0 & 0\\
0 & 2\sqrt{2}\gamma_1 & 0 & 0 & 0 & -iE_{10}-2\gamma_1-2\gamma_2 & 0 & 0 & 0\\
0 & 0 & 0 & 0 & 0 & 0 & -iE_{02}-\gamma_1-2\gamma_2 & 0 & 0\\
0 & 0 & 0 & 2\sqrt{2}\gamma_1 & 0 & 0 & 0 & -iE_{12}-2\gamma_1-2\gamma_2 & 0\\
0 & 0 & 0 & 0 & 4\gamma_1 & 0 & 0 & 0 & -4\gamma_2
\end{pmatrix}.
\label{eq:rho-evolution-operator}
\end{align}
Clearly, the matrix on the right has zero determinant, which implies  that the matrix $-iH_\times + \gamma_1 \mathcal{W}_1 + \gamma_2 \mathcal{W}_2$ has at least one zero-eigenvalue. Equation~\eqref{eq:Lindblad-vector} has the solution
\begin{align}
\vec{\rho}(t) = e^{(-iH_\times+\gamma_1 \mathcal{W}_1+\gamma_2 \mathcal{W}_2) t}\vec{\rho}(0),
\end{align}
from which it follows that the stationary state corresponds to the eigenvector with the zero eigenvalue. This yields the stationary-state density matrix as
\begin{align}
\hat{\rho}_\mathrm{ss}=\frac{1}{1+\frac{3\gamma_2}{\gamma_1}}
    \begin{pmatrix}
    \frac{2\gamma_2}{\gamma_1} & 0 & 0 \\
    0 & \frac{\gamma_2}{\gamma_1} & 0\\
    0 & 0 & 1
    \end{pmatrix}.
\end{align}
In the extreme quantum limit $\gamma_2 \to \infty$, we get the stationary state as
\begin{align}
\hat{\rho}_\mathrm{ss}=\begin{pmatrix}
    \frac{2}{3} & 0 & 0 \\
    0 & \frac{1}{3} & 0\\
    0 & 0 & 0
    \end{pmatrix}=\frac{2}{3}|0\rangle \langle0|+\frac{1}{3}|1\rangle \langle 1|.
    \label{eq:three-level-rho-ss}
\end{align}

Our aim now is to obtain the Wigner function corresponding to the state~\eqref{eq:three-level-rho-ss}. Using the displaced--parity representation of the Wigner function, Eq.~\eqref{eq:W-displaced-parity}, we get for the energy eigenstate $|n\rangle$ (equivalently, $\rho = |n\rangle\langle n|$) that
\begin{align}
W_n(\alpha)=\frac{2}{\pi}(-1)^n \langle n|\hat{D}(2\alpha)|n\rangle.
\label{eq:Wn-start}
\end{align}
Using $\hat{D}(\xi)=e^{-|\xi|^2/2}\, e^{\xi \hat{a}^\dagger} e^{-\xi^* \hat{a}}$ gives 
\begin{align}
\langle n|\hat{D}(\xi)|n\rangle
= e^{-|\xi|^2/2}
\sum_{k,\ell\ge0}
\xi^{k}(-\xi^*)^{\ell}/(k!\,\ell!)\,
\langle n|(\hat{a}^\dagger)^k \hat{a}^\ell|n\rangle.
\end{align}
Now, the matrix element \(\langle n|(\hat{a}^\dagger)^k \hat{a}^\ell|n\rangle\) vanishes unless \(k=\ell\), 
and for \(k=\ell\le n\), we have $\langle n|(\hat{a}^\dagger)^k \hat{a}^k|n\rangle = n!/(n-k)!$.
This gives
\begin{align}
 \langle n|\hat{D}(\xi)|n\rangle
= e^{-|\xi|^2/2}
\sum_{k=0}^{n}
(-1)^k \frac{n!}{(n-k)!}
\frac{|\xi|^{2k}}{(k!)^{2}}=e^{-|\xi|^2/2}
\sum_{k=0}^{n}(-1)^k
\binom{n}{k}\frac{(|\xi|^{2})^{k}}{k!},   
\end{align}
where we have used $n!/((n-k)!\,(k!)^{2})=\binom{n}{k}/k!$.
Next, recognizing the Laguerre polynomial, $L_n(x)=\sum_{k=0}^{n}(-1)^k\binom{n}{k}x^k/k!$,
we get
\begin{align}
\langle n|\hat{D}(\xi)|n\rangle
= e^{-|\xi|^2/2}\,L_n(|\xi|^2),
\label{eq:Dnn}
\end{align}
which on using Eq.~\eqref{eq:Wn-start} gives
\begin{align}
W_n(\alpha)
= \frac{2}{\pi}(-1)^n
e^{-2|\alpha|^2}\,
L_n\!\left(4|\alpha|^2\right).
\end{align}
Using $L_0(x)=1$, $L_1(x)=1-x$, we obtain for the state~\eqref{eq:three-level-rho-ss} the stationary-state Wigner function 
\begin{align}
    W^{(\mathrm{ss})}(\alpha)=\frac{2}{3\pi}(1+4|\alpha|^2)e^{-2|\alpha|^2}.
\end{align}

\section{Driven quantum limit cycles}
\label{sec:sec5}

Having established the occurrence of limit cycles under dynamical evolution~\eqref{eq:vdpo_lee}, we now proceed to discuss the case when the QHO is driven externally by a field that has amplitude $\mathcal{E}$ and frequency $\Omega \ne \omega$. The Hamiltonian of the system reads as
\begin{align}
\hat{H}_D(t)=\hat{H}+\hat{V}(t), \qquad \hat{H}=\hbar\omega\left(\hat{a}^\dagger \hat{a}+\frac{1}{2}\right),
\end{align}
where the time–dependent drive is
\begin{align}
\hat{V}(t)=\mathcal{E}\cos(\Omega t)\,\hat{x}.
\end{align}
Writing the position operator as
\begin{align}
\hat{x} = x_\mathrm{zp}(\hat{a}+\hat{a}^\dagger), \qquad
x_\mathrm{zp}=\sqrt{\frac{\hbar}{2m\omega}},
\end{align}
see~\ref{app:Coherent}, the Schr\"{o}dinger–picture drive Hamiltonian is
\begin{align}
\hat{V}(t)=\mathcal{E} x_\mathrm{zp}\cos(\Omega t)\,(\hat{a}+\hat{a}^\dagger).
\end{align}

We now want to transform to a rotating frame rotating at frequency $\Omega$ with respect to the inertial (laboratory) frame. In order to transform from the latter to the rotating frame, we define the unitary operator $\hat{R}(t)\equiv e^{i\Omega \hat{a}^\dagger \hat{a}\,t}$. To obtain the Hamiltonian in this rotating frame, we proceed as follows. Consider a system with a time-dependent Hamiltonian $\hat{H}(t)$. The Schr\"odinger-picture equation of motion reads as
\begin{align}
i\hbar \frac{d}{dt}\ket{\psi(t)} = \hat{H}(t)\ket{\psi(t)}.
\end{align}
Transforming to a rotating frame via a time-dependent unitary
transformation,
\begin{align}
\ket{\psi_\mathrm{rot}(t)} =\hat{R}(t)\ket{\psi(t)},
\end{align}
we obtain
\begin{align}
i\hbar \frac{d}{dt}\ket{\psi_\mathrm{rot}}
= 
\left[
\hat{R}\hat{H}(t) \hat{R}^\dagger
+ i\hbar \left(\frac{\partial \hat{R}}{\partial t}\right)\hat{R}^\dagger
\right]
\ket{\psi_\mathrm{rot}},
\end{align}
giving the Hamiltonian in the rotating frame as
\begin{align}
\hat{H}_\mathrm{rot}(t)
=
\hat{R}(t) \hat{H}(t) \hat{R}^\dagger(t)
+ i\hbar \Big(\frac{\partial \hat{R}(t)}{\partial t}\Big) \hat{R}^\dagger(t).
\end{align}

For the harmonic oscillator problem at hand, we choose
\begin{align}
\hat{R}(t)=e^{i\Omega \hat{a}^\dagger \hat{a}\, t}.
\end{align}
Using $\hat{R} \hat{a} \hat{R}^\dagger = \hat{a} e^{-i\Omega t}$, $\hat{R} \hat{a}^\dagger \hat{R}^\dagger = \hat{a}^\dagger e^{i\Omega t}$, 
$i\hbar\Big(\frac{\partial \hat{R}}{\partial t}\Big)\hat{R}^\dagger
= i\hbar(i\Omega \hat{a}^\dagger \hat{a})
= -\hbar\Omega a^\dagger a$, one obtains 
\begin{align}
\hat{R}\hat{H}\hat{R}^\dagger + i\hbar\Big(\frac{\partial \hat{R}}{\partial t}\Big)\hat{R}^\dagger
= \hbar(\omega-\Omega)a^\dagger a
\equiv \hbar\Delta a^\dagger a,
\end{align}
with detuning $\Delta$ defined as
\begin{align}
\Delta \equiv \omega-\Omega.
\end{align}
On the other hand, we have
\begin{align}
\hat{R} \hat{V}(t) \hat{R}^\dagger&= \mathcal{E} x_\mathrm{zp}\cos(\Omega t)
\left(\hat{a} e^{-i\Omega t}+\hat{a}^\dagger e^{i\Omega t}\right)\nonumber \\
&=\frac{\mathcal{E} x_\mathrm{zp}}{2}\Big[
\hat{a}
+ \hat{a}^\dagger
+ \hat{a} e^{-2i\Omega t}
+ \hat{a}^\dagger e^{2i \Omega t}
\Big].
\end{align}
Hence, we get
\begin{align}
\hat{H}_\mathrm{rot}(t)=\hbar\Delta\, \hat{a}^\dagger \hat{a}
+\frac{\mathcal{E} x_\mathrm{zp}}{2}(\hat{a}+\hat{a}^\dagger)
+\frac{\mathcal{E} x_\mathrm{zp}}{2}\left(
\hat{a} e^{-2i\Omega t}+\hat{a}^\dagger e^{2i\Omega t}
\right).
\end{align}
In the next step, it is customary to invoke the rotating wave approximation, whereby the fast–oscillating terms at frequency $2\Omega$ in the exponential on the right hand side of the above equation time-average to zero. We then obtain 
\begin{align}
\hat{H}_\mathrm{rot}(t)
= \hbar\Delta\, \hat{a}^\dagger\hat{a}+\frac{E}{2}\left(\hat{a}+\hat{a}^\dagger\right)=\hat{H}_\mathrm{rot},
\label{eq:Hrot simiplified}
\end{align}
where we have set $E=\mathcal{E}x_\mathrm{zp}$. In place of Eq.~\eqref{eq:vdpo_lee}, the Lindblad equation now reads as
\begin{align}
 \label{eq:vdpo_lee-drive}
    \frac{d\hat{\rho}}{dt}=-\frac{i}{\hbar}[\hat{H}_\mathrm{rot},\hat{\rho}]+\gamma_{1}\mathcal{D}[\hat{a}^\dagger]\hat{\rho}+\gamma_{2}\mathcal{D}[\hat{a}^2]\hat{\rho}. 
\end{align}

We now wish to obtain the evolution equation of the Wigner function $W(\alpha,\alpha^*)$. To this end, proceeding in a manner similar to what was done to obtain Eq.~\eqref{eq:Wigner-evolution}, the computation will involve addition of two terms $-(i/\hbar)\Tr[\hat{a}\hat{\rho}\hat{D}^\dagger(\eta)]$ and $(i/\hbar)\Tr[\hat{\rho} \hat{a}\hat{D}^\dagger(\eta)]$. Using the correspondences $a \rho
\longleftrightarrow
\left(\alpha + c \overline{D} \right)W$ and $\rho a
\longleftrightarrow
\,\!\left(\alpha - c\overline{D} \right)W$, we get 
\begin{align}
\mathcal{W}_{(-iE/(2\hbar))[\hat{a},\hat{\rho}]}
&=-\frac{iE}{2\hbar}\frac{\partial }{\partial \alpha^*}W.
\label{eq:zero-term-drive}
\end{align}
Similarly, using $a^\dagger\rho\longleftrightarrow\left(\alpha^* - cD\right)W$, $\rho a^\dagger\longleftrightarrow\left(\alpha^* +cD\right)W$, one gets
\begin{align}
\mathcal{W}_{(-iE/(2\hbar))[\hat{a}^\dagger,\hat{\rho}]}
&=\frac{iE}{2\hbar}\frac{\partial }{\partial \alpha}W.
\label{eq:zero-term-drive}
\end{align}
We finally get the evolution equation of the Wigner function as
\begin{align}
\frac{\partial W}{\partial t}&=i\Delta\left(\alpha \frac{\partial }{\partial \alpha}-\alpha^* \frac{\partial }{\partial \alpha^*}\right)W\nonumber +\frac{iE}{2\hbar}\left(\frac{\partial }{\partial \alpha}-\frac{\partial }{\partial \alpha^*}\right)W\\
&+\Big\{\left(\frac{\partial }{\partial \alpha}\alpha + \frac{\partial }{\partial \alpha^*}\alpha^*\right)\left[-\gamma_1+2\gamma_2(|\alpha|^2-1)\right]
+\frac{\partial^2}{\partial \alpha \partial \alpha^*}\left[\gamma_1+2\gamma_2(2|\alpha|^2-1)\right] \nonumber \\
&+\frac{\gamma_2}{2}\Big(\frac{\partial^2}{\partial \alpha^2}\frac{\partial}{\partial \alpha^*}\alpha
+ \frac{\partial }{\partial \alpha}\frac{\partial^2}{\partial (\alpha^*)^2} \alpha^*\Big)\Big\}W.
\label{eq:Wigner-evolution-drive}
\end{align}

In the semi-classical limit, the corresponding Fokker-Planck equation, obtained in much the same away as for the undriven case, reads as
\begin{align}
    \frac{\partial W}{\partial t}&=i\Delta\left(\alpha \frac{\partial }{\partial \alpha}-\alpha^* \frac{\partial }{\partial \alpha^*}\right)W\nonumber +\frac{iE}{2\hbar}\left(\frac{\partial }{\partial \alpha}-\frac{\partial }{\partial \alpha^*}\right)W\nonumber \\
    &-\left(\frac{\partial }{\partial \alpha}[-\alpha BW]+\frac{\partial }{\partial \alpha^*}[-\alpha^* BW]\right)+A_0\left(\frac{\partial^2}{\partial \alpha \partial \alpha^*}\right)W,
    \label{eq:FP-2-drive}
\end{align}
and the corresponding Langevin equation reads as
\begin{align}
    \dot{\alpha}=-i\Delta \alpha -\frac{iE}{2}+\alpha(\gamma_1+2\gamma_2-2\gamma_2|\alpha|^2)+\xi^R(t)+i\xi^I(t).
    \label{eq:Langevin-2}
\end{align}
Thus, with respect to the Langevin equation~\eqref{eq:Langevin-1} for the undriven case, one has an additional term $-i\Delta \alpha -iE/2$ on the rhs.

In the quantum limit, we have in place of Eq.~\eqref{eq:Lindblad-vector} that
\begin{align}
\dot{\vec{\rho}} = (-iH^\times_\mathrm{rot} + \gamma_1 \mathcal{W}_1+\gamma_2 \mathcal{W}_2)\vec{\rho},
\label{eq:Lindblad-vector-driven}
\end{align}
with
\begin{align}
H^\times_\mathrm{rot} =
\begin{pmatrix}
0 & \frac{E}{2} & 0 &
-\frac{E}{2} & 0 & 0 &
0 & 0 & 0 \\
\frac{E}{2} & \Delta & \frac{E}{\sqrt2} &
0 & -\frac{E}{2} & 0 &
0 & 0 & 0 \\
0 & \frac{E}{\sqrt2} & 2\Delta &
0 & 0 & -\frac{E}{2} &
0 & 0 & 0 \\
-\frac{E}{2} & 0 & 0 &
-\Delta & \frac{E}{2} & 0 &
-\frac{E}{\sqrt2} & 0 & 0 \\
0 & -\frac{E}{2} & 0 &
\frac{E}{2} & 0 & \frac{E}{\sqrt2} &
0 & -\frac{E}{2} & 0 \\
0 & 0 & -\frac{E}{2} &
0 & \frac{E}{\sqrt2} & \Delta &
0 & 0 & -\frac{E}{2} \\
0 & 0 & 0 &
-\frac{E}{\sqrt2} & 0 & 0 &
-2\Delta & \frac{E}{2} & 0 \\
0 & 0 & 0 &
0 & -\frac{E}{2} & 0 &
\frac{E}{2} & -\Delta & \frac{E}{\sqrt2} \\
0 & 0 & 0 &
0 & 0 & -\frac{E}{2} &
0 & \frac{E}{\sqrt2} & 0
\end{pmatrix}.
\end{align}
The expressions for $\mathcal{W}_1$ and $\mathcal{W}_2$ remain same as in Eqs.~\eqref{eq:W1} and~\eqref{eq:W2}, respectively. Equation~\eqref{eq:rho-evolution-operator} now reads as
\begin{align}
&-iH^\times_\mathrm{rot} + \gamma_1 \mathcal{W}_1 + \gamma_2 \mathcal{W}_2 \nonumber \\
&=
\scriptsize
\begin{pmatrix}
-2\kappa_1
&
-\frac{iE}{2}
&
0
&
\frac{iE}{2}
&
0
&
0
&
0
&
0
&
0
\\[2mm]
-\frac{iE}{2}
&
-3\kappa_1-i\Delta
&
-\frac{iE}{\sqrt2}
&
0
&
\frac{iE}{2}
&
0
&
0
&
0
&
0
\\[2mm]
0
&
-\frac{iE}{\sqrt2}
&
-\kappa_1-2\kappa_2-2i\Delta
&
0
&
0
&
\frac{iE}{2}
&
0
&
0
&
0
\\[2mm]
\frac{iE}{2}
&
0
&
0
&
-3\kappa_1+i\Delta
&
-\frac{iE}{2}
&
0
&
\frac{iE}{\sqrt2}
&
0
&
0
\\[2mm]
2\kappa_1
&
\frac{iE}{2}
&
0
&
-\frac{iE}{2}
&
-4\kappa_1
&
-\frac{iE}{\sqrt2}
&
0
&
\frac{iE}{\sqrt2}
&
0
\\[2mm]
0
&
0
&
\frac{iE}{2}
&
0
&
-\frac{iE}{\sqrt2}
&
 -5\kappa_1-2\kappa_2-i\Delta
&
0
&
0
&
\frac{iE}{\sqrt2}
\\[2mm]
0
&
0
&
0
&
\frac{iE}{\sqrt2}
&
0
&
0
&
-\kappa_1-2\kappa_2+2i\Delta
&
-\frac{iE}{2}
&
0
\\[2mm]
0
&
0
&
0
&
0
&
\frac{iE}{\sqrt2}
&
0
&
-\frac{iE}{2}
&
-5\kappa_1-2\kappa_2+i\Delta
&
-\frac{iE}{\sqrt2}
\\[2mm]
0
&
0
&
0
&
0
&
0
&
\frac{iE}{\sqrt2}
&
0
&
-\frac{iE}{\sqrt2}
&
-4\kappa_2
\end{pmatrix}.
\label{eq:rho-evolution-operator-driven}
\end{align}
The stationary-state density matrix reads as
\begin{align}
\hat{\rho}_{\rm ss}
=
\frac{1}
{4\Delta^{2}+36\gamma_{1}^{2}+3E^{2}}
\begin{pmatrix}
\Delta^{2}+9\gamma_{1}^{2}
&
E(\Delta-3i\gamma_{1})
&
0
\\[2mm]
E(\Delta+3i\gamma_{1})
&
3E^{2}+3\Delta^{2}+27\gamma_{1}^{2}
&
0
\\[2mm]
0 & 0 & 0
\end{pmatrix}. \nonumber\\
\end{align}

\section{Quantum synchronization}
\label{sec:sec6}
\subsection{Two interacting quantum limit cycles}
\label{sec:2-quantum}
Let us explore the quantum case.  Eq. \eqref{eq:van der Pol} generalizes to \cite{Lee2013}
\begin{align}
    \frac{d\hat{\rho}}{dt}=-\frac{i}{\hbar}[\hat{H}_\text{tot},\hat{\rho}]
    + \gamma_\text{ab}\left(\mathcal{D}[\hat{a}_1^\dagger] + \mathcal{D}[\hat{a}_2^\dagger]\right) \hat{\rho}+\gamma_\text{em}\left(\mathcal{D}[\hat{a}_1^2] + \mathcal{D}[\hat{a}_2^2]\right)\hat{\rho}, 
    \label{eq:Lindblad two interacting}
\end{align}
where $\hat H_\text{tot} = \hbar\omega_1(\hat a_1^\dagger \hat a_1 + 1/2) + \hbar\omega_2(\hat a_2^\dagger \hat a_2 + 1/2) + \hat{V}$, with the interaction term $\hat V(t)$ being given by
\begin{align}
    \hat{V} = V x_{1,zp}x_{2,zp}(a_1^\dagger a_2 + a_1 a_2^\dagger).
\end{align}
Following the steps leading to Eq. \eqref{eq:Hrot simiplified}, we get
\begin{align}
    \hat{H}_\text{tot}
= \hbar\omega_1(\hat a_1^\dagger \hat a_1 + 1/2) + \hbar\omega_2(\hat a_2^\dagger \hat a_2 + 1/2) + E\left(\hat{a}_1^\dagger\hat{a}_2+\hat{a}_1 \hat{a}_2^\dagger\right),
\end{align}
where $E = V x_{1,zp}x_{2,zp}$.

Using the above expression for $\hat{H}$, we obtain from Eq. \eqref{eq:Lindblad two interacting}
\begin{align}
    \frac{d\hat\rho}{dt} &= -i\omega_1[\hat{a}_1^\dagger\hat{a}_1,\hat{\rho}] - i\omega_2[\hat{a}_2^\dagger\hat{a}_2,\hat{\rho}] -\frac{iE}{\hbar}[\hat{a}_1^\dagger\hat{a}_2 + \hat{a}_1\hat{a}_2^\dagger,\hat{\rho}] \nonumber\\
    &+ \gamma_\text{ab}\left(\mathcal{D}[\hat{a}_1^\dagger] + \mathcal{D}[\hat{a}_2^\dagger]\right) \hat{\rho}+\gamma_\text{em}\left(\mathcal{D}[\hat{a}_1^2] + \mathcal{D}[\hat{a}_2^2]\right)\hat{\rho}.
\end{align}
We define $\mathcal{L}_\text{$i$,ab}[\hat{\rho}] := \gamma_\text{ab} \mathcal{D}[\hat{a}_i^\dagger]\hat{\rho}$ and
$\mathcal{L}_\text{$i$,em}[\hat{\rho}] := \gamma_\text{em} \mathcal{D}[\hat{a}_i^2]\hat{\rho}$, 
where $i=1,2$, so that the above equation can be rewritten as
\begin{align}
    \frac{d\hat\rho}{dt} &=-i\omega_1[\hat{a}_1^\dagger\hat{a}_1,\hat{\rho}] - i\omega_2[\hat{a}_2^\dagger\hat{a}_2,\hat{\rho}] -\frac{iE}{\hbar}[\hat{a}_1^\dagger\hat{a}_2 + \hat{a}_1\hat{a}_2^\dagger,\hat{\rho}] \nonumber\\
    &+ \mathcal{L}_\text{1,ab}[\hat{\rho}] + \mathcal{L}_\text{2,ab}[\hat{\rho}] 
    + \mathcal{L}_\text{1,em}[\hat{\rho}] + \mathcal{L}_\text{2,em}[\hat{\rho}].
    \label{eq:2quantum-Lindblad}
\end{align}

Using the correspondences between $\hat\rho$ and $W$, we readily obtain obtain for the first commutator on the right-hand side,
\begin{align}
    &\mathcal{W}_{[\hat a_1^\dagger \hat a_1,\hat \rho]} = \left( \alpha_1^* \frac{\partial}{\partial\alpha_1^*} - \alpha_1 \frac{\partial}{\partial\alpha_1} \right) W; \nonumber\\
    &\mathcal{W}_{[\hat a_2^\dagger \hat a_2,\hat \rho]} = \left( \alpha_2^* \frac{\partial}{\partial\alpha_2^*} - \alpha_2 \frac{\partial}{\partial\alpha_2} \right) W.
\end{align}
To get the Wigner term corresponding to commutator between the interaction term and $\hat{\rho}$, we first observe that
\begin{align}    
    &\mathcal{W}_{a_1^\dagger a_2 \rho}=\left( \alpha_1^* \alpha_2 + \frac{1}{2}\alpha_1^* \frac{\partial}{\partial\alpha_2^*} - \frac{1}{2}\alpha_2 \frac{\partial}{\partial\alpha_1} - \frac{1}{4} \frac{\partial^2}{\partial\alpha_1 \partial\alpha_2^*} \right) W; \nonumber\\
    &\mathcal{W}_{\rho a_1^\dagger a_2}=\left( \alpha_1^* \alpha_2 + \frac{1}{2}\alpha_2 \frac{\partial}{\partial\alpha_1} - \frac{1}{2}\alpha_1^* \frac{\partial}{\partial\alpha_2^*} - \frac{1}{4} \frac{\partial^2}{\partial\alpha_1 \partial\alpha_2^*} \right) W; \nonumber\\
     &\mathcal{W}_{a_1 a_2^\dagger \rho}=\left( \alpha_1 \alpha_2^* - \frac{1}{2}\alpha_1 \frac{\partial}{\partial\alpha_2} + \frac{1}{2}\alpha_2^* \frac{\partial}{\partial\alpha_1^*} - \frac{1}{4} \frac{\partial^2}{\partial\alpha_1^* \partial\alpha_2} \right) W; \nonumber\\
    &\mathcal{W}_{\rho a_1 a_2^\dagger}=\left( \alpha_1 \alpha_2^* - \frac{1}{2}\alpha_2^* \frac{\partial}{\partial\alpha_1^*} + \frac{1}{2}\alpha_1 \frac{\partial}{\partial\alpha_2} - \frac{1}{4} \frac{\partial^2}{\partial\alpha_1^* \partial\alpha_2} \right) W.
    \label{eq:two osc Liouville part 1}
\end{align}

Let us now move to the interaction picture. Using the above relations, we get
\begin{align}
    &\mathcal{W}_{[\hat a_1^\dagger \hat a_2,\hat \rho]} = \left( \alpha_1^* \frac{\partial}{\partial\alpha_2^*} - \alpha_2 \frac{\partial}{\partial\alpha_1} \right) W; \nonumber\\
    &\mathcal{W}_{[\hat a_1 \hat a_2^\dagger,\hat \rho]} = \left( \alpha_2^* \frac{\partial}{\partial\alpha_1^*} - \alpha_1 \frac{\partial}{\partial\alpha_2} \right) W.
    \label{eq:two osc Liouville part 2}
\end{align}
Using Eqs. \eqref{eq:first-term} and \eqref{eq:second-term}, we readily get
\begin{align}
&\mathcal{W}_{\mathcal{L}_\text{1,ab}[\rho]}=\gamma_\text{ab}\left[-\left(\frac{\partial }{\partial \alpha_1}\alpha_1 + \frac{\partial }{\partial \alpha^*_1}\alpha^*_1\right) \,W
+\frac{\partial^2W}{\partial \alpha_1 \partial \alpha^*_1} \right]. \nonumber \\
&\mathcal{W}_{\mathcal{L}_\text{2,ab}[\rho]}=\gamma_\text{ab}\left[-\left(\frac{\partial }{\partial \alpha_2}\alpha_2 + \frac{\partial }{\partial \alpha^*_2}\alpha^*_2\right) \,W
+\frac{\partial^2W}{\partial \alpha_2 \partial \alpha^*_2} \right]. \nonumber \\
&\mathcal{W}_{\mathcal L_\text{1,em}[\rho]}=2\gamma_\text{em}\Big[ \Big(\frac{\partial}{\partial \alpha_1}\alpha_1+\frac{\partial}{\partial \alpha_1^*}\alpha_1^*\Big)(|\alpha_1|^2-1)
+ \frac{\partial^2}{\partial \alpha_1\partial \alpha_1^*}(2|\alpha_1|^2-1)\Big]W\nonumber \\
&+\frac{\gamma_\text{em}}{2}\Big(\frac{\partial^2}{\partial \alpha_1^2}\frac{\partial}{\partial \alpha_1^*}\alpha_1
+ \frac{\partial }{\partial \alpha_1}\frac{\partial^2}{\partial (\alpha_1^*)^2} \alpha_1^*\Big)W. \nonumber\\
&\mathcal{W}_{\mathcal L_\text{2,em}[\rho]}=2\gamma_\text{em}\Big[ \Big(\frac{\partial}{\partial \alpha_2}\alpha_2+\frac{\partial}{\partial \alpha_2^*}\alpha_2^*\Big)(|\alpha_2|^2-1)
+ \frac{\partial^2}{\partial \alpha_2\partial \alpha_2^*}(2|\alpha_2|^2-1)\Big]W\nonumber \\
&+\frac{\gamma_\text{em}}{2}\Big(\frac{\partial^2}{\partial \alpha_2^2}\frac{\partial}{\partial \alpha_2^*}\alpha_2
+ \frac{\partial }{\partial \alpha_2}\frac{\partial^2}{\partial (\alpha_2^*)^2} \alpha_2^*\Big)W.
\label{eq:two osc dissipators}
\end{align}

The final equation for the Wigner function then becomes
\begin{align}
    \frac{\partial W}{\partial t} &=-i\omega_1 \mathcal{W}_{[a_1^\dagger a_1, \rho]} -i\omega_2 \mathcal{W}_{[a_2^\dagger a_2, \rho]}  
    - \frac{iE}{2\hbar} \left( \mathcal{W}_{[a_1^\dagger a_2, \rho]} + \mathcal{W}_{[a_1 a_2^\dagger, \rho]} \right) \nonumber\\
    & + \mathcal{W}_{\mathcal{L}_\text{1,ab}[\rho]} + \mathcal{W}_{\mathcal{L}_\text{2,ab}[\rho]} + \mathcal{W}_{\mathcal{L}_\text{1,em}[\rho]} + \mathcal{W}_{\mathcal{L}_\text{2,em}[\rho]},
    \label{eq:two osc wigner}
\end{align}
where the explicit expressions for the different terms are provided in Eqs. \eqref{eq:two osc Liouville part 1}, \eqref{eq:two osc Liouville part 2} and \eqref{eq:two osc dissipators}.

Next, we derive the Fokker-Planck equation for $W$. For this purpose, we remove the third order derivatives in Eq. \eqref{eq:two osc dissipators}, and write Eq. \eqref{eq:two osc wigner} as
\begin{align}
    \frac{\partial W}{\partial t} &=-i\omega_1\left( \alpha_1^* \frac{\partial}{\partial\alpha_1^*} - \alpha_1 \frac{\partial}{\partial\alpha_1} \right) W
    - i\omega_2 \left(\alpha_2^* \frac{\partial}{\partial\alpha_2^*} - \alpha_2 \frac{\partial}{\partial\alpha_2} \right) W \nonumber\\
   &- \frac{iE}{\hbar}\left(\alpha_2^* \frac{\partial}{\partial\alpha_1^*} - \alpha_2 \frac{\partial}{\partial\alpha_1} \right) W
   -\frac{iE}{\hbar} \left( \alpha_1^* \frac{\partial}{\partial\alpha_2^*} - \alpha_1 \frac{\partial}{\partial\alpha_2}\right) W
     \nonumber\\
   &+ \gamma_\text{ab}\left[-\left(\frac{\partial }{\partial \alpha_1}\alpha_1 + \frac{\partial }{\partial \alpha^*_1}\alpha^*_1\right) \,W
+\frac{\partial^2W}{\partial \alpha_1 \partial \alpha^*_1} \right] \nonumber\\
&+ \gamma_\text{ab}\left[-\left(\frac{\partial }{\partial \alpha_2}\alpha_2 + \frac{\partial }{\partial \alpha^*_2}\alpha^*_2\right) \,W
+\frac{\partial^2W}{\partial \alpha_2 \partial \alpha^*_2} \right] \nonumber\\
&+ 2\gamma_\text{em}\Big[ \Big(\frac{\partial}{\partial \alpha_1}\alpha_1+\frac{\partial}{\partial \alpha_1^*}\alpha_1^*\Big)(|\alpha_1|^2-1)
+ \frac{\partial^2}{\partial \alpha_1\partial \alpha_1^*}(2|\alpha_1|^2-1)\Big]W \nonumber\\
&+ 2\gamma_\text{em}\Big[ \Big(\frac{\partial}{\partial \alpha_2}\alpha_2+\frac{\partial}{\partial \alpha_2^*}\alpha_2^*\Big)(|\alpha_2|^2-1)
+ \frac{\partial^2}{\partial \alpha_2\partial \alpha_2^*}(2|\alpha_2|^2-1)\Big]W. \hspace{3cm}
\end{align}
Let us consider the first term on the right-hand side, which can be rewritten as
\begin{eqnarray*}
 -i\omega_1\left( \alpha_1^* \frac{\partial}{\partial\alpha_1^*} - \alpha_1 \frac{\partial}{\partial\alpha_1} \right) W &=& - \left( \frac{\partial}{\partial\alpha_1^*} [(-i\omega_1\alpha_1)^* W] + \frac{\partial}{\partial\alpha_1}[(-i\omega_1\alpha_1)W]\right) \nonumber\\
 &=& - \left( \frac{\partial}{\partial\alpha_1}[(-i\omega_1\alpha_1)W] + \text{c.c.}\right),  
\end{eqnarray*}
where ``c.c.'' represents complex conjugation. The above expression, when compared with the standard form of the drift terms in a Fokker-Planck equation (see Eq. \eqref{eq:FP-1}), readily gives the drift contribution of $-i\omega_1\alpha_1$. Similarly, all other drift contributions can be obtained. These are summarized here.
The first line on the right-hand side gives the drift contributions $\dot\alpha_1 = -i\omega_1\alpha_1$ and $\dot\alpha_2 = -i\omega_2\alpha_2$. The second line gives the drift contributions $\dot\alpha_1 = -(iE/\hbar)\alpha_2$ and $\dot\alpha_2 = -(iE/\hbar)\alpha_1$. The third and fourth lines contribute $\dot\alpha_1 = \gamma_\text{ab}\alpha_1$ and $\dot\alpha_2 = \gamma_\text{ab}\alpha_2$. The fifth and sixth lines contribute $\dot\alpha_1 = -2\gamma_\text{em}\alpha_1(|\alpha_1|^2-1)$ and $\dot\alpha_2 = -2\gamma_\text{em}\alpha_2(|\alpha_2|^2-1)$. Furthermore, the diffusion contributions to $\dot\alpha_1$ and $\dot\alpha_2$ will comprise complex noises $\xi_1(t) = \xi_1^R(t) + i\xi_1^I(t)$ and $\xi_2(t) = \xi_2^R(t) + i\xi_2^I(t)$, respectively.
Combining everything, we get
\begin{align}
    &\dot\alpha_1= (-i\omega_1 + \gamma_\text{ab} + 2\gamma_\text{em} - 2\gamma_\text{em}|\alpha_1|^2)\alpha_1 -(iE/\hbar)\alpha_2  + \xi_1^R(t) + i\xi_1^I(t); \nonumber\\
    &\dot\alpha_2=(-i\omega_2 + \gamma_\text{ab} + 2\gamma_\text{em} - 2\gamma_\text{em}|\alpha_2|^2) \alpha_2 -(iE/\hbar)\alpha_1 + \xi_2^R(t) + i\xi_2^I(t).
    \label{eq:two osc Langevin}
\end{align}
Here, the noise correlations for $n=1,2$ are given by (see Eq. \eqref{eq:noise correlations})
\begin{align}
    &\langle \xi_n^R(t)\rangle = \langle \xi_n^I(t)\rangle = \langle \xi_n^R(t)\xi_n^I(t')\rangle = 0; \nonumber\\
    &\langle \xi_n^R(t)\xi_n^R(t')\rangle =  \langle \xi_n^I(t)\xi_n^I(t')\rangle 
    = \left( \frac{3\gamma_\text{ab}}{2} + \gamma_\text{em}\right) \delta(t-t').
    \label{eq:two osc noise correlations}
\end{align}

To proceed as in Section~\ref{sec:2-interacting-classical}, we write the complex amplitudes in polar form, $\alpha_n=r_n e^{i\theta_n}$, with $r_n\ge0$. The Langevin equations~\eqref{eq:two osc Langevin} become
\begin{align}
    \dot{\alpha}_n=
    \left(-i\omega_n + \Gamma_n\right)\alpha_n
    -\frac{iE}{\hbar}\alpha_m
    +\eta_n,
    \label{eq:alphan}
\end{align}
with $\Gamma_n \equiv \gamma_{\rm ab}+2\gamma_{\rm em}-2\gamma_{\rm em}r_n^2$, and $\eta_n\equiv\xi_n^R+i\xi_n^I$, and $m\ne n$. Differentiating $\alpha_n=r_ne^{i\theta_n}$ gives
\begin{align}
    \dot{\alpha}_n
    =
    (\dot r_n+i r_n\dot\theta_n)e^{i\theta_n}.
\end{align}
Multiplying Eq.~\eqref{eq:alphan} by $e^{-i\theta_n}$ yields
\begin{align}
\dot r_n+i r_n\dot\theta_n
=
\Gamma_n r_n
-i\omega_n r_n
-\frac{iE}{\hbar}
r_m
e^{i(\theta_m-\theta_n)}
+\eta_ne^{-i\theta_n}.
\label{eq:polar_general}
\end{align}
Here, the transformed noise is
\begin{align}
\eta_ne^{-i\theta_n}
&=
(\xi_n^R+i\xi_n^I)
(\cos\theta_n-i\sin\theta_n) \nonumber \\
&=
\underbrace{\left(
\xi_n^R\cos\theta_n
+\xi_n^I\sin\theta_n
\right)}_{\eta_{nR}}
+i
\underbrace{\left(
-\xi_n^R\sin\theta_n
+\xi_n^I\cos\theta_n
\right)}_{\eta_{nI}}.
\end{align}

For oscillator $1$, Eq.~\eqref{eq:polar_general} becomes
\begin{align}
\dot r_1+i r_1\dot\theta_1
&=
\Gamma_1r_1
-i\omega_1 r_1
+\frac{E}{\hbar}r_2\sin\theta
-\frac{iE}{\hbar}r_2\cos\theta
+\eta_{1R}+i\eta_{1I},
\end{align}
where  we have defined the relative phase $\theta \equiv \theta_2-\theta_1$.
Equating real and imaginary parts gives
\begin{align}
&\dot r_1=\Gamma_1r_1+\frac{E}{\hbar}r_2\sin\theta+\eta_{1R},
\\
&r_1\dot\theta_1=-\omega_1 r_1-\frac{E}{\hbar}r_2\cos\theta+\eta_{1I},
\end{align}
yielding
\begin{align}
\dot\theta_1=-\omega_1-\frac{E}{\hbar}\frac{r_2}{r_1}\cos\theta+\frac{\eta_{1I}}{r_1}.
\end{align}
Similarly, for oscillator $2$, one gets
\begin{align}
\dot\theta_2
=
-\omega_2
-\frac{E}{\hbar}\frac{r_1}{r_2}\cos\theta
+\frac{\eta_{2I}}{r_2}.
\end{align}
We thus finally arrive at the following equation for $\theta$: 
\begin{align}
\dot\theta&= -\Delta\omega -\frac{E}{\hbar}\left(\frac{r_1}{r_2}-\frac{r_2}{r_1}\right)\cos\theta+\frac{\eta_{2I}}{r_2}-\frac{\eta_{1I}}{r_1},
\end{align}
where $\Delta\omega = \omega_2 - \omega_1$.
Thus, the amplitude-phase equations read as
\begin{align}
&\dot r_1=\left(\gamma_{\rm ab}+2\gamma_{\rm em}-2\gamma_{\rm em}r_1^2\right)r_1+\frac{E}{\hbar}r_2\sin\theta+\eta_{1R},\label{eq:r_1}\\
&\dot r_2=\left(\gamma_{\rm ab}+2\gamma_{\rm em}-2\gamma_{\rm em}r_2^2\right)r_2-\frac{E}{\hbar}r_1\sin\theta+\eta_{2R}, \label{eq:r_2}\\
&\dot\theta= -\Delta\omega -\frac{E}{\hbar}\left(\frac{r_1}{r_2}-\frac{r_2}{r_1}\right)\cos\theta+\frac{\eta_{2I}}{r_2}-\frac{\eta_{1I}}{r_1},\label{eq:theta}
\end{align}
where we have
\begin{align}
&\eta_{nR}=\xi_n^R\cos\theta_n+\xi_n^I\sin\theta_n,
\\
&\eta_{nI}=-\xi_n^R\sin\theta_n+\xi_n^I\cos\theta_n.
\end{align}
We have used the relation $\sin(\theta_2-\theta_1) = - \sin(\theta_1-\theta_2)$ in Eq. \eqref{eq:r_2}.

The synchronization conditions can be obtained by considering the
deterministic part of Eqs.~\eqref{eq:r_1} -- \eqref{eq:theta}. Defining
\begin{align}
K &\equiv \frac{E}{\hbar},~a \equiv \gamma_{\rm ab}+2\gamma_{\rm em},~b \equiv 2\gamma_{\rm em},
\label{eq:definitions_sync}
\end{align}
the equations become
\begin{eqnarray}
\dot r_1 &=& (a-br_1^2)r_1+Kr_2\sin\theta,\nonumber\\
\dot r_2 &=& (a-br_2^2)r_2-Kr_1\sin\theta, \nonumber\\
\dot\theta &=& -\Delta\omega
-K\left(\frac{r_1}{r_2}-\frac{r_2}{r_1}\right)\cos\theta.
\label{eq:deterministic_amp_phase}
\end{eqnarray}

Synchronization corresponds to a state in which the amplitudes become
stationary and the relative phase approaches a constant value. Thus, a
phase-locked stationary state satisfies
\begin{align}
\dot r_1=\dot r_2=\dot\theta=0.
\label{eq:sync_conditions}
\end{align}
For nonzero coupling, $E\neq0$, the condition $\dot\theta=0$ gives on using Eq.~\eqref{eq:deterministic_amp_phase} that
\begin{align}
\left(
\frac{r_1}{r_2}-\frac{r_2}{r_1}
\right)\cos\theta
=-\frac{\Delta\omega}{K}.
\label{eq:phase_lock_condition}
\end{align}
Given that $\Delta\omega\neq0$ (finite detuning) requires in general
$r_1\neq r_2$. Indeed, if $r_1=r_2$, the left-hand side of
Eq.~\eqref{eq:phase_lock_condition} vanishes, and therefore the phase-locking
condition can only be satisfied when $\Delta\omega=0$. Thus, for detuned
oscillators, frequency locking is accompanied by a stationary amplitude
imbalance.

\begin{figure}[!h]
    \centering
    \includegraphics[width=0.55\linewidth]{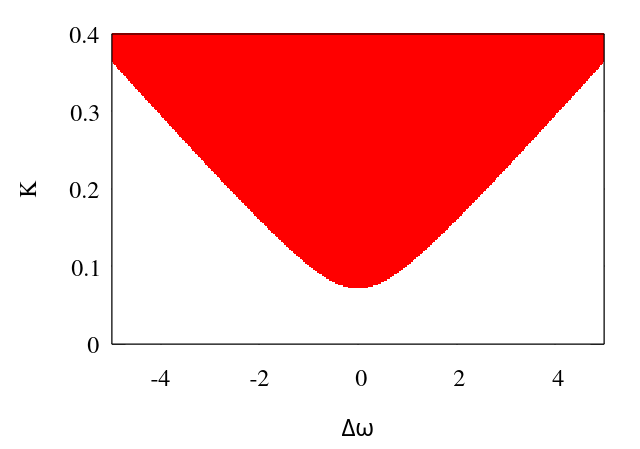}
    \caption{Arnold tongue as determined by Eq. \eqref{eq:locking_equation_z}, with $a = \gamma_\text{ab} + 2\gamma_\text{em} = 1$.}
    \label{fig:Arnold_quantum}
\end{figure}

To obtain the stationary solution for finite detuning, it is useful to
introduce the sum and difference of the squared amplitudes,
\begin{align}
S\equiv r_1^2+r_2^2,~
D\equiv r_1^2-r_2^2,
\label{eq:SD_definitions}
\end{align}
and the dimensionless amplitude-imbalance parameter
\begin{align}
z\equiv\frac{D}{S}
=\frac{r_1^2-r_2^2}{r_1^2+r_2^2},~ |z|<1.
\label{eq:z_definition}
\end{align}
The amplitude equations in the stationary state are obtained from Eq.~\eqref{eq:deterministic_amp_phase} as
\begin{align}
0=(a-br_1^2)r_1+Kr_2\sin\theta,~0=(a-br_2^2)r_2-Kr_1\sin\theta.
\label{eq:stationary_detuned_amplitudes}
\end{align}
Dividing the first equation by $r_1$ and the second by $r_2$, and then
adding and subtracting the resulting equations, gives
\begin{align}
2a-bS-\frac{bD^2}{S}&=0,
\label{eq:S_D_relation}
\end{align}
and
\begin{align}
K\sin\theta
=\frac{bD}{S}r_1r_2.
\label{eq:sin_theta_D}
\end{align}
Using $D=zS$, and $r_1r_2=\frac{S}{2}\sqrt{1-z^2}$, Eq.~\eqref{eq:S_D_relation} gives
\begin{align}
S=\frac{2a}{b(1+z^2)}.
\label{eq:S_stationary}
\end{align}
Consequently, the two stationary amplitudes are
\begin{align}
(r_1^{*})^2=\frac{a}{b}
\frac{1+z}{1+z^2},~~~~(r_2^{*})^2=\frac{a}{b}
\frac{1-z}{1+z^2}.
\label{eq:detuned_stationary_amplitudes}
\end{align}
The corresponding phase satisfies
\begin{align}
\sin\theta^*=
\frac{a}{K}
\frac{z\sqrt{1-z^2}}{1+z^2},~\cos\theta^*=-\frac{\Delta\omega}{2K}
\frac{\sqrt{1-z^2}}{z}.
\label{eq:cos_theta_detuned}
\end{align}
The parameter $z$ is determined by requiring that
$\sin^2\theta^*+\cos^2\theta^*=1$. This gives
\begin{align}
\frac{a^2z^2(1-z^2)}
{K^2(1+z^2)^2}
+
\frac{\Delta\omega^2(1-z^2)}
{4K^2z^2}
=1.
\label{eq:locking_equation_z}
\end{align}
Equivalently, the detuning corresponding to a given stationary amplitude
imbalance $z$ is
\begin{align}
\Delta\omega^2
=
\frac{4K^2z^2}{1-z^2}
-
\frac{4a^2z^4}{(1+z^2)^2}.
\label{eq:detuning_vs_imbalance}
\end{align}
Equation~\eqref{eq:detuning_vs_imbalance} provides the synchronization
condition for the detuned system. In particular, the detuning is balanced
by the amplitude imbalance $z$, while the coupling determines the
corresponding stationary phase difference. The Arnold tongue produced by the above synchronization condition is plotted in Fig. \ref{fig:Arnold_quantum}.

The identical-oscillator limit is recovered by setting
$\Delta\omega=0$. In this case Eq.~\eqref{eq:detuning_vs_imbalance} admits
the symmetric solution $z=0$, corresponding to
\begin{align}
r_1^*=r_2^*
=r_0,
\qquad
r_0^2=\frac{a}{b}
=
\frac{\gamma_{\rm ab}+2\gamma_{\rm em}}
{2\gamma_{\rm em}}.
\label{eq:identical_limit_amplitude}
\end{align}
For $z\rightarrow0$, the phase relation depends on which of the two
phase-locked branches is followed. The two branches continuously connect
to the in-phase and anti-phase solutions,
\begin{align}
\theta^*=0,\pi
\qquad
(\Delta\omega=0),
\label{eq:identical_phase_locked_solutions}
\end{align}
respectively. Thus, finite detuning deforms the two symmetric phase-locked
solutions into detuning-dependent phase-locked states with
$r_1^*\neq r_2^*$ and, in general, $\theta^*\neq0,\pi$.

The sign of the amplitude imbalance is correlated with the phase-locked
branch. For example, for small positive detuning, the branch continuously
connected to the anti-phase state has $z>0$, while the branch connected to
the in-phase state has $z<0$. To leading order for weak detuning,
Eq.~\eqref{eq:detuning_vs_imbalance} gives
\begin{align}
z\simeq \pm\frac{\Delta\omega}{2K},
\label{eq:weak_detuning_imbalance}
\end{align}
where the two signs correspond to the two phase-locked branches. Thus,
the frequency mismatch produces a small stationary amplitude difference
whose magnitude is proportional to the detuning-to-coupling ratio.

For completeness, the stability of the phase-locked state can be analyzed
by linearizing the full three-dimensional deterministic system around
\begin{align}
(r_1,r_2,\theta)
=(r_1^*,r_2^*,\theta^*).
\end{align}
Writing
\begin{align}
r_1&=r_1^*+\delta r_1,
\qquad
r_2=r_2^*+\delta r_2,
\qquad
\theta=\theta^*+\delta\theta,
\end{align}
the linearized dynamics takes the form
\begin{align}
\frac{d}{dt}
\begin{pmatrix}
\delta r_1\\
\delta r_2\\
\delta\theta
\end{pmatrix}
=
\mathsf{J}
\begin{pmatrix}
\delta r_1\\
\delta r_2\\
\delta\theta
\end{pmatrix},
\label{eq:detuned_jacobian_dynamics}
\end{align}
where the Jacobian evaluated at the stationary solution is
\begin{align}
\mathsf{J}
=
\begin{pmatrix}
a-3b r_1^{*2}
&
K\sin\theta^*
&
K r_2^*\cos\theta^*
\\[6pt]
-K\sin\theta^*
&
a-3b r_2^{*2}
&
-K r_1^*\cos\theta^*
\\[6pt]
-K\left(
\displaystyle\frac{1}{r_2^*}
+\displaystyle\frac{r_2^*}{r_1^{*2}}
\right)\cos\theta^*
&
K\left(
\displaystyle\frac{r_1^*}{r_2^{*2}}
+\displaystyle\frac{1}{r_1^*}
\right)\cos\theta^*
&
K\left(
\displaystyle\frac{r_1^*}{r_2^*}
-\displaystyle\frac{r_2^*}{r_1^*}
\right)\sin\theta^*
\end{pmatrix}.
\label{eq:detuned_jacobian}
\end{align}
The synchronized state is linearly stable when all three eigenvalues of
$\mathsf{J}$ have negative real parts,
\begin{align}
\operatorname{Re}\lambda_j<0,
\qquad j=1,2,3.
\label{eq:detuned_stability_condition}
\end{align}
Unlike the identical case, the common-amplitude and amplitude-difference
modes no longer decouple when $\Delta\omega\neq0$. Consequently, the
stability eigenvalues are, in general, obtained from the full
three-dimensional Jacobian in Eq.~\eqref{eq:detuned_jacobian}.

In the limit $\Delta\omega\rightarrow0$, one has
$r_1^*=r_2^*=r_0$ and $\theta^*=0$ or $\pi$, and the Jacobian separates
into the common-amplitude mode and the amplitude-difference/relative-phase
sector considered above. The common-amplitude eigenvalue becomes
\begin{align}
\lambda_{\rm com}=-2a,
\end{align}
while the remaining two eigenvalues are
\begin{align}
\lambda_\pm
=
-a\pm\sqrt{a^2-4K^2}.
\label{eq:eigenvalues_sync}
\end{align}
Thus, the results obtained for identical oscillators are recovered
continuously from the general detuned formulation.

For finite detuning, synchronization therefore corresponds to a
frequency-locked state with
\begin{align}
r_1^*\neq r_2^*,
\qquad
\theta^*=\text{constant},
\end{align}
where the amplitude imbalance compensates the frequency mismatch
according to Eq.~\eqref{eq:detuning_vs_imbalance}. The phase difference is
no longer restricted to $0$ or $\pi$, but is determined by
Eqs.~\eqref{eq:detuned_stationary_amplitudes} and
\eqref{eq:cos_theta_detuned}. The identical-oscillator result is recovered
as the special case $\Delta\omega=0$, in which the two phase-locked
branches reduce to the in-phase and anti-phase solutions.

Assuming $\gamma_\text{em}\gg\gamma_\text{ab},E/\hbar$ (quantum limit), let us solve Eq.~\eqref{eq:2quantum-Lindblad} perturbatively in $1/\gamma_\text{em}$. We first consider the
limit $\gamma_\text{em}\rightarrow\infty$. In this limit, the
two-phonon emission process rapidly removes any population in states
containing two or more excitations. Thus, to zeroth order in
$1/\gamma_\text{em}$, each oscillator occupies only the states
$\ket{0}$ and $\ket{1}$. For a single oscillator, the absorption process
$\mathcal{L}_{i,\text{ab}}$ induces the transitions
$\ket{0}\rightarrow\ket{1}$ and $\ket{1}\rightarrow\ket{2}$ with
rates $2\gamma_\text{ab}$ and $4\gamma_\text{ab}$, respectively (see the first term on the right hand side of Eq.~\eqref{eq:Lindblad-2}),
whereas $\mathcal{L}_{i,\text{em}}$ induces the transition
$\ket{2}\rightarrow\ket{0}$ with rate $4\gamma_\text{em}$ (see the second term on the right hand side of Eq.~\eqref{eq:Lindblad-2}). Hence,
in the limit $\gamma_\text{em}\rightarrow\infty$, the population of
$\ket{2}$ vanishes and the stationary populations of $\ket{0}$ and
$\ket{1}$ satisfy
\begin{equation}
2\gamma_\text{ab}\rho_{00}
=
4\gamma_\text{ab}\rho_{11},
\qquad
\rho_{00}+\rho_{11}=1.
\end{equation}
It follows that
\begin{equation}
\rho_{00}=\frac{2}{3},
\qquad
\rho_{11}=\frac{1}{3}.
\end{equation}
Consequently, the zeroth-order stationary state of the two
oscillators is given by 
\begin{align}
\hat{\rho}^{(0)}=&\left(\frac{2}{3}\ket{0}\bra{0}+\frac{1}{3}\ket{1}\bra{1}\right)\otimes\left(\frac{2}{3}\ket{0}\bra{0}+\frac{1}{3}\ket{1}\bra{1}\right)\nonumber\\
=&\frac{4}{9}\ket{00}\bra{00}+\frac{2}{9}\ket{01}\bra{01}+\frac{2}{9}\ket{10}\bra{10}+\frac{1}{9}\ket{11}\bra{11}.
\label{eq:rho0}
\end{align}

The state in Eq.~\eqref{eq:rho0} is diagonal in the number basis. Hence, it is
invariant under independent phase rotations of the two oscillators. To see this explicitly, consider a number state $\ket{n_i}$ of
oscillator $i$. Under a phase rotation, $\ket{n_i}\rightarrow e^{in_i\phi_i}\ket{n_i}$, the corresponding diagonal density operator transforms as $\ket{n_i}\bra{n_i}\rightarrow e^{in_i\phi_i}\ket{n_i}\bra{n_i}e^{-in_i\phi_i}=\ket{n_i}\bra{n_i}$. Thus, a diagonal number-state density matrix is invariant under a phase rotation. For the two oscillators, a diagonal element $\ket{n_1n_2}\bra{n_1n_2}$ transforms according to $\ket{n_1n_2}\bra{n_1n_2}
\rightarrow  e^{i(n_1\phi_1+n_2\phi_2)}\ket{n_1n_2}\bra{n_1n_2}e^{-i(n_1\phi_1+n_2\phi_2)}=
\ket{n_1n_2}\bra{n_1n_2}$. Hence, none of the diagonal terms in $\hat{\rho}^{(0)}$ can depend
on $\phi_1$ or $\phi_2$, and therefore $\hat{\rho}^{(0)}$ contains no
information about the relative phase $\phi_1-\phi_2$.

In contrast, an off-diagonal density-matrix element can carry
information about the relative phase. For example, consider
$\ket{02}\bra{20}$. Under a phase rotation, $\ket{02}\rightarrow e^{2i\phi_2}\ket{02}$, 
we have $\ket{02}\bra{20} \rightarrow e^{-2i(\phi_1-\phi_2)} \ket{02}\bra{20}$. Consequently, the Hermitian combination $\ket{02}\bra{20}+\ket{20}\bra{02}$ contains the phase dependence $e^{2i(\phi_1-\phi_2)}+e^{-2i(\phi_1-\phi_2)}=2\cos\left[2(\phi_1-\phi_2)\right]$. Thus, the diagonal zeroth-order state gives a uniform relative-phase
distribution, whereas the off-diagonal coherences
$\ket{02}\bra{20}$ and $\ket{20}\bra{02}$ generate a
phase-dependent term $\cos[2(\phi_1-\phi_2)]$. Such a dependence appears at finite
$\gamma_\text{em}$ through the coherent coupling term in Eq.~\eqref{eq:2quantum-Lindblad}.
Writing $\hat{H}_{12}=\frac{E}{\hbar}\left(\hat{a}_1^\dagger\hat{a}_2+
\hat{a}_1\hat{a}_2^\dagger\right)$, we have
\begin{align}
\hat{H}_{12}\ket{11}=\frac{E}{\hbar}\left(\hat{a}_1^\dagger\hat{a}_2+\hat{a}_1\hat{a}_2^\dagger\right)\ket{11}=
\sqrt{2}\frac{E}{\hbar}\left(\ket{20}+\ket{02}\right).
\label{eq:H12_on_11}
\end{align}
Thus, the coherent coupling generates off-diagonal elements
connecting $\ket{11}$ with $\ket{02}$ and $\ket{20}$.

To obtain the leading phase-dependent contribution, we first
calculate the corresponding first-order coherences. Define $\rho_{n_i'n_j',n_i n_j} \equiv  \bra{n_i'n_j'}\hat{\rho}\ket{n_i n_j}$, so that
\begin{equation}
\rho_{02,11}\equiv\bra{02}\hat{\rho}\ket{11},
\qquad\rho_{11,20} \equiv
\bra{11}\hat{\rho}\ket{20}.
\end{equation}
We note the following relations:
\begin{eqnarray}
\hspace{-1.5cm}\bra{02}a_1^\dagger a_1 + a_2^\dagger a_2\ket{11} = 2;~\bra{11}a_1^\dagger a_1 + a_2^\dagger a_2\ket{02} = 2;~\bra{02}a_1^\dagger a_2 + a_1 a_2^\dagger\ket{11} = \sqrt{2},
\label{eq:amplitudes 1}
\end{eqnarray}
whence we obtain, using the completeness of basis states and retaining terms up to the linear order in $1/\gamma_\text{em}$,
\begin{eqnarray}
    \bra{02}[a_1^\dagger a_1 + a_2^\dagger a_2,\hat{\rho}]\ket{11} = 0,\qquad\bra{02}[a_1^\dagger a_2 + a_1 a_2^\dagger,\hat{\rho}]\ket{11} \simeq \sqrt{2}\rho_{11,11}^{(0)}.
    \label{eq:amplitudes 2}
\end{eqnarray}
Further, it can be shown, using the relation ${a_i^\dagger}^2a_i^2 = \hat{n}_i^2 - \hat{n}_i$, that
\begin{eqnarray}
    \bra{02}\mathcal{D}[a_1^2]\rho\ket{11} &=& 0; \nonumber\\
    \bra{02}\mathcal{D}[a_2^2]\rho\ket{11} &=& -2\rho_{02,11}^{(0)}.
    \label{eq:dissipator contribution}
\end{eqnarray}
Using Eq. \eqref{eq:amplitudes 2} and \eqref{eq:dissipator contribution} in Eq.~\eqref{eq:2quantum-Lindblad}, the final equations of motion take the form 
\begin{align}
\frac{d\rho_{02,11}}{dt}&=-2\gamma_\text{em}\rho_{02,11}-i\sqrt{2}\frac{E}{\hbar}
\rho^{(0)}_{11,11}+O(\gamma_\text{em}^{-1}),\\
\frac{d\rho_{11,20}}{dt}
&=-2\gamma_\text{em}\rho_{11,20}+i\sqrt{2}\frac{E}{\hbar}\rho^{(0)}_{11,11}+O(\gamma_\text{em}^{-1}).
\end{align}
Since we have $\rho^{(0)}_{11,11}=\frac{1}{9}$, the stationary values are
\begin{align}
\rho_{02,11}=-\frac{i}{18\gamma_\text{em}}\sqrt{2}\frac{E}{\hbar}+O(\gamma_\text{em}^{-2}),\qquad\rho_{11,20}=\frac{i}{18\gamma_\text{em}}\sqrt{2}\frac{E}{\hbar}+O(\gamma_\text{em}^{-2}).
\label{eq:first_order_coherences}
\end{align}

The phase-dependent term appears at the next order through the
coherence
\begin{equation}
\rho_{02,20}\equiv\bra{02}\hat{\rho}\ket{20}.
\end{equation}
Using
\begin{equation}
\hat{H}_{12}\ket{20}=\sqrt{2}\frac{E}{\hbar}\ket{11},\qquad\bra{02}\hat{H}_{12}=\sqrt{2}\frac{E}{\hbar}\bra{11},
\end{equation}
the equation of motion for $\rho_{02,20}$ becomes, to the linear order in $1/\gamma_\text{em}$,
\begin{equation}
\frac{d\rho_{02,20}}{dt}=-4\gamma_\text{em}\rho_{02,20}+i\sqrt{2}\frac{E}{\hbar}\left(
\rho_{02,11}-\rho_{11,20}\right)+O(\gamma_\text{em}^{-2}).
\end{equation}
Substituting Eq.~\eqref{eq:first_order_coherences} gives
\begin{equation}
\frac{d\rho_{02,20}}{dt}=-4\gamma_\text{em}\rho_{02,20}+\frac{2}{9\gamma_\text{em}}\left(\frac{E}{\hbar}\right)^2+O(\gamma_\text{em}^{-2}).
\end{equation}
Therefore, in the stationary state, we have
\begin{equation}
\rho_{02,20}=\frac{1}{18\gamma_\text{em}^2}\left(\frac{E}{\hbar}\right)^2+O(\gamma_\text{em}^{-3}),
\label{eq:rho0220}
\end{equation}
and, from Hermiticity,
\begin{equation}
\rho_{20,02}=\rho_{02,20}^{*}.
\end{equation}

We now determine the dependence of these coherences on the relative
phase. Writing the complex phase-space variables as
\begin{equation}
\alpha_i=r_i e^{i\phi_i},
\end{equation}
the Wigner representation of the coherence
$\ket{0}\bra{2}$ contains a factor $\alpha_i^2$, while that of
$\ket{2}\bra{0}$ contains $(\alpha_i^*)^2$. Hence, the two-mode
coherence $\ket{02}\bra{20}$ gives a phase factor $\alpha_1^2(\alpha_2^*)^2=
r_1^2r_2^2e^{2i\phi_1}e^{-2i\phi_2}=r_1^2r_2^2 e^{2i(\phi_1-\phi_2)}$.
Similarly, the Hermitian-conjugate coherence $\ket{20}\bra{02}$
gives $r_1^2r_2^2 e^{-2i(\phi_1-\phi_2)}$. The two contributions therefore combine according to $e^{2i(\phi_1-\phi_2)}+e^{-2i(\phi_1-\phi_2)}=2\cos\left[2(\phi_1-\phi_2)\right]$.

After integrating out $r_1$, $r_2$, and the overall phase
$\phi_1+\phi_2$, the diagonal part of $\hat{\rho}^{(0)}$ gives a
uniform relative-phase distribution, $1/(2\pi)$. The off-diagonal
elements in Eq.~\eqref{eq:rho0220} give the leading correction.
Consequently, the quantum phase distribution is
\begin{equation}
W_q(\phi_1-\phi_2)=\frac{1}{2\pi}+\frac{E^2}
{9\pi\hbar^2\gamma_\text{em}^2}\cos\left[2(\phi_1-\phi_2)\right]+O\left(\gamma_\text{em}^{-3}\right).
\label{eq:13}
\end{equation}

\subsection{Many interacting quantum limit cycles}
We now generalize the analysis of two coupled quantum limit cycles to an ensemble of $N$ interacting quantum oscillators. Building upon the classical all-to-all coupled model discussed in Section~\ref{sec:many-classical} and the two-oscillator master equation in Section~\ref{sec:2-quantum}, we consider a network of $N$ quantum van der Pol oscillators globally coupled through linear exchange interactions.

The open quantum dynamics of the $N$-oscillator network is described by the Lindblad master equation for the density operator $\hat{\rho}(t)$ on the product Hilbert space $\mathcal{H} = \bigotimes_{n=1}^N \mathcal{H}_n$:
\begin{equation}
\frac{d\hat{\rho}}{dt} = -\frac{i}{\hbar}[\hat{H}_{\text{tot}}, \hat{\rho}] + \sum_{n=1}^N \mathcal{L}_n[\hat{\rho}],
\label{eq:many_lindblad}
\end{equation}
where the dissipator for the $n$-th oscillator contains single-photon gain at rate $\gamma_\text{ab}$ and two-photon loss at rate $\gamma_\text{em}$:
\begin{equation}
\mathcal{L}_n[\hat{\rho}] = \gamma_\text{ab}\mathcal{D}[\hat{a}_n^\dagger]\hat{\rho} + \gamma_\text{em}\mathcal{D}[\hat{a}_n^2]\hat{\rho}.
\end{equation}
The total Hamiltonian describes non-identical oscillators subject to all-to-all coherent coupling:
\begin{equation}
\hat{H}_{\text{tot}} = \sum_{n=1}^N \hbar\omega_n \left(\hat{a}_n^\dagger\hat{a}_n + \frac{1}{2}\right) + \frac{E}{N}\sum_{n=1}^N\sum_{m=1}^N \hat{a}_n^\dagger\hat{a}_m,
\end{equation}
where $\omega_n$ are the bare natural frequencies drawn from a distribution $P(\omega)$, and the factor $1/N$ ensures that the mean-field interaction energy per oscillator remains finite in the thermodynamic limit $N \to \infty$.

\subsubsection{Multi-Mode Wigner Equation and Semiclassical Langevin Dynamics}
To derive the semiclassical dynamics, we use the multimode characteristic function:
\begin{equation}
\chi(\{\eta_n, \eta_n^*\}) = \text{Tr}\left[ \hat{\rho} \prod_{n=1}^N \hat{D}_n(\eta_n) \right],
\end{equation}
and define the $2N$-dimensional Wigner function via the Fourier transform:
\begin{equation}
W(\{\alpha_n, \alpha_n^*\}) = \frac{1}{\pi^{2N}} \int \prod_{n=1}^N d^2\eta_n \, \chi(\{\eta_n, \eta_n^*\}) \exp\left[ \sum_{n=1}^N (-\eta_n\alpha_n^* + \eta_n^*\alpha_n) \right].
\end{equation}

Using the operator-to-Wigner correspondences established in Section~\ref{sec:wigner}, we get
\begin{align}
\hat{a}_n \hat{\rho} &\leftrightarrow \left(\alpha_n + \frac{1}{2}\frac{\partial}{\partial\alpha_n^*}\right)W, \quad \hat{\rho}\hat{a}_n \leftrightarrow \left(\alpha_n - \frac{1}{2}\frac{\partial}{\partial\alpha_n^*}\right)W, \\
\hat{a}_n^\dagger \hat{\rho} &\leftrightarrow \left(\alpha_n^* - \frac{1}{2}\frac{\partial}{\partial\alpha_n}\right)W, \quad \hat{\rho}\hat{a}_n^\dagger \leftrightarrow \left(\alpha_n^* + \frac{1}{2}\frac{\partial}{\partial\alpha_n}\right)W,
\end{align}
we evaluate the drift and diffusion terms generated by each Hamiltonian and dissipator commutator:

\begin{itemize}
    \item \textbf{Free Oscillation Terms:}
    \begin{equation}
    \mathcal{W}_{-i\omega_n [\hat{a}_n^\dagger\hat{a}_n, \hat{\rho}]} = -i\omega_n \left( \alpha_n^*\frac{\partial}{\partial\alpha_n^*} - \alpha_n\frac{\partial}{\partial\alpha_n} \right)W = -\left( \frac{\partial}{\partial\alpha_n}[-i\omega_n\alpha_n W] + \text{c.c.} \right).
    \end{equation}

    \item \textbf{All-to-All Coupling Commutator:}
    \begin{equation}
    \mathcal{W}_{-\frac{iE}{\hbar N} [\hat{a}_n^\dagger\hat{a}_m + \hat{a}_m^\dagger\hat{a}_n, \hat{\rho}]} = -\frac{iE}{\hbar N} \left( \alpha_m^*\frac{\partial}{\partial\alpha_n^*} - \alpha_m\frac{\partial}{\partial\alpha_n} + \alpha_n^*\frac{\partial}{\partial\alpha_m^*} - \alpha_n\frac{\partial}{\partial\alpha_m} \right)W.
    \end{equation}
    Summing over $m$, the net coherent drift acting on mode $n$ is:
    \begin{equation}
    \mathcal{W}_{\text{drift}, n}^{\text{int}} = -\frac{\partial}{\partial\alpha_n}\left( -\frac{iE}{\hbar N}\sum_{m\ne n} \alpha_m \, W \right) + \text{c.c.}
    \end{equation}

    \item \textbf{Dissipator Drift and Diffusion:}
    Applying Eqs. (126) and (130) to each site $n$, we obtain:
    \begin{equation}
    \mathcal{W}_{\mathcal{L}_n[\hat{\rho}]} = -\left( \frac{\partial}{\partial\alpha_n}\Big[(\gamma_\text{ab} + 2\gamma_\text{em} - 2\gamma_\text{em}|\alpha_n|^2)\alpha_n W\Big] + \text{c.c.} \right) + A_0 \frac{\partial^2 W}{\partial\alpha_n\partial\alpha_n^*},
    \end{equation}
    where $A_0 = 3\gamma_\text{ab} + 2\gamma_\text{em}$ in the vicinity of the limit-cycle orbit.
\end{itemize}

Combining all drift and diffusion contributions and dropping third-order quantum corrections, the Fokker-Planck equation for $W(\{\alpha_n, \alpha_n^*\})$ reads:
\begin{equation}
\frac{\partial W}{\partial t} = \sum_{n=1}^N \left[ -\frac{\partial}{\partial\alpha_n}(F_n W) - \frac{\partial}{\partial\alpha_n^*}(F_n^* W) + A_0 \frac{\partial^2 W}{\partial\alpha_n\partial\alpha_n^*} \right],
\end{equation}
with the complex drift coefficients:
\begin{equation}
F_n(\{\alpha_m\}) = \left(-i\omega_n + \gamma_\text{ab} + 2\gamma_\text{em} - 2\gamma_\text{em}|\alpha_n|^2\right)\alpha_n - \frac{iE}{\hbar N}\sum_{m\ne n} \alpha_m.
\end{equation}

The corresponding set of coupled It\^o-Langevin equations for the complex amplitudes 
$\alpha_n(t)$ is:
\begin{equation}
\dot{\alpha}_n = \left(-i\omega_n + \gamma_\text{ab} + 2\gamma_\text{em} - 2\gamma_\text{em}|\alpha_n|^2\right)\alpha_n - \frac{iE}{\hbar N}\sum_{m\ne n} \alpha_m + \xi_n^R(t) + i\xi_n^I(t),
\end{equation}
with Gaussian white noise correlations:
\begin{equation}
\langle \xi_n^R(t)\xi_m^R(t') \rangle = \langle \xi_n^I(t)\xi_m^I(t') \rangle = \left(\frac{3\gamma_\text{ab}}{2} + \gamma_\text{em}\right)\delta_{nm}\delta(t-t'), \quad \langle \xi_n^R(t)\xi_m^I(t') \rangle = 0.
\end{equation}

\subsubsection{Amplitude-Phase Separation and the Noisy Kuramoto Limit}
Writing $\alpha_n = r_n e^{i\theta_n}$, defining $a \equiv \gamma_\text{ab} + 2\gamma_\text{em}$, $b \equiv 2\gamma_\text{em}$, and $K \equiv E/\hbar$, the deterministic part of the Langevin equations separates into radial and phase components:
\begin{align}
\dot{r}_n &= (a - b r_n^2)r_n + \frac{K}{N}\sum_{m\ne n} r_m \sin(\theta_m - \theta_n) + \eta_{nR}(t), \\
\dot{\theta}_n &= -\omega_n - \frac{K}{N}\sum_{m\ne n} \frac{r_m}{r_n}\cos(\theta_m - \theta_n) + \frac{\eta_{nI}(t)}{r_n},
\end{align}
where $\eta_{nR}(t)$ and $\eta_{nI}(t)$ are the transformed noise components:
\begin{equation}
\eta_{nR} = \xi_n^R\cos\theta_n + \xi_n^I\sin\theta_n, \quad \eta_{nI} = -\xi_n^R\sin\theta_n + \xi_n^I\cos\theta_n.
\end{equation}

In the weak-coupling regime ($K \ll a$), the amplitude variables are strongly damped toward their stationary limit-cycle radius $r_n \approx r_0 = \sqrt{a/b} = \sqrt{1 + \gamma_\text{ab}/(2\gamma_\text{em})}$. Substituting $r_n \to r_0$, the phase equation reduces to:
\begin{equation}
\dot{\theta}_n = -\omega_n - \frac{K}{N}\sum_{m=1}^N \cos(\theta_m - \theta_n) + \xi_n^\theta(t),
\end{equation}
where $\xi_n^\theta(t) \equiv \eta_{nI}(t)/r_0$ satisfies $\langle \xi_n^\theta(t)\xi_m^\theta(t') \rangle = 2D_Q \delta_{nm}\delta(t-t')$ with quantum diffusion constant:
\begin{equation}
D_Q = \frac{3\gamma_\text{ab} + 2\gamma_\text{em}}{4 r_0^2} = \frac{3\gamma_\text{ab} + 2\gamma_\text{em}}{4\left(1 + \frac{\gamma_\text{ab}}{2\gamma_\text{em}}\right)}.
\end{equation}

Defining the classical complex Kuramoto order parameter $Z \equiv R e^{i\Theta} = \frac{1}{N}\sum_{m=1}^N e^{i\theta_m}$ as in Section 3.3, the phase dynamics becomes:
\begin{equation}
\dot{\theta}_n = -\omega_n - K R \cos(\Theta - \theta_n) + \xi_n^\theta(t).
\end{equation}
Under a phase shift $\theta_n \to \theta_n + \pi/2$, this maps directly onto the standard Kuramoto equation with noise:
\begin{equation}
\dot{\theta}_n = -\omega_n + K R \sin(\Theta - \theta_n) + \xi_n^\theta(t).
\end{equation}

In the continuum limit ($N \to \infty$), the probability density $\rho(\theta; \omega, t)$ satisfies the non-linear Fokker-Planck (Smoluchowski) equation:
\begin{equation}
\frac{\partial\rho}{\partial t} = -\frac{\partial}{\partial\theta}\left[ \Big(-\omega + K R \sin(\Theta - \theta)\Big)\rho \right] + D_Q \frac{\partial^2\rho}{\partial\theta^2}.
\end{equation}
For a unimodal symmetric natural frequency distribution $P(\omega)$, self-consistency leads to the critical synchronization threshold (compare with the classical threshold value given in Eq. \eqref{eq:critical coupling classical}):
\begin{equation}
K_c = \frac{2}{\pi P(0)} + 2D_Q,
\end{equation}
where quantum fluctuations directly increase the coupling strength required to achieve collective phase locking.

\subsubsection{Mean-Field Master Equation Formulation}
In the thermodynamic limit ($N \to \infty$), the all-to-all coupling can be handled via a Hartree-type mean-field decoupling ansatz for the density operator:
\begin{equation}
\hat{\rho}(t) \approx \bigotimes_{n=1}^N \hat{\rho}_n(t).
\end{equation}
Under this ansatz, the interaction Hamiltonian seen by the $n$-th oscillator is replaced by coupling to a self-consistent collective mean field:
\begin{equation}
\langle \hat{a} \rangle \equiv \frac{1}{N}\sum_{m=1}^N \text{Tr}[\hat{a}_m \hat{\rho}_m] = \int d\omega \, P(\omega) \langle \hat{a}(\omega) \rangle.
\end{equation}
The full $N$-body Lindblad equation decouples into $N$ single-site non-linear master equations:
\begin{equation}
\frac{d\hat{\rho}_n}{dt} = -i\omega_n [\hat{a}_n^\dagger\hat{a}_n, \hat{\rho}_n] - \frac{iE}{\hbar}\left[ \langle\hat{a}\rangle^* \hat{a}_n + \langle\hat{a}\rangle \hat{a}_n^\dagger, \hat{\rho}_n \right] + \gamma_\text{ab}\mathcal{D}[\hat{a}_n^\dagger]\hat{\rho}_n + \gamma_\text{em}\mathcal{D}[\hat{a}_n^2]\hat{\rho}_n.
\end{equation}
The second term acts as a self-consistent coherent drive of effective amplitude $E|\langle\hat{a}\rangle|/\hbar$. For $K < K_c$, the only self-consistent solution is the incoherent state $\langle\hat{a}\rangle = 0$. For $K > K_c$, a spontaneous $U(1)$ symmetry breaking occurs, yielding a non-vanishing stationary mean field $|\langle\hat{a}\rangle| > 0$ that continuously locks the individual quantum oscillators into a collective synchronous state.

\subsubsection{Extreme Quantum Limit and Perturbation in $1/\gamma_\text{em}$}
In the deep quantum regime ($\gamma_\text{em} \gg \gamma_\text{ab}, E/\hbar$), each oscillator is confined predominantly to the lowest two Fock states $\{|0\rangle_n, |1\rangle_n\}$. To zeroth order in $1/\gamma_\text{em}$, the stationary density matrix is completely diagonal (see Eq. \eqref{eq:rho0}):
\begin{equation}
\hat{\rho}^{(0)} = \bigotimes_{n=1}^N \left( \frac{2}{3}|0\rangle_n\langle 0| + \frac{1}{3}|1\rangle_n\langle 1| \right).
\end{equation}
Because $\hat{\rho}^{(0)}$ is a product of phase-invariant number states, all pairwise and collective off-diagonal coherences vanish identically: $\langle \hat{a}_n^\dagger \hat{a}_m \rangle^{(0)} = \langle \hat{a}_n\rangle^{*(0)} \langle\hat{a}_m \rangle^{(0)} = 0$ for all $n \neq m$, where we have used 
\[
\langle a_n\rangle^{(0)} = \frac{2}{3} \langle 0|a_n|0\rangle_n + \frac{1}{3}\langle 1|a_n|1\rangle_n = 0.
\]

At finite $\gamma_\text{em}$, the interaction term $\hat{V} = \frac{E}{N}\sum_{n, m} \hat{a}_n^\dagger \hat{a}_m$ couples the state $|1\rangle_n|1\rangle_m$ to the two-excitation states $|0\rangle_n|2\rangle_m$ and $|2\rangle_n|0\rangle_m$ with matrix element $\sqrt{2}E/N$ (compare with Eq. \eqref{eq:H12_on_11}). Following the perturbation steps of Section 6.1, the steady-state inter-site coherence is generated at second order in $1/\gamma_\text{em}$:
\begin{equation}
\rho_{02,20}^{(nm)} \equiv \langle 0_n 2_m | \hat{\rho} | 2_n 0_m \rangle = \frac{1}{18\gamma_\text{em}^2}\left(\frac{E}{\hbar N}\right)^2 + \mathcal{O}(\gamma_\text{em}^{-3}).
\end{equation}

Integrating out the amplitudes and individual reference phases, the marginal joint phase distribution for any pair $(n, m)$ takes the form:
\begin{equation}
W_q(\phi_n - \phi_m) = \frac{1}{2\pi} + \frac{E^2}{9\pi \hbar^2 N^2 \gamma_\text{em}^2}\cos[2(\phi_n - \phi_m)] + \mathcal{O}(\gamma_\text{em}^{-3}).
\end{equation}

\subsubsection{Quantum Synchronization Observables}
In an $N$-body quantum network, collective synchronization cannot be determined from single classical trajectories; it is quantified using operator expectation values and correlation functions:

\begin{itemize}
    \item \textbf{Pairwise Coherence:} The mutual phase locking between sites $n$ and $m$ is measured by the normalized off-diagonal correlation:
    \begin{equation}
    \mathcal{C}_{nm} \equiv \frac{|\langle \hat{a}_n^\dagger \hat{a}_m \rangle|}{\sqrt{\langle \hat{a}_n^\dagger \hat{a}_n \rangle \langle \hat{a}_m^\dagger \hat{a}_m \rangle}}.
    \end{equation}

    \item \textbf{Macroscopic Quantum Order Parameter:} Generalizing the classical Kuramoto order parameter $R^2 = \frac{1}{N^2}\sum_{n, m} e^{i(\theta_n - \theta_m)}$, macroscopic synchronization across the network is captured by the normalized collective observable:
    \begin{equation}
    S_N \equiv \frac{1}{N(N-1)}\sum_{n \neq m}^N \langle \hat{a}_n^\dagger \hat{a}_m \rangle = \frac{\langle \hat{J}_+ \hat{J}_- \rangle - \sum_{n=1}^N \langle \hat{a}_n^\dagger \hat{a}_n \rangle}{N(N-1)},
    \end{equation}
    where $\hat{J}_- = \sum_{n=1}^N \hat{a}_n$ is the collective annihilation operator. In the unsynchronized phase, independent phase fluctuations yield $S_N \sim \mathcal{O}(1/N) \to 0$ as $N \to \infty$. Above the synchronization threshold ($K > K_c$), long-range phase coherence builds across the network, and $S_N$ settles into a non-zero steady-state value in the thermodynamic limit.
\end{itemize}

\section{Quantum synchronization in spin systems}
\label{sec:sec7}
The smallest possible quantum system that can undergo synchronization, in terms of the space spanned in Hilbert space, is the spin of a particle. A qubit should have been the ideal choice, but it turns out that qubits do not possess a valid limit cycle \cite{Roulet2018}. The next best choice comprises spin-1 particles, which indeed exhibit limit cycles. We begin by  discussing why qubits cannot show synchronization.

In order to have a proper limit cycle, the set of vectors that are taken into account must be phase-invariant \cite{Pikovsky}. For identical classical oscillators, this means that when there is no coupling between them, even if they oscillate with different phases, the measured physical properties remain the same. Therefore, these properties cannot be used to distinguish one oscillator from another. Now, a qubit in a pure state can be represented as (see \ref{app:Bloch-sphere}):
\begin{align}
    |\psi\rangle = \cos\left(\frac{\theta}{2}\right) |0\rangle +  \sin\left(\frac{\theta}{2}\right) e^{i\phi} |1\rangle.
\end{align}
In this context, if the vector $|\psi\rangle$ is rotated about an axis $\vec{n}$ through an angle $\phi_r$, then this rotation angle is the phase associated with the qubit. The question that now arises is: does a set of vectors $\{|\psi_i\rangle\}$ exist, whose elements retain all their physical properties despite being rotated through $\phi_r$? This must be the condition for the state to be phase-invariant.  If the state is not pure, i.e., it is a statistical mixture of multiple pure states, then each of the pure states in the mixture must follow phase-invariance.
The latter statement can be mathematically formulated in terms of the density matrix rather than the state vector, as shown below.

As explained in detail in \ref{app:Bloch-sphere}, the density matrix can be represented in terms of the Pauli matrices $\vec\sigma = (\sigma_x, \sigma_y, \sigma_z)$ as
\begin{align}
    \hat\rho = \frac{1}{2}(I + \vec{a}\cdot \vec{\sigma}),
\end{align}
with $a_x,~a_y$ and $a_z$ being real quantities, and $|\vec a|\le 0$, the equality holding for pure states that lie on the surface of the Bloch sphere. For pure states, the vector $\vec a$  is given by
\begin{align}
    a_x = \sin 2\theta \cos\phi, \hspace{1cm} a_y = \sin 2\theta \sin\phi, \hspace{1cm} a_z = \cos 2\theta.
\end{align}
We say that a state is invariant under rotation through an angle $\phi_r$ about an axis with unit vector $\hat n = (n_x, n_y, n_z)$ if the following condition is met:
\begin{align}
    \hat{\rho}' = e^{-i\phi_r \hat n \cdot \vec \sigma/2} \hat{\rho} e^{+i\phi_r \hat n \cdot \vec \sigma/2} = \hat{\rho}.
\end{align}
This is achieved if and only if $\vec a = \lambda \hat{n}$, where $\lambda$ is a real number. In particular, for an infinitesimal rotation $\phi_r\to 0$, it is easy to check that (ignoring the terms quadratic in $\phi_r$)
\begin{align}
    \left(1 - \frac{i}{2}\phi_r \hat{n}\cdot\vec\sigma\right) (\vec a \cdot \vec\sigma) \left(1 + \frac{i}{2}\phi_r \hat{n}\cdot\vec\sigma\right) = \vec a \cdot \vec\sigma
\end{align}
only when $\vec a = \lambda\hat{n}$, from which the conclusion mentioned above readily follows. Now, this implies that each of these vectors that can be considered for the synchronization can only be statistical mixtures of vectors that are either parallel or anti-parallel to $\vec n$, i.e., of the vectors $|\pm \vec n\rangle$. However, precisely for these vectors, the phase variable becomes undefined, since the rotation axis lies along the direction of the vector itself.  So the condition for synchronization between multiple quantum states cannot be attained for qubits.

\subsection{Synchronization of spin-1 systems}

Consider a large-spin system, where $S\gg 1/2$. The Holstein-Primakoff (HP) transformation is a handy way to represent the spin operators of this system in terms of the harmonic oscillator  creation and annihilation operators. Due to its mapping with the harmonic oscillator, a large-spin system should be able to exhibit synchronization. The problem is to identify what is the smallest value of $S$ for which this is possible.
Since qubits are out of contention, the next higher spin system has $S=1$. These systems indeed exhibit valid limit cycles and synchronization. To study this problem, one needs to extend the well-known Bloch sphere representation of qubits to qutrits. 

Consider the state $|S,M\rangle$, where the magnetic spin quantum number $M$ equals $S$. This extremal state can therefore be represented by $|S,S\rangle$, which forms the ground state of the harmonic oscillator in the HP transformation. The extended Bloch sphere representation is obtained by subjecting the extremal state to rotation operators as follows \cite{Loh2015}:
\begin{align}
    |\theta,\phi\rangle = e^{-i\phi \hat{S}_z} e^{-i\theta \hat{S}_y} |S,S\rangle.
    \label{eq:Bloch S>1/2}
\end{align}
Let the Hamiltonian of the system be
\begin{align}
    H_0 = \hbar\omega_0 \hat{S}_z.
    \label{eq:Hamiltonian S>1/2}
\end{align}
Evolution of the state of the system under this Hamiltonian is given by the standard relation
\begin{align}
    |\theta,\phi'\rangle = e^{-iH_0t/\hbar} |\theta,\phi\rangle.
\end{align}
Note that since the Hamiltonian depends only on $\hat{S}_z$ and not $\hat{S}_y$, the above evolution can only change $\phi$ (see Eq. \eqref{eq:Bloch S>1/2}), while $\theta$ remains constant.

Now, it is well known that the expectation of values of a spin operator behaves like a classical vector precessing about the $z$-axis, with the components (see \cite{Griffiths})
\begin{align}
    \langle S_x\rangle=S \sin\theta\cos\phi; \hspace{1cm}
    \langle S_y\rangle = S \sin\theta\sin\phi; \hspace{1cm}
    \langle S_z\rangle = S\cos\theta.
\end{align}
In other words, we can define the direction of $\langle \hat S\rangle$ by the vector
\begin{align}
\vec n(\theta,\phi) = (\cos\phi\sin\theta) \hat x + (\sin\theta\sin\phi) \hat y + (\cos\theta)\hat z.
\label{eq:n vector}
\end{align}
The vector $|\vec n\rangle = |\theta,\phi\rangle$ can be written by making use of the Wigner $D$-matrix \cite{Loh2015,Sakurai},
given by
\begin{align}
    D^S_{M'M} (\alpha,\beta,\gamma)=\langle S,M'| e^{-i\alpha S_z/\hbar} e^{-i\beta S_y/\hbar} e^{-i\gamma S_z/\hbar} |S,M\rangle,
\end{align}
where $\alpha,~\beta$ and $\gamma$ are the Euler angles of rotation, executed by the rotation  operator $D^S_{M'M} (\alpha,\beta,\gamma)$.
It is convenient to define the Wigner's small $d$ matrix, $d^S_{M'M}$, such that
\begin{align}
    D^S_{M'M} (\alpha,\beta,\gamma) = e^{-i(M'\alpha + M\gamma)} d^S_{M'M}(\beta),
\end{align}
where 
\begin{align}
    d^S_{M'M}(\beta) = \langle S,M'| e^{-iS_y\beta/\hbar} |S,M\rangle.
\end{align}
In the present case, $\beta=\theta$ and $\gamma=\phi$. Making use of the properties of the above matrix, one can arrive at
\begin{align}
    |\theta,\phi\rangle=\sum_{M=-S}^S \sqrt{ \frac{(2S)!}{(S+M)!(S-M)!} }~\left( \cos\frac{\theta}{2}\right)^{S+M} \left( \sin\frac{\theta}{2}\right)^{S-M} e^{-iM\phi} |S,M\rangle. \nonumber\\
    \label{eq:state expansion}
\end{align}
This allows us to find the overlap between two such states, which on simplification (making use of Eq. \eqref{eq:n vector}) gives
\begin{align}
    |\langle \theta',\phi'|\theta,\phi\rangle|^2 = \left( \frac{1+\vec n \cdot \vec n'}{2} \right)^{2S}.
    \label{eq:overlap}
\end{align}
As the spin quantum number becomes very large, $S\to\infty$, the vectors generated from Eq. \eqref{eq:Bloch S>1/2} induce substantial differences in the vector $|\theta,\phi\rangle$, even for small changes in $\theta$ and $\phi$, so that the overlap given by Eq. \eqref{eq:overlap} becomes negligible.

In order to show the completeness of the $|\theta,\phi\rangle$ states, we define the below integral and simplify:
\begin{align}
    I_{M'M}&=\int d\Omega ~\langle S,M'|\theta,\phi\rangle\langle\theta,\phi|S,M\rangle \nonumber\\
    &=\int_0^\pi d\theta~ \sin\theta \int_0^{2\pi} d\phi~ \langle S,M'|\theta,\phi\rangle \langle\theta,\phi|S,M\rangle \nonumber\\
    &=\int_0^\pi d\theta~ \sin\theta \int_0^{2\pi} d\phi~ \langle S,M'| e^{-i\phi \hat{S}_z} e^{-i\theta \hat{S}_y} |S,S\rangle \langle S,S| e^{i\phi \hat{S}_z} e^{i\theta \hat{S}_y} |S,M\rangle\nonumber\\
    &=\int_0^\pi d\theta~ \sin\theta \int_0^{2\pi} d\phi~ e^{i(M-M')\phi} \langle S,M'| e^{-i\theta \hat{S}_y} |S,S\rangle \langle S,S| e^{i\theta \hat{S}_y} |S,M\rangle\nonumber\\
    &=\int_0^{2\pi} d\phi~ e^{i(M-M')\phi} \int_0^\pi d\theta~ \sin\theta ~d^S_{M'S}(\theta) ~d^S_{MS}(\theta) \nonumber\\
    &=2\pi\delta_{M'M} \int_0^\pi d\theta~ \sin\theta ~[d^S_{MS}(\theta)]^2.
\end{align}
In the last line, we have used the definition for delta function in polar coordinates while integrating over $\phi$, and the fact that the Kr\"onecker delta enforces $M'=M$ in the integral over $\theta$. The orthogonality of the Wigner's $d$-function is given by \cite{}
\begin{align}
    \int_0^\pi d\theta\sin\theta ~d^S_{M'S}(\theta) ~d^S_{MS}(\theta) = \frac{2}{2S+1}\delta_{M'M},
\end{align}
so that the integral $I_{M'M}$ reduces to
\begin{align}
    I_{M'M} = \frac{4\pi}{2S+1} \delta_{M'M}.
\end{align}
Thus, from the definition of the integral $I_{M'M}$, we readily get the completeness relation
\begin{align}
   \frac{2S+1}{4\pi} \int_0^\pi d\theta~ \sin\theta \int_0^{2\pi} d\phi~ |\theta,\phi\rangle \langle\theta,\phi| = \hat I.
\end{align}

It is simpler to use the Husimi $Q$-function as compared to the Wigner function to look for synchronization phenomenon in this case (see \ref{app:PQW}):
\begin{align}
    Q(\theta,\phi) = \frac{2S+1}{4\pi} \langle\theta,\phi|\hat\rho|\theta,\phi\rangle.
    \label{eq:Q any S}
\end{align}
The prefactor is essential in order to ensure the normalization: $\int d\Omega~ Q(\Omega)=1$. For a spin-1 particle, the values of $M$ can be $-1,~0$ and 1. Since $S=1$, we have from Eq. \eqref{eq:Q any S},
\begin{align}
    Q(\theta,\phi) = \frac{3}{4\pi} \langle\theta,\phi|\hat\rho|\theta,\phi\rangle.
    \label{eq:Q function spin 1}
\end{align}
The $Q$-function for any $\theta$ and $\phi$ can be computed using the expansion provided by Eq. \eqref{eq:state expansion} for the state $|\theta,\phi\rangle$:
\begin{align}
|\theta,\phi\rangle&=
\cos^2\!\left(\frac{\theta}{2}\right)e^{-i\phi}\,|1,1\rangle +
\sqrt{2}\,\sin\left(\frac{\theta}{2}\right)\cos\left(\frac{\theta}{2}\right)\,|1,0\rangle+ \sin^2\!\left(\frac{\theta}{2}\right)e^{i\phi}\,|1,-1\rangle.
\end{align}
Now, let us consider the state $|S,M\rangle = |1,0\rangle$, which is independent of the extremal basis states $|1,\pm 1\rangle$. Thus, the invariance of $\hat\rho$ for a state involving $|1,0\rangle$ does not automatically preclude the presence of a limit cycle, as was the case for a qubit. 

\begin{figure}
    \centering
    \begin{subfigure}{0.48\linewidth}
    \includegraphics[width=\linewidth]{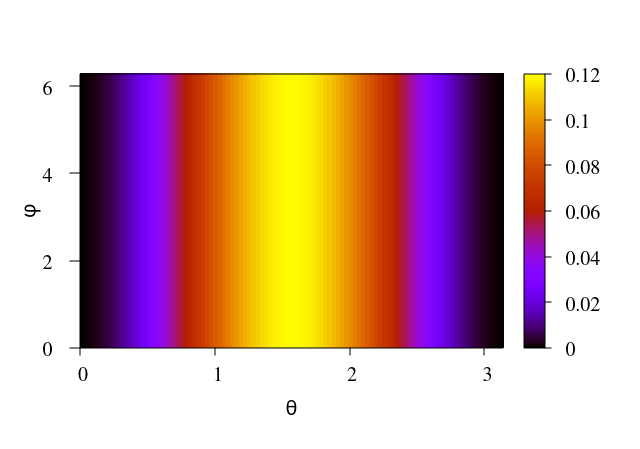}
    \caption{}
    \end{subfigure}
    \begin{subfigure}{0.48\linewidth}
    \includegraphics[width=\linewidth]{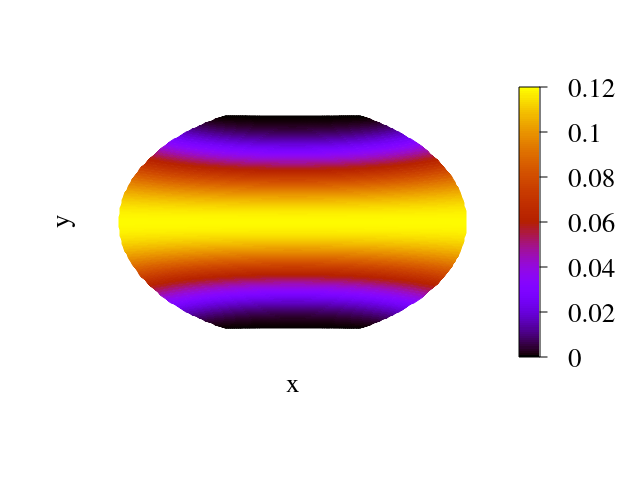}
    \caption{}
    \end{subfigure}

    \caption{(a) Figure showing dependence of the $Q$-function for the $|0,1\rangle$ state on the angles $\theta$ and $\phi$. The radial symmetry about the $z$-axis is clearly visible. (b) The associated Winkel tripel diagram, extending from $-\pi$ to $\pi$ along the $x$-axis, and $\pi$ (south pole) to 0 (north pole) along the $y$-axis.}

    \label{fig:Q theta phi}
\end{figure}

As a special case, we consider the system to be entirely in the pure state $|1,0\rangle$. The corresponding density matrix is $\hat{\rho} = |1,0\rangle \langle 1,0|$. 
Thus, we get
\begin{align}
    \langle\theta,\phi|\hat\rho|\theta,\phi\rangle = 2\sin^2\left(\frac{\theta}{2}\right)\cos^2\left(\frac{\theta}{2}\right) = \frac{1}{2}\sin^2\theta.
\end{align}
The associated $Q$-function then be written down using Eq. \eqref{eq:Q function spin 1}:
\begin{align}
    Q(\theta,\phi) = \frac{3}{8\pi} \sin^2\theta.
\end{align}
We note that the value of the $Q$-function does not depend on $\phi$, so that it is rotationally symmetric about the $z$-axis for a given $\theta$.
A density plot of  $Q(\theta,\phi)$ has been shown in Fig. \ref{fig:Q theta phi}(a), as a function of $\theta$ and $\phi$. However, a better visualization is obtained by using the Winkel tripel projection \cite{}, which involves the following transformations:
\begin{align}
    x &=\frac{1}{2} \left[ \lambda\cos\theta_1 + \frac{2\cos(\theta-\pi/2) \sin(\lambda/2)}{\sin\alpha/\alpha}\right]; \nonumber\\
    y &=\frac{1}{2} \left[ \theta - \frac{\pi}{2} + \frac{\sin(\theta-\pi/2)}{\sin\alpha/\alpha}\right],
\end{align}
where
\begin{align}
    \lambda &= \phi - \pi; \nonumber\\
    \alpha &=\cos^{-1}\left[ \cos\left(\frac{\theta-\pi/2}{2}\right) \cos\left( \frac{\lambda}{2} \right)\right].
\end{align}
The outcome of the Winkel tripel projection has been shown in Fig. \ref{fig:Q theta phi}(b).

Introduction of dissipation helps to stabilize the orbit. In a rotating with angular velocity $\omega_0$ (co-moving frame), the corresponding master equation is given by \cite{Roulet2018}
\begin{align}
    \dot{\hat{\rho}} = \frac{\gamma_{\text{ab}}}{2} \mathcal{D}[\hat{ \mathcal{O} }_{\text{ab}}] \hat{\rho} + \frac{\gamma_{\text{em}}}{2} \mathcal{D}[\hat{ \mathcal{O} }_{\text{em}}] \hat{\rho},
\end{align}
where the dissipator takes the form 
\[
\mathcal{D}[\mathcal{O}]\hat{\rho} = \frac{1}{2}\left(\hat{\mathcal{O}} \hat{\rho}\hat{\mathcal{O}}^\dagger - \frac{1}{2}\{\hat{\mathcal{O}}^\dagger \mathcal{O}, \hat\rho\}\right).
\]
Here, $\mathcal{O}_{\text{ab}} = \hat{S}_+ \hat{S}_z$ and $\mathcal{O}_{\text{em}} = \hat{S}_- \hat{S}_z$, and $S_{\pm}$ are the ladder operators. The authors in  \cite{Roulet2018} use the following measure for synchronization:
\begin{align}
    S(\phi) = \int_0^\pi d\theta \sin\theta Q(\theta,\phi) - \frac{1}{2\pi}.
\end{align}
With this measure, they could show that synchronization exists when either of the two rates, $\gamma_\text{ab}$ or $\gamma_\text{em}$, dominates over the other. The authors used the signal strength, signal detuning and the phase $\phi$ as parameters to show how the synchronization depends on the interplay between them. In particular, they could demonstrate the presence of Arnold's tongue on a density plot for $S(\phi)$ as a function of signal strength and detuning. In a nutshell, a parallel with the results from classical synchronization could be obtained.

\section{Conclusions and perspectives}
\label{sec:sec9}
In this review, we have developed a unified picture of synchronization from classical to quantum systems. We introduced classical limit cycles, phase locking, and the Arnold tongue, and then extended these concepts to interacting oscillators and the Kuramoto model. We subsequently discussed quantum limit cycles and their phase-space description, followed by driven quantum oscillators and synchronization between quantum limit cycles. We also considered synchronization in spin systems and noise-induced synchronization, highlighting how dissipation, quantum fluctuations, and noise can play an essential role in stabilizing synchronized states. We hope that the connection established between classical and quantum synchronization offers a useful framework for both newcomers and experts to appreciate how familiar concepts such as phase locking and collective dynamics persist and acquire new manifestations in the quantum regime.

%%%%%%%%%%%%%%%%%%%%%%%%%%%%%%%%%%%%%%%%%%%%%%
\setcounter{section}{1}

\appendix

%%%%%%%%%%%%%%%%%%%%%%%%%%%%%%%%%%%%%%%%%%%%%%
\addcontentsline{toc}{section}{Appendix A: Brief recap of coherent states for QHO}
\section*{Brief recap of coherent states for QHO}
\setcounter{equation}{0}
\renewcommand{\theequation}{A\arabic{equation}}
\label{app:Coherent}

Here, we review some basic features of coherent states for a QHO. We also summarize some important well-known results for a QHO. 

The Hamiltonian of a single quantum harmonic oscillator in one dimension is given by
\begin{align}
    \hat{H}=\frac{\hat{p}^2}{2m}+\frac{1}{2}m\omega^2\hat{x}^2;~~[\hat{x},\hat{p}]=\mathrm{i}\hbar\hat{I},
    \label{eq:H-SHO}
\end{align}
where $\hat{I}$ is the identity operator, while $x$ and $p$ are respectively the position and the momentum, $m$ is the mass, and $\omega$ is the frequency of the oscillator. 

The creation operator $\hat{a}^\dagger$ and the annihilation operator $\hat{a}$ are defined as
\begin{align}
\hat{a}^\dagger\equiv \frac{1}{\sqrt{2}}\Big(\sqrt{\frac{m\omega}{\hbar}}\hat{x}-\mathrm{i}\frac{1}{\sqrt{m\omega \hbar}}\hat{p}\Big),~\hat{a}\equiv \frac{1}{\sqrt{2}}\Big(\sqrt{\frac{m\omega}{\hbar}}\hat{x}+\mathrm{i}\frac{1}{\sqrt{m\omega \hbar}}\hat{p}\Big),
    \label{eq:aadagger}
    \end{align}
which satisfy
\begin{align}
    [\hat{a},\hat{a}^\dagger]=\hat{I},
\end{align}
as may be derived using $[\hat{x},\hat{p}]=\mathrm{i}\hbar\hat{I}$.
Equation~(\ref{eq:aadagger}) gives
\begin{align}
\hat{x}=\sqrt{\frac{\hbar}{2m\omega}}(\hat{a}^\dagger+\hat{a}),~\hat{p}=\mathrm{i}\sqrt{\frac{m\omega \hbar}{2}}(\hat{a}^\dagger-\hat{a}).
\label{eq:xp-SHO}
\end{align}
Note that $\hat{a}^\dagger \ne \hat{a}$, which means that the creation and the annihilation operator are not Hermitian and hence do not represent physical observables.

The number operator is defined as
\begin{align}
    \hat{N}\equiv\hat{a}^\dagger \hat{a}=\frac{1}{2}\Big(\frac{m\omega}{\hbar}\hat{x}^2+\frac{1}{m\omega \hbar}\hat{p}^2-1\Big),
    \label{eq:Ndefn}
\end{align}
where we have used Eq.~(\ref{eq:aadagger}). One may check that
\begin{align}
    [\hat{N},\hat{a}]=-\hat{a},~[\hat{N},\hat{a}^\dagger]=\hat{a}^\dagger.
    \label{eq:Ncommutator}
\end{align}

With these definitions, we now enumerate some important results for a quantum harmonic oscillator.
\begin{enumerate}
\item Hamiltonian 
$\hat{H}=\hbar \omega\Big(\hat{N}+\frac{1}{2}\Big)$.
\item Operation of $\hat{a}$ and $\hat{a}^\dagger$:
$\hat{a}|n\rangle=\sqrt{n}|n-1\rangle,~\hat{a}^\dagger|n\rangle=\sqrt{n+1}|n+1\rangle,~\hat{a}|0\rangle=0$.
\item When expressed in terms of the creation and the annihilation operator, we have $\hat{x}=\sqrt{\frac{\hbar}{2m\omega}}(\hat{a}^\dagger+\hat{a}),~\hat{p}=\mathrm{i}\sqrt{\frac{m\omega \hbar}{2}}(\hat{a}^\dagger-\hat{a})$.
\item Energy eigenstate
$|n\rangle=\frac{1}{\sqrt{n!}}(\hat{a}^\dagger)^n|0\rangle,~E_n=\hbar \omega\Big(n+\frac{1}{2}\Big);~n=0,1,2,\ldots$.
\item Energy eigenfunction
$\psi_n(x)=\frac{1}{\sqrt{2^n n!}}\Big(\frac{m\omega}{\pi \hbar}\Big)^{1/4}H_n\Big(\sqrt{\frac{m\omega}{\hbar}}x\Big)e^{-m\omega x^2/(2\hbar)}$, where $H_n(x)$ is the Hermite function: $H_0(x)=1, H_1(x)=2x, H_2(x)=4x^2-2,~\mathrm{etc.}$
\item Coherent~states are eigenstates of the annihilation operator: $\hat{\alpha}|\alpha\rangle=\alpha|\alpha\rangle;~\alpha \in \mathbb{C}$; they may be expressed in terms of the energy eigenstates as
\begin{align}
|\alpha\rangle=e^{-|\alpha|^2/2}\sum_{n=0}^\infty \frac{\alpha^n}{\sqrt{n!}}|n\rangle.
\label{eq:coherent-expansion}
\end{align}
It then follows that $\langle n|\alpha\rangle=e^{-|\alpha|^2/2}\frac{\alpha^n}{\sqrt{n!}}$.
\item If the system is in a coherent state, the probability that on measuring the energy, one gets the energy eigenvalue $E_n$ is given by a
Poisson distribution
$P(E_n)=\frac{|\alpha|^{2n}}{n!}e^{-|\alpha|^2}$.
\item The oscillator system when
prepared in a coherent state stays coherent at all times: $|\psi(0)\rangle=|\alpha\rangle \implies |\psi(t)\rangle=e^{-\mathrm{i}\omega t/2}|\alpha e^{-\mathrm{i}\omega t}\rangle$.
\item The motion of the coherent states most closely resembles the oscillatory motion that we see in the position of the classical harmonic oscillator, and this is the reason why coherent states are also called quasi-classical states. We now go on to demonstrate this fact. For the classical motion, we have 
\begin{align}
    &x(t)=x_0\cos(\omega t-\phi),\\
    &p(t)=m\dot{x}=-m\omega x_0\sin(\omega t-\phi),\\
    &E(t)=\frac{1}{2}m\omega^2x_0^2,
\end{align}
where the amplitude $x_0$ of the oscillatory motion and the phase $\phi$ are fixed by initial conditions on the position and the velocity of the oscillator, and where $E(t)$ is the total energy of the system at any instant of time $t$. The expectation value of $\hat{x}$ in an energy eigenstate $|n\rangle$ is
\begin{align}
    \langle \hat{x}\rangle_n=\langle n|\hat{x}|n\rangle=\sqrt{\frac{\hbar}{2m\omega}}\langle n|(\hat{a}+\hat{a}^\dagger)|n\rangle=0.
\end{align}
The above result holds for all times, and so we have $\langle \hat{x}\rangle_n(t)=0$. In the same manner, one may show that $\langle \hat{p}\rangle_n(t)=0$, and $\langle \hat{H}\rangle_n(t)=(n+1/2)\hbar \omega$. On the other hand, using
\begin{align}
    \langle \alpha|(\hat{a})^n|\alpha\rangle=\alpha^n,~\langle \alpha|(\hat{a}^\dagger)^n|\alpha\rangle=(\alpha^*)^n,~\langle \alpha|\hat{a}^\dagger\hat{a}|\alpha\rangle=|\alpha|^2,~\langle\alpha|\hat{a}\hat{a}^\dagger|\alpha\rangle=1+|\alpha|^2,
\end{align}
where $n$ is any positive integer and $\star$ denotes complex conjugation, we get
\begin{align}
    \langle \hat{x}\rangle_\alpha=\sqrt{\frac{\hbar}{2m\omega}}\langle \alpha|\hat{a}+\hat{a}^\dagger|\alpha\rangle=\sqrt{\frac{\hbar}{2m\omega}}(\alpha+\alpha^*)=\sqrt{\frac{2\hbar}{m\omega}}\mathrm{Re}(\alpha),
    \label{eq:xavg-alpha}
\end{align}
where $\mathrm{Re}$ denotes the real part. One may also obtain
\begin{align}
    \langle \hat{x}^2\rangle_\alpha=\frac{\hbar}{2m\omega}[(\alpha+\alpha^*)^2+1],
\end{align}
and hence the following result for the root-mean-squared deviation:
\begin{align}
    \Delta \hat{x}_\alpha\equiv \sqrt{\langle \hat{x}^2\rangle_\alpha-\langle \hat{x}\rangle_\alpha^2}=\sqrt{\frac{\hbar}{2m\omega}},
    \label{eq:x-rms}
\end{align}
which is time independent. One may also derive the following results:
\begin{align}
    &\langle \hat{p}\rangle_\alpha=-\mathrm{i}\sqrt{\frac{m\hbar \omega}{2}}(\alpha-\alpha^*)=\sqrt{2m\hbar \omega}\mathrm{Im}(\alpha),\\
    &\langle \hat{p}^2\rangle_\alpha=\frac{m\hbar \omega}{2}[1-(\alpha-\alpha^*)^2],~\Delta \hat{p}_\alpha=\sqrt{\frac{m\hbar \omega}{2}},
    \label{eq:pavg-alpha}
\end{align}
where $\mathrm{Im}$ denotes the imaginary part.

Suppose the system is initially in a coherent state $|\alpha\rangle$: $|\psi(0)\rangle=|\alpha\rangle$, which we know to be an eigenstate of the lowering operator $\hat{a}$. The corresponding eigenvalue is $\alpha$, which being complex may be written as $\alpha=|\alpha|e^{\mathrm{i}\phi}$. We have seen that in this case, the system remains in a coherent state. Thus, the state at a later time is
\begin{align}
    |\psi(t)\rangle=e^{-\mathrm{i}\omega t/2}|\alpha(t)\rangle;~~\alpha(t)=\alpha e^{-\mathrm{i}\omega t}=|\alpha|e^{-\mathrm{i}(\omega t-\phi)}.
\end{align}
This suggests that we have 
\begin{align}
    \langle \hat{x}\rangle_{\alpha}(t)=\sqrt{\frac{2\hbar}{m\omega}}\mathrm{Re}(\alpha e^{-\mathrm{i}\omega t})=\sqrt{\frac{2\hbar}{m\omega}}|\alpha|\cos(\omega t-\phi).
    \label{eq:xavg-1}
\end{align}
One has similarly that
\begin{align}
    \langle \hat{p}\rangle_{\alpha}(t)=-\sqrt{2m\hbar \omega}|\alpha|\sin(\omega t-\phi).
\end{align}
Further, using the above results, one has
\begin{align}
    &\langle \hat{H}\rangle_\alpha=\hbar \omega\Big(|\alpha|^2+\frac{1}{2}\Big),~\langle \hat{H}^2\rangle_\alpha=\hbar^2\omega^2\Big(|\alpha|^4+2|\alpha|^2+\frac{1}{4}\Big),\\
    &\Delta  \hat{H}_\alpha=\hbar \omega |\alpha|=\Delta \hat{H}_\alpha(t).
\end{align}

Let us note in passing that we have on the basis of the derived results that
\begin{align}
\Delta \hat{x}_\alpha  \Delta \hat{p}_\alpha=\frac{\hbar}{2},   
\end{align}
while the Heisenberg's uncertainty principle states that one has $\Delta \hat{x} \Delta \hat{p} \ge \hbar/2$. We thus see that the coherent states are states with minimum uncertainty.

On the basis of the above discussions, we have
\begin{align}
&\mathrm{Classical}:x(t)=x_0\cos(\omega t-\phi),~~\mathrm{Quantum}:\langle \hat{x}\rangle_\alpha(t)=\sqrt{\frac{2\hbar}{m\omega}}|\alpha|\cos(\omega t-\phi),\\
&\mathrm{Classical}:p(t)=-m\omega x_0\sin(\omega t-\phi),~~\mathrm{Quantum}:\langle \hat{p}\rangle_\alpha(t)=-\sqrt{2m\hbar \omega}|\alpha|\sin(\omega t-\phi),\\
&\mathrm{Classical}:E(t)=\frac{1}{2}m\omega^2x_0^2,~~\mathrm{Quantum}:\langle \hat{H}\rangle_\alpha(t)=\hbar \omega\Big(|\alpha|^2+\frac{1}{2}\Big). 
\end{align}
Then, effecting the identification $x_0=\sqrt{\frac{2\hbar}{m\omega}}|\alpha|$, we get
\begin{align}
    x(t)=\langle \hat{x}\rangle_\alpha,~p(t)=\langle \hat{p}\rangle_\alpha,~E(t)=\langle \hat{H}\rangle_\alpha-\frac{1}{2}\hbar \omega.
\end{align}
Thus, when the quantum harmonic oscillator is in a coherent state, the expectation value of the position and the momentum in such a state oscillates in exactly the same manner as in the corresponding classical motion. The difference between the quantity $E(t)$ and the quantity $\langle \hat{H}\rangle_\alpha$ is the zero-point energy $\hbar \omega/2$, which has a purely quantum origin in the sense this energy contribution is nonzero provided $\hbar\ne 0$.

\item \textit{The displacement operator:} 
The displacement operator is defined as
\begin{align}
    \hat{D}(\alpha)\equiv e^{\alpha \hat{a}^\dagger-\alpha^* \hat{a}};~\alpha \in \mathbb{C}.
    \label{eq:displacement-operator}
\end{align}
The Baker-Campbell-Hausdorff formula gives
\begin{align}
    &\hat{D}(\alpha)=e^{-|\alpha|^2/2}e^{\alpha \hat{a}^\dagger}e^{-\alpha^* \hat{a}}, \\
    &\hat{D}^\dagger(\alpha)
=e^{\alpha^* \hat{a}-\alpha \hat{a}^\dagger}=e^{|\alpha|^2/2}e^{\alpha^* \hat{a}}e^{-\alpha \hat{a}^\dagger},\\
    &\hat{D}(\alpha)\hat{D}^\dagger(\alpha)=e^{-|\alpha|^2/2}e^{\alpha \hat{a}^\dagger}e^{-\alpha^* \hat{a}}e^{|\alpha|^2/2}e^{\alpha^* \hat{a}}e^{-\alpha \hat{a}^\dagger}=\hat{I}=\hat{D}^\dagger(\alpha)\hat{D}(\alpha).
    \label{eq:Dunitary}
\end{align}
Thus, we have that $\hat{D}(\alpha)$ is unitary: $\hat{D}^\dagger(\alpha)=\hat{D}^{-1}(\alpha)$. Also, we have $\hat{D}(-\alpha)=\hat{D}^\dagger(\alpha)$.

The following identities can be easily derived:
\begin{align}
    &\hat{D}(\alpha)\hat{D}(\beta)=e^{(\alpha \beta^*-\alpha^* \beta)/2}\hat{D}(\alpha+\beta),\label{eq:DaDb} \\
    &[\hat{a},\hat{D}(\alpha)]=e^{-|\alpha|^2/2}[\hat{a},\hat{a}^\dagger]\alpha e^{\alpha\hat{a}^\dagger} e^{-\alpha^*\hat{a}}=\alpha\hat{D}(\alpha), \\
    &[\hat{a}^\dagger,\hat{D}(\alpha)]=\alpha^*\hat{D}(\alpha), \\
&\hat{D}^\dagger(\alpha) \hat{a}\hat{D}(\alpha)=\hat{D}^\dagger(\alpha)\hat{D}(\alpha)\hat{a}+\alpha \hat{D}^\dagger(\alpha)\hat{D}(\alpha)=\hat{a}+\alpha, \\
&\hat{D}^\dagger(\alpha) \hat{a}^\dagger\hat{D}(\alpha)=\hat{a}^\dagger+\alpha^*.
\end{align}

We want to now show that the coherent state $|\alpha\rangle$ may be obtained as 
\begin{align}
    |\alpha\rangle=\hat{D}(\alpha)|0\rangle,
    \label{eq:coherent-Dalpha}
\end{align}
which provides an alternative definition of the coherent states.
Indeed, we have
\begin{align}
    \hat{a}\hat{D}(\alpha)|0\rangle=(\hat{D}(\alpha)\hat{a}+\alpha \hat{D}(\alpha))|0\rangle=\alpha(\hat{D}(\alpha)|0\rangle),
\end{align}
which shows that $\hat{D}(\alpha)|0\rangle$ is an eigenstate of $\hat{a}$ with eigenvalue $\alpha$, and this proves Eq.~(\ref{eq:coherent-Dalpha}).
Acting on the state $|\alpha\rangle$ in Eq.~(\ref{eq:coherent-Dalpha}) by $\hat{D}(\beta)$ and using Eq.~(\ref{eq:DaDb}), we get 
\begin{align}
    \hat{D}(\beta)|\alpha\rangle=e^{(\beta \alpha^*-\beta^* \alpha)/2}|\alpha+\beta\rangle,
    \label{eq:Dbetaalpha-1}
\end{align}
which means acting on the coherent state $|\alpha\rangle$ by $\hat{D}(\beta)$ produces, up to an irrelevant phase factor, another coherent state displaced by an amount equal to $\beta$ compared to the starting state $|\alpha\rangle$. This explains the use of the word ``displacement operator" for the operator $\hat{D}(\alpha)$.

\item Let us now discuss some properties of the coherent states.
\begin{enumerate}
    \item The coherent states are not orthogonal, and one has
    \begin{align}
        |\langle \alpha|\beta\rangle|^2=e^{-|\alpha-\beta|^2}.
    \end{align}
    Indeed, we have 
    \begin{align}
        \langle \alpha|\beta\rangle\nonumber&=e^{-|\alpha|^2/2}e^{-|\beta|^2/2}\sum_{n,m=0}^\infty \frac{(\alpha^*)^n\beta^m}{\sqrt{n!m!}}\langle n|m\rangle\nonumber \\
        &=e^{-|\alpha|^2/2}e^{-|\beta|^2/2}\sum_{n=0}^\infty \frac{(\alpha^* \beta)^n}{n!}\nonumber \\
        &=e^{-|\alpha|^2/2}e^{-|\beta|^2/2}e^{\alpha^*\beta}.
        \label{eq:coherent-state-orthonormal-app}
    \end{align}
    We thus get 
\begin{align}
    |\langle \alpha|\beta \rangle|^2=e^{-|\alpha|^2}e^{-|\beta|^2}e^{\alpha^*\beta+\beta^*\alpha}=e^{-|\alpha-\beta|^2}.
    \label{eq:coherent-prop00}
\end{align}
\item The coherent states are complete:
\begin{align}
    \frac{1}{\pi}\int \mathrm{d}^2\alpha~|\alpha\rangle \langle \alpha|=\hat{I},
    \label{eq:coherent-prop0}
\end{align}
where the integration is over the whole of the complex-$\alpha$ plane. This may be shown as follows: We have
\begin{align}
    \frac{1}{\pi}\int \mathrm{d}^2\alpha~|\alpha\rangle\langle \alpha|&=\frac{1}{\pi}\sum_{n,m=0}^\infty \frac{|n\rangle\langle m|}{\sqrt{n!m!}}\int \mathrm{d}^2\alpha~e^{-|\alpha|^2}(\alpha^*)^m \alpha^n.
\end{align}
Writing $\alpha$ in terms of polar coordinates, as $\alpha=re^{\mathrm{i}\phi}$, with $r,\phi \in \mathbb{R};~0\le r \le 1,\phi \in [0,2\pi)$, we get
\begin{align}
    \frac{1}{\pi}\int \mathrm{d}^2\alpha~|\alpha\rangle\langle \alpha|&=\frac{1}{\pi}\sum_{n,m=0}^\infty \frac{|n\rangle\langle m|}{\sqrt{n!m!}}\int_0^\infty \mathrm{d}r~e^{-r^2}r^{n+m+1} \int \mathrm{d}\phi~e^{\mathrm{i}(n-m)\phi}\nonumber \\
    &=2\sum_{n=0}^\infty \frac{|n\rangle\langle n|}{n!}\int_0^\infty \mathrm{d}r~e^{-r^2}r^{2n+1},
    \end{align}
    \label{eq:coherent-prop1}
where we have used $\int_0^{2\pi}\mathrm{d}\phi~e^{-\mathrm{i}(n-m)\phi}=2\pi \delta_{n,m}$.
Now, we have 
\begin{align}
&2\int_0^\infty \mathrm{d}r~e^{-r^2}r^{2n+1}\nonumber \\
&=\int_0^\infty \mathrm{d}y~e^{-y}y^{n}=[-y^{n}e^{-y}]_0^\infty+n\int_0^\infty \mathrm{d}y~y^{n-1}e^{-y}=n\int_0^\infty \mathrm{d}y~y^{n-1}e^{-y}.   
\end{align}
Integrating the last integral by parts, we get 
\begin{align}
&2\int_0^\infty \mathrm{d}r~e^{-r^2}r^{2n+1}\nonumber \\
&=n[-y^{n-1}e^{-y}]_0^\infty+n(n-1)\int_0^\infty \mathrm{d}y~y^{n-2}e^{-y}=n(n-1)\int_0^\infty \mathrm{d}y~y^{n-2}e^{-y}.   
\end{align}
We thus see that an integration of the term $2\int_0^\infty \mathrm{d}r~e^{-r^2}r^{2n+1}$ an $n$ number of times equals $n!$. Hence, we get from Eq.~(\ref{eq:coherent-prop1}) that 
\begin{align}
    \frac{1}{\pi}\int \mathrm{d}^2\alpha~|\alpha\rangle\langle \alpha|=\sum_{n=0}^\infty\frac{|n\rangle \langle n|}{n!}{n!}=\hat{I},
    \end{align}
    proving Eq.~(\ref{eq:coherent-prop0}). 
    Note that we have
    \begin{align}
        1=\mathrm{Tr}[\hat{\rho}]=\mathrm{Tr}\left[\frac{1}{\pi}\int \mathrm{d}^2\alpha~|\alpha\rangle \langle \alpha|\hat{\rho}\right]=\frac{1}{\pi}\int \mathrm{d}^2\alpha\,\mathrm{Tr}[|\alpha\rangle \langle \alpha|\hat{\rho}].
    \end{align}
    \end{enumerate}
    Now, we have $\mathrm{Tr}[|\alpha\rangle \langle \alpha|\hat{\rho}]=\sum_n\langle n |\alpha\rangle \langle \alpha|\hat{\rho}|n\rangle=\sum_n \langle \alpha|\hat{\rho}|n\rangle\langle n |\alpha\rangle=\langle \alpha|\hat{\rho}|\alpha \rangle$. Hence, it follows that 
    \begin{align}
        1=\mathrm{Tr}[\hat{\rho}]=\frac{1}{\pi}\int \mathrm{d}^2\alpha\,\langle \alpha|\hat{\rho}|\alpha \rangle.
        \label{eq:tr-rho-app}
    \end{align}

\end{enumerate}    
%%%%%%%%%%%%%%%%%%%%%%%%%%%%%%%%%%%%%%%%%%%%%%
\addcontentsline{toc}{section}{Appendix B: Phase-space representations in quantum mechanics: Glauber-Sudarshan $P$ function, Husimi $Q$ function, the Wigner function}
\section*{Phase-space representations in quantum mechanics: the Glauber-Sudarshan $P$ function, the Husimi $Q$ function, the Wigner function}
\setcounter{equation}{0}
\renewcommand{\theequation}{B\arabic{equation}}
\label{app:PQW}

Quantum phase-space distributions offer a way to represent quantum states using functions over position–momentum space, analogous to classical phase space, but enriched with inherently-quantum features. The most common of them is the Wigner function, which captures full quantum information yet can take on negative values, reflecting non-classical behavior such as interference. Other examples are the Husimi $Q$ and the Glauber–Sudarshan $P$ functions. Together, they provide tools to bridge classical and quantum descriptions, enabling visualization of quantum dynamics in phase space. In order to introduce the three functions, we first digress to discuss the various possible ordering of the non-commuting creation and annihilation operators.

\subsection*{Normal Ordering}
Normal ordering places all creation operators to the left of all annihilation operators:
\begin{align}
:\hat{a} \hat{a}^\dagger:= \hat{a}^\dagger \hat{a} + 1.
\end{align}
In general, one has
\begin{align}
:\hat{a}^{m} (\hat{a}^\dagger)^{n}:
= \sum_{k=0}^{\min(m,n)}
k!\,\binom{m}{k}\binom{n}{k}\,
(\hat{a}^\dagger)^{\,n-k} \hat{a}^{\,m-k}.
\end{align}

\subsection*{Antinormal Ordering}
Antinormal ordering places all annihilation operators to the left of the creation operators:
\begin{align}
\vdots \hat{a}^\dagger \hat{a} = \hat{a} \hat{a}^\dagger - 1.
\end{align}
In general, we have 
\begin{align}
\vdots(\hat{a}^\dagger)^{n} \hat{a}^{m}\vdots
= \sum_{k=0}^{\min(m,n)}
k!\,\binom{m}{k}\binom{n}{k}\,
\hat{a}^{\,m-k}(\hat{a}^\dagger)^{\,n-k}.
\end{align}

\subsection*{Symmetric (Weyl) Ordering}
Symmetric ordering averages over all possible permutations of creation and annihilation operators:
\begin{align}
\{\hat{a}^\dagger \hat{a}\}_\mathrm{symm} = \frac12 (\hat{a}^\dagger \hat{a} + \hat{a} \hat{a}^\dagger).
\end{align}
Here, the general formula is
\begin{align}
\left\{\hat{a}^{m}(\hat{a}^\dagger)^{n}\right\}_\mathrm{symm}
=\frac{1}{2^{m+n}}
\sum_{k=0}^{\min(m,n)}
k!\,\binom{m}{k}\binom{n}{k}\,
(\hat{a}^\dagger)^{\,n-k} \hat{a}^{\,m-k}.
\end{align}

Corresponding to a density operator $\hat{\rho}$, the Husimi $Q$-function is defined as
\begin{align}
    Q(\alpha,\alpha^*)\equiv\frac{1}{\pi}\langle \alpha |\hat{\rho}|\alpha\rangle.
\end{align}
We have 
\begin{align}
    1=\Tr[\hat{\rho}]=\Tr\left[\frac{1}{\pi}\int d^2\alpha~|\alpha\rangle\langle \alpha|\hat{\rho}|\right]=\frac{1}{\pi}\int d^2\alpha~\sum_n \langle n|\alpha\rangle\langle \alpha|\hat{\rho}|n\rangle=\frac{1}{\pi}\int d^2\alpha~\sum_n \langle \alpha|\hat{\rho}|n\rangle\langle n|\alpha\rangle,
\end{align}
where $|n\rangle$ are the harmonic oscillator eigenstates.
Using $\sum_n |n\rangle \langle n|=1$, we get 
\begin{align}
    1=\frac{1}{\pi}\int d^2\alpha~\langle \alpha|\hat{\rho}|\alpha\rangle,
\end{align}
implying
\begin{align}
    \int d^2\alpha~Q(\alpha,\alpha^*)=1.
\end{align}
It may be shown that the averages of antinormally-ordered products of creation and destruction operators may be expressed in terms of the $Q$ function, as
\begin{align}
    \langle \hat{a}^r(\hat{a}^\dagger)^s\rangle=\Tr[\hat{a}^r(\hat{a}^\dagger)^s\hat{\rho}]=\int d^2\alpha~\alpha^r(\alpha^*)^sQ(\alpha,\alpha^*). 
\end{align}
For a coherent state with $\hat{\rho}=|\alpha_0\rangle \langle \alpha_0|$, one has
\begin{align}
    Q(\alpha,\alpha^*)=\frac{1}{\pi}\exp(-|\alpha-\alpha_0|^2);
\end{align}
The exponential form implies that $Q$ is significantly different from zero only when $\alpha$ is different from $\alpha_0$.

In contrast to the $Q$ function, the Glauber-Sudarshan $P$ function is defined by the following expression:
\begin{align}
  \hat{\rho}\equiv\int d^2\alpha~P(\alpha,\alpha^*)|\alpha\rangle \langle \alpha|.  
\end{align}
Then, we have
\begin{align}
    1=\Tr[\hat{\rho}]=\int d^2\alpha~P(\alpha,\alpha^*)\sum_n\langle n|\alpha\rangle \langle \alpha|n\rangle=\int d^2\alpha~P(\alpha,\alpha^*)\langle \alpha|\alpha\rangle=\int d^2\alpha~P(\alpha,\alpha^*).
\end{align}
 Using Eq.~\eqref{eq:coherent-prop00}, we get $\langle\alpha|\alpha\rangle=1$, yielding
\begin{align}
    \int d^2\alpha~P(\alpha,\alpha^*)=1.
\end{align}
Here, the averages of normally-ordered products of creation and destruction operators may be expressed in terms of the $P$ function, as
\begin{align}
    \langle (\hat{a}^\dagger)^r\hat{a}^s\rangle=\Tr[(\hat{a}^\dagger)^r\hat{a}^s\hat{\rho}]=\int d^2\alpha~(\alpha^*)^r\alpha^sP(\alpha,\alpha^*). 
\end{align}
For a coherent state with $\hat{\rho}=|\alpha_0\rangle\langle \alpha_0|$, one has
\begin{align}
    P(\alpha,\alpha^*)=\delta^2(\alpha-\alpha_0).
\end{align}

In order to establish a relation between the $Q$ and the $P$ function, we define the so-called normally-ordered characteristic function
\begin{align}
    \widetilde{\chi}(\eta,\eta^*)\equiv \Tr[\hat{\rho}e^{\eta \hat{a}^\dagger}e^{-\eta^*\hat{a}}].
\end{align}
Let us recall that in probability theory, the characteristic function for a real random variable $X$ with probability distribution $P_X(x)$ is defined as $\varphi_X(t)\equiv \langle e^{itX}\rangle=\int dx~P_X(x)e^{itx}$. In other words, $\varphi_X(t)$ is the Fourier transform of the probability distribution, which therefore is obtained as the inverse Fourier transform of the characteristic function: $P_X(x)=1/(2\pi)\int_{-\infty}^\infty dt~e^{-itx}\varphi_X(t)$.
The characteristic function has the property that if the $n$-th moment of $P_X(x)$ exists, it is given by the $n$-th derivative of the characteristic function at $t=0$:
\begin{align}
\langle X^n\rangle = \frac{1}{i^n}\,\varphi_X^{(n)}(0),
\end{align}
where $\varphi_X^{(n)}(0)$ is the $n$-th derivative of $\varphi_X(t)$ evaluated at $t=0$. Similar to this role of the classical characteristic function, the derivatives of $\widetilde{\chi}$ at $\eta=0$ give the normally-ordered moments:
\begin{align}
    \langle (\hat{a}^\dagger)^r \hat{a}^s\rangle=(-1)^s\frac{\partial^{r+s}}{\partial^r \eta \partial^s\eta^*}\widetilde{\chi}(\eta,\eta^*)\Big|_{\eta=0}.
\end{align}

Using the Baker-Campbell-Hausdorff formula and the completeness formula $\hat{I}=(1/\pi)\int d^2|\alpha\rangle\langle\alpha|$, see Eq.~\eqref{eq:coherent-prop0}, one gets
\begin{align}
    \widetilde{\chi}(\eta,\eta^*)=\Tr\left[\frac{1}{\pi}\int d^2\alpha~\hat{\rho}e^{|\eta|^2}e^{-\eta^*\hat{a}}|\alpha\rangle \langle \alpha|e^{\eta\hat{a}^\dagger}\right],
\end{align}
implying
\begin{align}
    \widetilde{\chi}(\eta,\eta^*)=e^{|\eta|^2}\int d^2\alpha~e^{\eta \alpha^*-\alpha \eta^*}Q(\alpha,\alpha^*).
    \label{eq:Q}
\end{align}
One may also show that
\begin{align}
    \widetilde{\chi}(\eta,\eta^*)=\int d^2\alpha~e^{\eta \alpha^*-\alpha \eta^*}P(\alpha,\alpha^*).
    \label{eq:P}
\end{align}
Equations~\eqref{eq:Q} and~\eqref{eq:P} establish the connection between the $Q$ and the $P$ function:
\begin{align}
    Q(\alpha,\alpha^*)=\frac{1}{\pi}\int d^2\beta~e^{-|\alpha-\beta|^2}P(\beta,\beta^*).
\end{align}

Next, we come to the Wigner function. In terms of the characteristic function
\begin{align}
    \chi(\eta,\eta^*)=e^{-|\eta|^2/2}\widetilde{\chi}(\eta,\eta^*)=\Tr[e^{-|\eta|^2/2}\hat{\rho}e^{\eta \hat{a}^\dagger}e^{-\eta^*\hat{a}}]=\Tr[\hat{\rho}\hat{D}(\eta)],
\end{align}
where $\hat{D}(\eta)=e^{-|\eta|^2/2}e^{\eta \hat{a}^\dagger}e^{-\eta^*\hat{a}}=e^{\eta \hat{a}^\dagger-\eta^*\hat{a}}$ is the displacement operator, see Eq.~\eqref{eq:displacement-operator}, the Wigner function is defined as a Fourier transform:
\begin{align}
    W(\alpha,\alpha^*) \equiv \frac{1}{\pi^2}\int d^2\alpha~e^{-\eta \alpha^*+\eta^*\alpha}\chi(\eta,\eta^*).
    \label{eq:Wigner-app-0}
\end{align}
The averages of symmetrically-ordered products of creation and destruction operators may be expressed in terms of the Wigner function, as
\begin{align}
    \langle \{\hat{a}^r(\hat{a}^\dagger)^s\}_\mathrm{symm}\rangle=\Tr[\{\hat{a}^r(\hat{a}^\dagger)^s\}_\mathrm{symm}\hat{\rho}]=\int d^2\alpha~\alpha^r(\alpha^*)^sW(\alpha,\alpha^*). 
\end{align}
For a coherent state $\hat{\rho}=|\alpha_0\rangle \langle \alpha_0|$, we have
\begin{align}
    W(\alpha,\alpha^*)=\frac{2}{\pi}e^{-2|\alpha-\alpha_0|^2}.
\end{align}

%%%%%%%%%%%%%%%%%%%%%%%%%%%%%%%%%%%%%%%%%%%%%%%%%%%%%%%%%%%%%%%%%%%
\addcontentsline{toc}{section}{Appendix C: Dynamical equations for the real and imaginary parts of Wigner function}
\section*{Dynamical equations for the real and imaginary parts of Wigner function}
\setcounter{equation}{0}
\renewcommand{\theequation}{C\arabic{equation}}
\label{sec:Separation of FP}

\setcounter{equation}{0}
\renewcommand{\theequation}{D\arabic{equation}}

We recall that the dynamics of the Wigner function (see Eq. \eqref{eq:FP-1}) is given by a form similar to the Fokker-Planck equation:
\begin{align}
    \frac{\partial W}{\partial t}=-\left(\frac{\partial }{\partial \alpha}[-\alpha BW]+\frac{\partial }{\partial \alpha^*}[-\alpha^* BW]\right)+\left(\frac{\partial^2}{\partial \alpha \partial \alpha^*}\right)[AW],
    \label{eq:app-FP-1}
\end{align}
where
\begin{eqnarray}
    A(\alpha,\alpha^*) &=& \gamma_1 + 2\gamma_2(2|\alpha|^2-1),\nonumber\\
    B(\alpha,\alpha^*) &=& -\gamma_1 + 2\gamma_2(|\alpha|^2-1).
\end{eqnarray}
Simulations of the above equation can be carried out by separating it into real and imaginary parts, and each of these new equations can be readily subjected to the standard numerical integration algorithm to solve Fokker-Planck equations. Let us define $\alpha \equiv \alpha_x + i \alpha_y$.
We further note for any complex function $f(\alpha,\alpha^*)$ that
\begin{eqnarray}
    df &=& \frac{\partial f}{\partial\alpha} d\alpha + \frac{\partial f}{\partial\alpha^*} d\alpha^* \nonumber\\
    &=&   \frac{\partial f}{\partial \alpha_x} d\alpha_x + \frac{\partial f}{\partial \alpha_y} d\alpha_y  \nonumber\\
    &=& \frac{\partial f}{\partial \alpha_x} \frac{d\alpha + d\alpha^*}{2} + \frac{\partial f}{\partial \alpha_y} \frac{d\alpha - d\alpha^*}{2i}.
\end{eqnarray}
By comparing the coefficients of $d\alpha$ and $d\alpha^*$ appearing in the first and the third line, we find
\begin{eqnarray}
    \frac{\partial}{\partial\alpha} &=& \frac{1}{2}\left( \frac{\partial }{\partial \alpha_x} - i \frac{\partial}{\partial \alpha_y}\right),\nonumber\\ \\
    \frac{\partial}{\partial\alpha^*} &=& \frac{1}{2}\left( \frac{\partial }{\partial \alpha_x} + i \frac{\partial}{\partial \alpha_y}\right).\nonumber
\end{eqnarray}

Equation~\eqref{eq:app-FP-1} now becomes
\begin{eqnarray}
    \frac{\partial}{\partial t}W &=& \frac{1}{2}\left( \frac{\partial }{\partial \alpha_x} - i \frac{\partial}{\partial \alpha_y}\right)\left[(\alpha_x + i \alpha_y)B W\right] + \frac{1}{2}\left( \frac{\partial }{\partial \alpha_x} + i \frac{\partial}{\partial \alpha_y}\right)\left[(\alpha_x - i \alpha_y)B W\right] \nonumber\\
    && + \frac{1}{4}\left( \frac{\partial^2}{\partial\alpha_x^2} + \frac{\partial^2}{\partial\alpha_y^2} \right)(A W)\nonumber\\
    &=& \frac{1}{2}\left( \frac{\partial }{\partial \alpha_x} - i \frac{\partial}{\partial \alpha_y}\right)(X_1 + iY_1) 
    + \frac{1}{2}\left( \frac{\partial }{\partial \alpha_x} + i \frac{\partial}{\partial \alpha_y}\right)(X_2 + iY_2) \nonumber\\
    && + \frac{1}{4}\left( \frac{\partial^2}{\partial\alpha_x^2} + \frac{\partial^2}{\partial\alpha_y^2} \right) (A W), 
    \label{eq:FP-split-0}
\end{eqnarray}
where we have
\begin{eqnarray}
    X_1 &=& \alpha_x BW; \hspace{1cm} X_2 = \alpha_x BW; \hspace{1cm}  Y_1 = \alpha_y BW; 
    \hspace{1cm} Y_2 =- \alpha_y BW.
    \label{eq:X and Y}
\end{eqnarray}
%
% %
Equation~\eqref{eq:FP-split-0} then simplifies to
\begin{eqnarray}
    \frac{\partial W}{\partial t} &=& \frac{1}{2}\left[ \frac{\partial }{\partial \alpha_x}(X_1+X_2) + \frac{\partial }{\partial \alpha_y}(Y_1-Y_2)\right] + \frac{1}{4}\nabla^2 (AW) \nonumber\\
    && + \frac{i}{2}\left[ \frac{\partial }{\partial \alpha_x}(Y_1+Y_2) - \frac{\partial }{\partial \alpha_y}(X_1-X_2)\right],
    \label{eq:Wigner 1}
\end{eqnarray}
where we have
\begin{eqnarray}
    \nabla^2 \equiv \frac{\partial^2}{\partial\alpha_x^2} + \frac{\partial^2}{\partial\alpha_y^2}.
\end{eqnarray}
From Eq. \eqref{eq:X and Y} we readily obtain
\begin{eqnarray}
    X_1 + X_2 &=& 2\alpha_x BW; \hspace{1cm} Y_1 - Y_2 = 2\alpha_y BW; 
    \hspace{1cm} X_1 - X_2 = Y_1 + Y_2 = 0; \hspace{1cm} \nonumber
\end{eqnarray}
Note that the imaginary part of \eqref{eq:Wigner 1} becomes identically zero. Then we arrive at the following equation for $W$:
\begin{eqnarray}
    \frac{\partial W}{\partial t} &=& \frac{\partial }{\partial \alpha_x}(\alpha_x BW) + \frac{\partial }{\partial \alpha_y} (\alpha_y BW) + \frac{1}{4}\nabla^2(A W).
\end{eqnarray}
The above equation can be further simplified to
\begin{eqnarray}
    \frac{\partial W}{\partial t} &=& \left( 2 + \alpha_x\frac{\partial}{\partial\alpha_x} + \alpha_y\frac{\partial}{\partial\alpha_y} \right) (B W) + \frac{1}{4}\nabla^2(A W).
\end{eqnarray}
%%%%%%%%%%%%%%%%%%%%%%%%%%%%%%%%%%%%

\addcontentsline{toc}{section}{Appendix D: Derivation of the Lindblad equation}
\section*{Derivation of the Lindblad equation}
\setcounter{equation}{0}
\renewcommand{\theequation}{D\arabic{equation}}
\label{app:Lindblad}

This appendix is devoted to a derivation of the Gorini–Kossakowski–Sudarshan–Lindblad (GKSL) equation or simply the Lindblad equation, an example of which is Eq.~\eqref{eq:vdpo_lee}. Understanding how a quantum system interacts with its environment is essential for understanding the phenomenon of quantum synchronization, as will be exemplified in the main text.

In the standard formulation of quantum mechanics, a system evolves in time according to unitary dynamics. Consequently, phase coherence is perfectly maintained. In quantum mechanics, a system can exist in a superposition state $|\psi\rangle=c_1|\psi_1\rangle$ and $c_2|\psi_2\rangle$, where the $c_i$'s
 are complex numbers, with the relative phase between them  
 determining how the two states $|\psi_1\rangle$ and $|\psi_2\rangle$ interfere. By phase coherence we mean that this relative phase remains constant or predictable over time, and it is this coherence that leads to quantum interference patterns such as those seen in the double-slit experiment. Unitary dynamics is a result of the underlying quantum system being completely isolated from the environment, and the energy of the system remains conserved at all times. A realistic system is however never completely isolated, and interactions with external degrees of freedom constituting the environment lead to energy dissipation into the environment, resulting in decay processes and loss of phase coherence.
To model the aforementioned effects, one particularly convenient approach that takes into account the coupling between the quantum system and its environment involves writing down the master equation for the time evolution of the density matrix or the density operator of the system alone. The Gorini–Kossakowski–Sudarshan–Lindblad (GKSL) equation (named after Vittorio Gorini, Andrzej Kossakowski, George Sudarshan and G\"{o}ran Lindblad) is a general form of Markovian master equation describing the dynamics of open quantum systems. The equation generalizes the Schr\"{o}dinger equation to open quantum systems. The Lindblad dynamics is no longer unitary. Under Lindblad evolution, the trace of the density matrix remains constant (equal to unity) over time, and moreover, the density matrix remains completely positive semi-definite for any initial condition\footnote[1]{A square matrix $A$ is positive semidefinite if $\langle \psi|A|\psi\rangle$ is a nonnegative real number
for every vector $|\psi\rangle$. An equivalent definition is that $A$ is positive semi-definite if (i) $A$ is Hermitian, and (ii) all eigenvalues of $A$ are nonnegative real numbers.}. 

Our derivation of the Lindblad equation gets simplified in the so-called interaction picture (different from the well-known Schr\"{o}dinger and the Heisenberg picture), which we briefly describe below. Suppose the total Hamiltonian $\hat{H}_\text{tot}$ of a quantum system can be split into two parts:
\begin{eqnarray}
    \hat{H}_\text{tot} = \hat{H}_0 + \hat{H}_1,
\end{eqnarray}
where $\hat{H}_0$ is the part whose eigenfunctions and eigenvalues are known, and $\hat{H}_1$ may be interpreted as a perturbation (not necessarily small) applied to the system. Then, an operator $\hat{\mathcal{O}}$ in the Schr\"odinger picture is related to the corresponding operator $\hat{\widetilde{\mathcal{O}}}$ in the interaction picture by
\begin{eqnarray}
    \hat{\widetilde{\mathcal{O}}} = e^{i\hat{H}_0t/\hbar} \hat{\mathcal{O}} e^{-i\hat{H}_0t/\hbar}.
    \label{eq:interaction picture}
\end{eqnarray}
The advantage is that in this picture, the Liouville-von Neumann evolution equation $i\hbar d\hat{\rho}_\text{tot}/dt = [\hat{H}_\text{tot},\hat{\rho}_\text{tot}]$ for the total density operator $\hat{\rho}_\text{tot}$ of the system becomes
\begin{eqnarray}
    i\hbar\frac{d\hat{\widetilde{\rho}}_\text{tot}}{dt} = [\hat{\widetilde H}_1,\hat{\widetilde{\rho}}_\text{tot}],
    \label{eq:evolution interaction picture}
\end{eqnarray}
i.e., the part $\hat{H}_0$ does not appear explicitly in the time-evolution equation. 

In order to derive the Lindblad equation, it is usual to model the environment as a heat bath in a stationary state. For a quantum system in contact with such a heat bath, we begin by writing the total Hamiltonian of the system and the heat bath as
\begin{eqnarray}
    \hat{H}_\text{tot} = \hat{H}_\text{S} + \hat{H}_\text{B} + \hat{H}_\text{SB},
    \label{eq:Hamiltonian}
\end{eqnarray}
where $\hat{H}_\text{S}$ is the system Hamiltonian, $\hat{H}_\text{B}$ is the bath Hamiltonian, and $\hat{H}_\text{SB}$ models the system-bath interaction. The Hamiltonians $\hat{H}_\text{S}$ and $\hat{H}_\text{B}$ commute with each other, while the interaction term will be considered to be of the form \cite{Breuer,Manzano2020}
\begin{eqnarray}
    \hat{H}_\text{SB} = \sum_i \hat{A}_i\otimes \hat{B}_i,
    \label{eq:interaction Hamiltonian}
\end{eqnarray}
where $\hat{A}_i$ and $\hat{B}_i$ are Hermitian operators in the Hilbert space of the system and the bath, respectively.
Consider $\hat{H}_0=\hat{H}_\text{S}+\hat{H}_\text{B}$, and $\hat{H}_1=\hat{H}_\text{SB}$, so that Eq.~\eqref{eq:interaction picture} becomes
\begin{eqnarray}
\hat{\widetilde{\mathcal{O}}} = e^{i(\hat{H}_\text{S}+\hat{H}_\text{B})t/\hbar} \hat{\mathcal{O}} e^{-i(\hat{H}_\text{S}+\hat{H}_\text{B})t/\hbar}.
\label{eq:relevant interaction picture}
\end{eqnarray}
The interaction Hamiltonian $\hat{\widetilde{H}}_\text{SB}$ and the quantity  $\hat{\widetilde{\rho}}_{\text{tot}}$ can be obtained by replacing $\hat{\mathcal{O}}$ in Eq. \eqref{eq:relevant interaction picture} by $\hat{H}_\text{SB}$ and $\hat{\rho}_\text{tot}$, respectively.
Within this setting,  Eq. \eqref{eq:evolution interaction picture} becomes
\begin{eqnarray}
     i\hbar\frac{d\hat{\widetilde{\rho}}_\text{tot}}{dt} = [ \hat{\widetilde{H}}_\text{SB},\hat{\widetilde{\rho}}_\text{tot}],
    \label{eq:int evol 0}
\end{eqnarray}
expanding which we get
\begin{eqnarray}
    \hat{\widetilde{\rho}}_\text{tot}(t) = \hat{\widetilde{\rho}}_\text{tot}(0) - \frac{i}{\hbar} \int_0^t dt' [\hat{\widetilde{H}}_\text{SB}(t'),\hat{\widetilde{\rho}}_{\text{tot}}(t')].
    \label{eq:int evol 1}
\end{eqnarray}
Substituting Eq.~\eqref{eq:int evol 1} in Eq.~\eqref{eq:int evol 0}, we obtain
\begin{eqnarray}
    i\hbar\frac{d\hat{\widetilde{\rho}}_\text{tot}(t)}{dt} = [\hat{\widetilde{H}}_\text{SB}, \hat{\widetilde{\rho}}_\text{tot}(0)] - \frac{i}{\hbar}\int_0^t dt'~ [~\hat{\widetilde{H}}_\text{SB}(t),[\hat{\widetilde{H}}_\text{SB}(t'),\hat{\widetilde{\rho}}_\text{tot}(t')]~].
\end{eqnarray}
 We may now trace over the bath degrees of freedom, to arrive at the following equation for the density operator of the system alone:
\begin{eqnarray}
    \frac{d\hat{\widetilde{\rho}}(t)}{dt} &=& -\frac{i}{\hbar}\text{Tr}_\text{B} [\hat{\widetilde{H}}_\text{SB}, \hat{\widetilde{\rho}}_\text{tot}(0)] \nonumber\\
    &&-\frac{1}{\hbar^2} \int_0^t dt' ~\text{Tr}_\text{B}[~\hat{\widetilde{H}}_\text{SB}(t),[\hat{\widetilde{H}}_\text{SB}(t'),\hat{\widetilde{\rho}}_\text{tot}(t')]~].
    \label{eq:Lindblad step 1}
\end{eqnarray}
    We further assume that initially, i.e., at time $t=0$, the system and the bath were uncoupled, so that the total density operator is a tensor product of the system density operator and the bath density operator, i.e., $\hat{\widetilde{\rho}}_\text{tot}(0) = \hat{\widetilde{\rho}}(0)\otimes \hat{\rho}_\text{B}(0)$ (note that from Eq. \eqref{eq:interaction picture}, we have $\hat{\widetilde{\rho}}_\text{B}(0)=\hat{\rho}_\text{B}(0)$). This gives
\begin{align}
\text{Tr}_\text{B}[\hat{\widetilde{H}}_\text{SB}, \hat{\widetilde{\rho}}_\text{tot}(0)] &= \sum_i \text{Tr}_\text{B} [\hat{A}_i\otimes \hat{B}_i, \hat{\widetilde{\rho}}(0)\otimes\hat{\rho}_\text{B}(0)] \nonumber\\
&= \sum_i [\hat{A}_i,\hat{\widetilde{\rho}}(0)] \text{Tr}_\text{B}[\hat{\rho}_\text{B}(0) \hat{B}_i] = \sum_i [\hat{A}_i,\hat{\widetilde{\rho}}(0)] \langle \hat{B}_i\rangle_\text{B}(0).
\end{align}
In the last equality, we have used the definition for the expectation value written in terms of the density operator, as $\text{Tr}_\text{B}[\hat{\rho}(0)_\text{B} \hat{B}_i]= \langle \hat{B}_i\rangle_\text{B}(0)$. We next assume that $\langle \hat{B}_i\rangle_\text{B}(0) = 0$, so that the first term on the right hand side of Eq.~\eqref{eq:Lindblad step 1} vanishes\footnote[1]{This may be justified by noting that the system and the interaction Hamiltonian can always be redefined so as to satisfy this condition: set $\hat{H} \to \hat{H} + \sum_i \langle \hat{B}_i(0)\rangle_\text{B} \hat{A}_i$, and $\hat{H}_\text{SB}\to \sum_i \hat{A}_i \otimes (\hat{B}_i - \langle \hat{B}_i(0)\rangle_\text{B})$, so that $\hat{H}_\text{tot}$ in Eq.~\eqref{eq:Hamiltonian} remains unchanged. The condition $\langle \hat{B}_i(0)\rangle_\text{B}=0$ now becomes $\langle \hat{B}_i(0) - \langle \hat{B}_i(0)\rangle_\text{B}\rangle_\text{B} = 0$, which is always true.}.
Also, we have $\hat{\widetilde{\rho}}(t)=\text{Tr}_\text{B} [\hat{\widetilde{\rho}}_\text{tot}(t)]$, by definition. 

To proceed, one makes the ``Born approximation'', whereby we assume that the bath state is negligibly affected by the evolution of the system. In other words, by the virtue of the bath being much larger than the system, it remains in a stationary state at all times.  This implies that we have $\hat{\widetilde{\rho}}_\text{tot}(t') \simeq \hat{\widetilde{\rho}}(t')\otimes \hat{\rho}_\text{B}(0)$. Setting $\text{Tr}_\text{B}[\hat{\widetilde{H}}_\text{SB}, \hat{\widetilde{\rho}}_\text{tot}(0)] = 0$ and using the Born approximation, Eq. \eqref{eq:Lindblad step 1} becomes
\begin{eqnarray}
    \frac{d\hat{\widetilde{\rho}}(t)}{dt} =  -\frac{1}{\hbar^2} \int_0^t dt' ~\text{Tr}_\text{B}[~\hat{\widetilde{H}}_\text{SB}(t),[\hat{\widetilde{H}}_\text{SB}(t'),\hat{\widetilde{\rho}}(t')\otimes \hat{\rho}_\text{B}(0)]~].
    \label{eq:Lindblad step 1.1}
\end{eqnarray}

We further assume that the system evolution lacks memory (the ``Markov approximation''), so that the time derivative $d\hat{\widetilde{\rho}}(t)/d t$ depends only on $\hat{\widetilde{\rho}}(t)$ and not on $\hat{\widetilde{\rho}}(t')$ with $t'<t$. To understand the source of this approximation, let us first redefine the variable $t'\to t-t'$ and rewrite Eq. \eqref{eq:Lindblad step 1.1} as 
\begin{eqnarray}
    \frac{d\hat{\widetilde{\rho}}(t)}{dt} =  -\frac{1}{\hbar^2} \int_0^t dt' ~\text{Tr}_\text{B}[~\hat{\widetilde{H}}_\text{SB}(t),[\hat{\widetilde{H}}_\text{SB}(t-t'),\hat{\widetilde{\rho}}(t-t')\otimes \hat{\rho}_\text{B}(0)~].
\end{eqnarray}
Now, since the integral involves the commutator between $\hat{\widetilde{H}}_\text{SB}(t)$ and $\hat{\widetilde{H}}_\text{SB}(t-t')$, taking the trace would involve factors containing time-correlations between bath operators in the integrand. Defining $\tau_B$ to be the finite correlation time for the bath operators, almost the entire contribution to the integral comes from times $t'\lesssim \tau_B$. Typically, the system correlation time $\tau_S$ is much larger than $\tau_B$. Due to this, for $t'\lesssim \tau_B$, the system state remains almost unaffected. Thus, we can approximate $\hat{\widetilde{\rho}}(t-t')$ by $\hat{\widetilde{\rho}}(t)$:
\begin{eqnarray}
    \frac{d\hat{\widetilde{\rho}}(t)}{dt} =  -\frac{1}{\hbar^2} \int_0^t dt' ~\text{Tr}_\text{B}[~\hat{\widetilde{H}}_\text{SB}(t),[\hat{\widetilde{H}}_\text{SB}(t-t'),\hat{\widetilde{\rho}}(t)\otimes \hat{\rho}_\text{B}(0)~].
    \label{eq:Redfield}
\end{eqnarray}
This is known as the Redfield equation.
Next, defining $\mathcal{G}(t-t')$ to be the integrand appearing in Eq. \eqref{eq:Redfield}, we observe that
\begin{eqnarray*}
    \int_0^t dt' ~\mathcal{G}(t-t') = \int_0^\infty dt' ~\mathcal{G}(t-t') - \int_t^\infty dt' ~ \mathcal{G}(t-t') \simeq \int_0^\infty dt' ~ \mathcal{G}(t-t'),
\end{eqnarray*}
since in the integral $\int_t^\infty dt' ~ \mathcal{G}(t-t')$, we have $t'\gg \tau_B$, as our observation time $t$ must be larger than the time scale over which the system changes, and can be neglected. Thus, one can safely replace the upper limit $t$ of the integral in Eq. \eqref{eq:Redfield} by infinity:
\begin{align}
    \frac{d\hat{\widetilde\rho}(t)}{dt} = -\frac{1}{\hbar^2} \int_0^\infty dt' ~\text{Tr}_\text{B}[~\hat{\widetilde H}_\text{SB}(t),[\hat{\widetilde H}_\text{SB}(t-t'),\hat{\widetilde\rho}(t)\otimes \hat{\rho}_\text{B}(0)]~].
\end{align}
This completes the imposition of the Markov approximation.

Let us define the operators $\hat A_i(\omega)$ that are related to $\hat A_i$ in Eq. \eqref{eq:interaction Hamiltonian} via
\begin{align}
    \hat A_i(\omega) = \sum_{\varepsilon'-\varepsilon=\omega} |\varepsilon\rangle\langle\varepsilon|\hat A_i|\varepsilon'\rangle\langle\varepsilon'| = \sum_{\varepsilon'-\varepsilon=\omega} (\langle\varepsilon|\hat A_i|\varepsilon'\rangle)|\varepsilon\rangle\langle\varepsilon'| ,
    \label{eq:A definition}
\end{align}
where $\varepsilon$ and $\varepsilon'$ are pairs of energy eigenvalues of $H_S$ whose difference equals $\omega$. The summation thus includes only transitions that take place at a given frequency $\omega$. 

The commutation relations of this operator with the system Hamiltonian are given by
\begin{align}
    [\hat H_\text{S},\hat{A}_i(\omega)] = -\omega\hat{A}_i(\omega); 
    ~~~~ [\hat H_\text{S},\hat{A}_i^\dagger(\omega)] = +\omega\hat{A}_i^\dagger(\omega).
    \label{eq:eigenoperators}
\end{align}
These operators therefore behave like ladder operators of a quantum harmonic oscillator. Operators having properties \eqref{eq:eigenoperators} are said to be eigenoperators of $H_\text{S}$. From Eq. \eqref{eq:relevant interaction picture}, it is clear that since $\hat A_i$ belongs to the system's Hilbert space, the interaction-picture operator becomes $\hat{\widetilde{A}}_i
= e^{i(\hat{H}_\text{S}+\hat{H}_\text{B})t/\hbar} \hat{A}_i e^{-i(\hat{H}_\text{S}+\hat{H}_\text{B})t/\hbar} 
= e^{i\hat{H}_\text{S}t/\hbar} \hat A_i e^{-i\hat{H}_\text{S}t/\hbar}$, 
because  $e^{i\hat H_\text{B}t/\hbar}$ and $e^{-i\hat H_\text{B}t/\hbar}$ cancel each other. Similarly, we get $\hat{\widetilde{B}}_i = e^{i\hat{H}_\text{B}t/\hbar} \hat B_i e^{-i\hat{H}_\text{B}t/\hbar}$. One can readily show that
\begin{align}
\sum_\omega \hat{A}_i(\omega) &= \sum_\omega \hat{A}_i^\dagger(\omega) =\hat{A}_i; \hspace{0.2cm}  \nonumber\\
    ~~~~\hat{\widetilde A}_i(\omega) &= e^{i\hat{H}_\text{S}t/\hbar}\hat{A}_i(\omega) e^{-iH_\text{S}t/\hbar} = e^{-i\omega t/\hbar}\hat{A}_i(\omega); \nonumber\\
    ~~~~\hat{\widetilde A}_i^\dagger(\omega) &= e^{iH_\text{S}t/\hbar}\hat{A}_i^\dagger(\omega) e^{-iH_\text{S}t/\hbar} = e^{i\omega t/\hbar}\hat{A}_i^\dagger(\omega).
    \label{eq:A properties}
\end{align}
We further note from the Eq. \eqref{eq:A definition} that $\hat A^\dagger(\omega) = \hat A(-\omega)$, which, when combined with  Eq. \eqref{eq:A properties}, leads to $\hat{\widetilde A}_i(\omega) = \hat{\widetilde A}_i^\dagger(\omega)$.
 Also, from definition (see discussion above Eq. \eqref{eq:A properties}), we get $\hat{\widetilde B}_i = \hat{\widetilde B}_i^\dagger$.
Using these relations and Eq. \eqref{eq:A properties} in Eq. \eqref{eq:interaction Hamiltonian}, we arrive at
\begin{align}
    \hat{\widetilde H}_\text{SB} = \sum_i \hat{\widetilde A}_i\otimes \hat{\widetilde B}_i = \sum_{i,\omega} e^{-i\omega t/\hbar}\hat{A}_i(\omega)\otimes \hat{\widetilde B}_i = \sum_{i,\omega} e^{+i\omega t/\hbar}\hat{A}_i^\dagger(\omega)\otimes \hat{\widetilde B}_i^\dagger,
    \label{eq:H_SB interaction picture}
\end{align}
 Using this form of $\hat{\widetilde H}_\text{SB}$, after some algebra, Eq. \eqref{eq:Redfield} simplifies to
\begin{eqnarray}
    \hspace{-1cm}\frac{d\hat{\widetilde\rho}}{dt} &=& \sum_{i,j}\sum_{\omega,\omega'}e^{i(\omega'-\omega)t/\hbar} \Gamma_{ij}(\omega) [\hat{A}_j(\omega)\hat{\widetilde\rho}\hat{A}_i^\dagger(\omega') 
    -\hat{A}_i^\dagger(\omega')\hat{A}_j(\omega)\hat{\widetilde\rho}] + \text{h.c.},
    \label{eq:Lindblad step 2}
\end{eqnarray}
where h.c. implies Hermitian conjugate, and
\begin{align}
     \Gamma_{ij}(\omega) = \frac{1}{\hbar^2}\int_0^\infty dt'~ e^{i\omega t'/\hbar} \langle \hat{\widetilde B}_i^\dagger(t)\hat{\widetilde B}_j(t-t')\rangle_\text{B}.
     \label{eq:Gamma full}
\end{align}
Here, since we are considering the bath to be in a stationary state, the correlation appearing inside the integrand depends only on the time difference $t'$. The integration over $t'$ ensures that $\Gamma_{ij}(\omega)$ depends only on $\omega$ and not on $t$. 

Let $\tau_\text{S}$ be the time scale over which the system state varies appreciably,  and $\tau_\text{relax}$ be the time scale over which the system relaxes to a stationary state. The time scale over which the relative phase between different transitions in the state of the system oscillates is proportional to the inverse of the typical frequency difference: $\tau_\text{S} \sim |\omega'-\omega|^{-1}$, for $\omega'\ne\omega$. Typically, in a quantum optical system, the condition $\tau_\text{relax} \gg \tau_\text{S} \gg  \tau_\text{B}$ is satisfied, where $\tau_\text{B}$ is the bath correlation time. In this case, only the contributions coming from $\omega'\approx\omega$ (that produce a slowly varying expression on the right-hand side) are significant. The other contributions, where $\omega$ and $\omega'$ are significantly different, can be neglected due to the highly oscillatory nature of $e^{i(\omega'-\omega)t/\hbar}$. This approximation of neglecting the $\omega\ne\omega'$ terms is called the ``rotating wave approximation''\footnote[1]{The approximation derives its name from the fact that the phase factor  in the summand on the right-hand side of Eq. \eqref{eq:Lindblad step 2} ``rotates'' in the complex plane with frequency $\omega'-\omega$, and we are discarding the fast-rotating terms.}, and  simplifies Eq. \eqref{eq:Lindblad step 2} to 
\begin{align}
    \frac{d\widetilde\rho}{dt} = \sum_{ij\omega} \Gamma_{ij}(\omega) [\hat A_j(\omega)\hat{\widetilde\rho}\hat{A}_i^\dagger(\omega) 
    -\hat{A}_i^\dagger(\omega)\hat A_j(\omega)\hat{\widetilde\rho}] + \text{h.c.}
    \label{eq:rotating wave}
\end{align}
    Without this approximation, it is known that one or more of the eigenvalues of the density operator can become negative, depending on the chosen set of parameters. This would render the density operator unphysical \cite{Manzano2020}.

Let us now write $\Gamma_{ij}(\omega)$ as the sum of a symmetric part $\Gamma^\text{s}_{ij}$ and an antisymmetric part $\Gamma^\text{as}_{ij}$:
\begin{align}
    \Gamma_{ij}(\omega) = \frac{1}{2}\Gamma^\text{s}_{ij}(\omega) + \frac{i}{\hbar}\Gamma^\text{as}_{ij}(\omega),
    \label{eq:Gamma split}
\end{align}
where
\begin{align}
    \Gamma^\text{s}_{ij}(\omega) &= \Gamma_{ij}(\omega) + \Gamma^*_{ij}(\omega) = 2\mathrm{Re}[\Gamma_{ij}(\omega)];\nonumber\\
    \Gamma^\text{as}_{ij}(\omega) &= \frac{\hbar}{2i} (\Gamma_{ij}(\omega) - \Gamma^*_{ij}(\omega)) = \hbar~\mathrm{Im}[\Gamma_{ij}(\omega)].
    \label{eq:Gamma sym asym}
\end{align}
Using Eq. \eqref{eq:Gamma full} in \eqref{eq:Gamma sym asym}, we obtain
\begin{align}
   \Gamma^\text{s}_{ij}(\omega) = \frac{1}{\hbar^2}\int_{-\infty}^\infty dt'~e^{i\omega t'}\langle\hat{\widetilde B}_i^\dagger(t)\hat{\widetilde B}_j(t-t')\rangle_\text{B}.
   \label{eq:Gamma sym}
\end{align}
Substituting Eq. \eqref{eq:Gamma split}  into Eq. \eqref{eq:rotating wave}, straightforward algebra leads to
\begin{align}
    \frac{d\widetilde\rho}{dt} = -\frac{i}{\hbar}[\hat H_\text{LS},\hat{\widetilde\rho}] + \mathcal{D}[\hat{\widetilde\rho}],
    \label{eq:Lindblad step 3}
\end{align}
where
\begin{align}
    ~~~~~~\hat H_\text{LS} &= \sum_{ij\omega} \Gamma^\text{as}_{ij}(\omega)\hat A_i^\dagger(\omega)\hat A_j(\omega); \nonumber\\
    \mathcal{D}[\hat{\widetilde\rho}] &= \sum_{ij\omega} \Gamma^\text{s}_{ij}(\omega)\left[\hat A_j(\omega)\hat{\widetilde\rho}\hat{A}_i^\dagger(\omega) - \frac{1}{2}\{\hat A_i^\dagger(\omega)\hat A_j(\omega),\hat{\widetilde\rho}\}\right].
\end{align}
The operator $H_\text{LS}$ is called the Lamb shift Hamiltonian.
% caused by the interaction of the atom with vacuum fluctuations \cite{Wiseman}. 
The superoperator $\mathcal{D}[\hat{\widetilde\rho}]$ is known as the dissipator. The anticommutator $\{\cdots\}$  appearing on the right-hand side is defined with respect to two arbitrary operators $\hat \alpha_1$ and $\hat \alpha_2$ as: $\{\hat \alpha_1,\hat \alpha_2\} = \hat \alpha_1\hat \alpha_2+\hat \alpha_2\hat \alpha_1$ .

One may now transform to the Schr\"odinger picture, by replacing $\widetilde\rho$ by $e^{i\hat H_\text{S}t/\hbar}\hat\rho e^{-i\hat H_\text{S}t/\hbar}$ (see Eq. \eqref{eq:relevant interaction picture}). Then, the left-hand side of Eq. \eqref{eq:Lindblad step 3} becomes
\[
\frac{d\hat{\widetilde\rho}}{dt} = e^{i\hat H_\text{S}t/\hbar}\left(\frac{i}{\hbar}[\hat H_\text{S},\hat\rho] + \frac{d\hat\rho}{dt}\right)e^{-i\hat H_\text{S}t/\hbar}.
\]
With the help of the above relation, we find that Eq. \eqref{eq:Lindblad step 3} transforms to
\begin{align}
    \frac{d\hat\rho}{dt} = -\frac{i}{\hbar}[\hat H_\text{S}+\hat H_\text{LS},\hat\rho] + \sum_{ij\omega} \Gamma^\text{s}_{ij}(\omega)\left[\hat A_j(\omega)\hat\rho\hat{A}_i^\dagger(\omega) - \frac{1}{2}\{\hat A_i^\dagger(\omega)\hat A_j(\omega),\hat\rho\}\right].
    \label{eq:Lindblad step 4}
\end{align}
Now, let $\hat U$ be the unitary operator that diagonalizes $\Gamma^\text{s}_{ij}(\omega)$, so that $\Gamma^\text{s}(\omega) = \hat U \gamma(\omega) \hat U^\dagger$, or $\Gamma^\text{s}_{ij}(\omega) = U_{ik}\gamma_{k}(\omega)U^\dagger_{kj}$, where $\gamma(\omega)$ is the diagonalized matrix, and $\gamma_{kk}$ has been written simply as $\gamma_k$. Substituting in \eqref{eq:Lindblad step 4} and defining the operators $L_k(\omega) = \sum_j U^\dagger_{kj}A_j(\omega)$, we get
\begin{align}
    \frac{d\hat \rho}{dt} =  -\frac{i}{\hbar}[\hat H_\text{S}+\hat H_{LS},\hat \rho] + \sum_{k}\sum_{\omega} \gamma_k(\omega)\left[\hat L_k(\omega)\hat \rho \hat L_k^\dagger(\omega) - \frac{1}{2}\{\hat L_k^\dagger(\omega)\hat L_k(\omega),\hat\rho\}\right].
    \label{eq:Lindblad Schrodinger}
\end{align}
It is easy to check that $\hat{L}_k(\omega)$ bear similar commutation relations with $\hat H_S$ as $\hat A_i(\omega)$ (see Eq. \eqref{eq:eigenoperators}). Eq. \eqref{eq:Lindblad Schrodinger} is the general form of the Lindblad master equation in the Schr\"odinger picture.  Often, only a single frequency $\omega$ is of interest, like for an open quantum harmonic oscillator, or a two-level system. In that case, the sum over $\omega$ can be dropped, and Eq. \eqref{eq:Lindblad Schrodinger} gets further simplified:
\begin{align}
    \frac{d\hat\rho}{dt} =  -\frac{i}{\hbar}[\hat H_\text{S}+\hat H_{LS},\hat \rho] + \sum_{k} \gamma_k\left[\hat L_k\hat \rho \hat L_k^\dagger - \frac{1}{2}\{\hat L_k^\dagger \hat L_k,\hat\rho\}\right].
    \label{eq:Lindblad Schrodinger simplified}
\end{align}
The operators $L_k$ and $L_k^\dagger$ are referred to as \textit{jump operators}. Often, the $H_{LS}$ term is small compared to $H_\text{S}$, and is dropped from the commutator.

\paragraph{Lindblad equation for open quantum harmonic oscillator} This formalism may be used to study the damping of an electron oscillating under the effect of a radiation fields inside a cavity \cite{Breuer}.  For such a mode of frequency $\omega_0$ interacting with radiation modes of the bath field, the system Hamiltonian is known to be given by
\begin{align}
    \hat H_\text{S} = \hbar\omega_0 (a^\dagger a + 1/2),
\end{align}
where $a^\dagger$ and $a$ are the creation and annihilation operators of the QHO. The bath Hamiltonian $H_\text{B}$ consists of the sum of the Hamiltonians of all the modes of oscillations of its electric field, the latter being given by
\begin{align}
    \hat{\vec{E}} &= i \sum_{\textbf{k}}\sum_{\lambda=1,2}\left(\sqrt{\frac{2\pi\hbar\omega_k}{V}}\right)\vec{e}_\lambda(\vec{k})\left(\hat b_\lambda(\vec{k})-\hat b_\lambda^\dagger(\vec{k})\right) \nonumber\\
    &= \sum_{j=\textbf{k},\lambda} (\vec \gamma_j \hat b_j + \vec \gamma_j^* \hat b_j^\dagger),
    \label{eq:E}
\end{align}
where $\textbf{k}$ is the wave vector, $\hat{\vec{e}}_\lambda$ gives the two polarization directions for $\lambda=1,2$, the volume of the container is $V$, the bath oscillator frequencies are $\omega_j$, and $\vec\gamma_j = i(\sqrt{2\pi\hbar\omega_j/V})\hat{\vec{e}}_\lambda(\vec{k})$. We can then write
\begin{align}
    H_\text{B} = \sum_j \hbar\omega_j (b_j^\dagger b_j + 1/2).
\end{align}
The interaction Hamiltonian for the oscillating charge takes the form
\begin{align}
    \hat H_\text{SB} = -\hat{\vec\mu}\cdot\hat{\vec E} = -Q\hat{\vec{r}}\cdot\hat{\vec E},
    \label{eq:H_SB dipole}
\end{align}
where $\hat{\vec\mu} = Q\hat{\vec{r}}$ is the electric dipole moment of the system, and $Q$ is the charge of the particle. 
Thus, the interaction Hamiltonian, from Eq. \eqref{eq:H_SB dipole}, is given by
\begin{align}
    \hat H_\text{SB} = (\hat a + \hat a^\dagger)\sum_{j} (g_j \hat b_j + g_j^* \hat b_j^\dagger),
    \label{eq:H_SB dipole Schrod}
\end{align}
where $g_j = [Q \sqrt{\hbar/(2m\omega_0)}] \textbf{n}\cdot \vec{e}_j$, with $\textbf{n}$ being the unit position vector in the direction of the dipole moment. In other examples, the overall form of the interaction Hamiltonian remains the same, with the specific system-bath coupling being captured by the parameters $g_j$. In the interaction picture, we have
\begin{align}
    \hat{\widetilde{a}} &= e^{iH_\text{S}t/\hbar}\hat a e^{-iH_\text{S}t/\hbar} =  \hat a e^{-i\omega_0 t};\nonumber\\
    \hat{\widetilde{a}}^\dagger &= e^{iH_\text{S}t/\hbar}\hat a^\dagger e^{-iH_\text{S}t/\hbar} = \hat a^\dagger e^{+i\omega_0 t};\nonumber\\
    \hat{\widetilde{b}}_j &= e^{iH_\text{B}t/\hbar}\hat b_j e^{-iH_\text{B}t/\hbar} = \hat b_j e^{-i\omega_j t}; \nonumber\\
    \hat{\widetilde{b}}^\dagger_j &= e^{iH_\text{B}t/\hbar}\hat b_j^\dagger e^{-iH_\text{B}t/\hbar} =\hat b_j^\dagger e^{+i\omega_j t}.
\end{align}
Thus, we obtain from Eq. \eqref{eq:H_SB dipole Schrod},
\begin{align}
    \hat{\widetilde{H}}_\text{SB} &= \left(\hat a e^{-i\omega_0 t} + \hat a^\dagger e^{+i\omega_0 t}\right)\sum_j\left(g_j\hat b_j e^{-i\omega_j t} + g_j^*\hat b_j^\dagger e^{+i\omega_j t}\right)\nonumber\\
    &\simeq \sum_j \left(g_j^*\hat a\hat b_j^\dagger e^{i(\omega_j-\omega_0) t} + g_j\hat a^\dagger\hat b_j e^{-i(\omega_j-\omega_0) t}\right),
\end{align}
where we have ignored the highly oscillating terms of frequencies $(\omega_j+\omega_0)$. This approximation tantamounts to writing Eq. \eqref{eq:H_SB dipole Schrod} as $\hat{H}_\text{SB} = \sum_i \hat{A}_i\otimes \hat{B}_i$ with the operators
\begin{align}
    \hat{A}_1 &= \hat{a}; ~~~~ \hat{A}_2 = 
    \hat{a}^\dagger; \nonumber\\
    \hat{B}_1 &= \sum_j g_j^* \hat{b}_j^\dagger; ~~~~ \hat{B}_2 = \sum_j g_j \hat{b}_j.
\end{align}
The rotating wave approximation, given by \eqref{eq:rotating wave}, is automatically satisfied, since the coefficients $\langle\varepsilon| \hat A_i |\varepsilon'\rangle$ appearing in the definition \eqref{eq:A definition} vanishes for all $|\varepsilon' - \varepsilon| \ne \omega_0$.
The coefficients $\Gamma_{ij}(\omega_0)$, appearing in Eq. \eqref{eq:Lindblad step 2} are given by Eq. \eqref{eq:Gamma full}:
\begin{align}
    \Gamma  _{ij}(\omega_0) = \frac{1}{\hbar^2}\int_{0}^\infty dt'~e^{i\omega_0 t'}\langle\hat{\widetilde B}_i^\dagger(t)\hat{\widetilde B}_j(t-t')\rangle_\text{B}.
    \label{eq:QHO Gamma full}
\end{align}
To evaluate the expectation values appearing inside the integral, we first note the identities (obtained by using the definition $\langle \hat{\mathcal{O}}\rangle_\text{B} = \langle N| \hat{\mathcal{O}}|N\rangle$ for any number state $|N\rangle$ of the bath)
\begin{eqnarray}
    \langle \hat b_j \hat b_{j'} \rangle_\text{B} &=&  \langle \hat b_j^\dagger \hat b_{j'}^\dagger\rangle_\text{B} = 0; \nonumber\\
    \langle \hat b_j^\dagger \hat b_{j'}\rangle_\text{B} &=& \delta_{jj'}N(\omega_j);\nonumber\\
    \langle \hat b_j \hat b_{j'}^\dagger\rangle_\text{B} &=& \delta_{jj'}(1+N(\omega_j)).
\end{eqnarray}
Here, $N(\omega_j)$ is the Planck occupation probability at the inverse temperature $\beta=(k_B T)^{-1}$, given by
\begin{align}
    N(\omega_j) = \frac{1}{e^{\beta\hbar\omega_j}-1}.
\end{align}
Using these identities, we obtain
\[
 \langle \hat{\widetilde B}_1(t) \hat{\widetilde B}_1(t-t') \rangle_\text{B} =  \sum_{j,j'} g_j^* g_{j'}^* e^{i\omega_j t}  e^{i\omega_{j'}(t-t')} \langle \hat b^\dagger_j \hat b^\dagger_{j'}\rangle_\text{B} = 0.
\]
The other combinations can be similarly calculated, and the full list is given below:
\begin{eqnarray}
    \langle \hat{\widetilde B}_1(t) \hat{\widetilde B}_1(t-t') \rangle_\text{B} &=& \langle \hat{\widetilde B}_2(t) \hat{\widetilde B}_2(t-t') \rangle_\text{B} = 0; \nonumber\\
    \langle \hat{\widetilde B}_1(t) \hat{\widetilde B}_2(t-t') \rangle_\text{B} &=& \sum_j |g_j|^2 e^{i\omega_j t'} N(\omega_j); \nonumber\\
    \langle \hat{\widetilde B}_2(t) \hat{\widetilde B}_1(t-t') \rangle_\text{B} &=& \sum_j |g_j|^2 e^{-i\omega_j t'} (1+N(\omega_j)).
\end{eqnarray}
We can substitute these expressions into Eq. \eqref{eq:QHO Gamma full}, and convert the summation to an integral as follows:
\begin{align}
    \sum_j \to \int \frac{d^3 k_j}{(2\pi/L)^3} = \frac{V}{(2\pi c)^3}\int_0^\infty d\omega_j\omega_j^2\int d\Omega.
\end{align}
Here, $L^3 = V$ is the volume of the cavity, and $d\Omega$ is the solid angle subtended by a volume element in $\vec k$ space at the origin. 
After carrying out the integral, we find that the symmetric part (see Eq. \eqref{eq:Gamma split}) of $\Gamma_{ij}(\omega)$ is given by \cite{Breuer}
\begin{align}
    \Gamma^\text{s}_{ij}(\omega_0) = \frac{1}{2}\delta_{ij}\gamma(\omega_0),
\end{align}
where
\begin{align}
    \gamma(\omega_0) &=& \left\{ 
    \begin{array}{cc}
      \gamma_0 \omega_0^3(1+N(\omega_0)),    & \mbox{if } \omega_0 > 0 ~~\mbox{(emission)}; \\ \\
    \gamma_0 |\omega_0|^3 N(|\omega_0|),      & \mbox{if } \omega_0 < 0 ~~ \mbox{(absorption)}.
    \end{array}
    \right.
\end{align}
Here, the constant $\gamma_0$ depends on the system and bath parameters that determine the coupling strength.
Putting everything together, we find that Eq. \eqref{eq:Lindblad step 4} takes the form
\begin{eqnarray}
    \frac{d\hat\rho}{dt} 
     &=& -\frac{i}{\hbar}[\hat{H}_\text{S} + \hat{H}_\text{LS},\hat\rho] + \gamma_\text{ab} \left[\hat a^\dagger \hat\rho \hat a - \frac{1}{2}(\hat a \hat a^\dagger \hat\rho + \hat\rho \hat a \hat a^\dagger)\right] \nonumber\\
     && + \gamma_\text{em} \left[\hat a \hat\rho \hat a^\dagger - \frac{1}{2}(\hat a^\dagger \hat a \hat\rho + \hat\rho \hat a^\dagger \hat a)\right].
\end{eqnarray}
Here, $\gamma_\text{ab} = \gamma_0 \omega_0^3 N(\omega_0)$ and $\gamma_\text{em} = \gamma_0  \omega_0^3(1+N(\omega_0))$ (with $\omega_0 = |\omega_0| > 0$ in both terms, by convention) are the transition rates associated with the absorption and emission processes of the system, respectively. 

It can be checked that higher powers of $a$ and $a^\dagger$ (that correspond to jumps between non-consecutive energy levels) are also follow the properties \eqref{eq:eigenoperators} of the system Hamiltonian, so that these transitions can form a part of the Lindblad equation if the coupling between the system and bath is nonlinear (as opposed to Eq. \eqref{eq:H_SB dipole Schrod}).

%%%%%%%%%%%%%%%%%%%%%%%%%%%%%%%%%%%%%%%%%%%%%%
\addcontentsline{toc}{section}{Appendix E: Vectorization of the von Neumann and Lindblad equations}
\label{app:Vectorization}
\section*{Vectorization of the von Neumann and Lindblad equations}
\setcounter{equation}{0}
\renewcommand{\theequation}{E\arabic{equation}}

An $m\times n$ matrix $A$ may be written as an $mn\times 1$ column vector by stacking its columns:
\begin{align}
\vec{A} = [a_{11}, \ldots, a_{m1}, a_{12}, \ldots, a_{m2}, \ldots, a_{1n}, \ldots, a_{mn}]^T.
\end{align}
Given two matrices $A$ and $B$ of sizes $k\times l$ and $l\times m$, it can be shown that
\begin{align}
\overrightarrow{AB} = (I_m \otimes A)\vec{B} = (B^T \otimes I_k)\vec{A}.
\end{align}

Treating the Hamiltonian $\hat{H}$ and the density matrix $\hat{\rho}$ as matrices $H$ and $\rho$, respectively, and using $\overrightarrow{H\rho} = (I\otimes H)\vec{\rho}$ and $\overrightarrow{\rho H} = (H^T\otimes I)\vec{\rho}$, we get
\begin{align}
\overrightarrow{[H,\rho]} = (I\otimes H - H^T\otimes I)\vec{\rho},
\end{align}
yielding the von Neumann equation $d\hat{\rho}/dt=-(i/\hbar)[\hat{H},\hat{\rho}]$, as
\begin{align}
\dot{\vec{\rho}} = -iH_\times \vec{\rho}, \quad H_\times = I\otimes H - H^T\otimes I.
\end{align}
The solution reads as
\begin{align}
\vec{\rho}(t) = e^{-iH_\times t}\vec{\rho}(0).
\end{align}

For the Lindblad equation $d\hat{\rho}/dt=-(i/\hbar)[\hat{H},\hat{\rho}]+\mathcal{D}[\hat{O}]\hat{\rho}$ with collapse operator $\mathcal{D}[\hat{O}]\hat{\rho} = 2\hat{O}\hat{\rho} \hat{O}^\dagger - \hat{O}^\dagger \hat{O}\hat{\rho} - \hat{\rho} \hat{O}^\dagger \hat{O}$, using
\begin{align}
(A\otimes B)\vec{V} = \overrightarrow{BVA^T},
\end{align}
we have
\begin{align}
\overrightarrow{\mathcal{D}[O]\rho} = 2(O^\ast \otimes O)\vec{\rho} - (I\otimes O^\dagger O)\vec{\rho} - ((O^\dagger O)^T \otimes I)\vec{\rho} = \mathcal{W}\vec{\rho}.
\end{align}
The Lindblad equation then reads as
\begin{align}
\dot{\vec{\rho}} = (-iH_\times + \mathcal{W})\vec{\rho},
\end{align}
which has the solution
\begin{align}
\vec{\rho}(t) = e^{(-iH_\times+\mathcal{W}) t}\vec{\rho}(0).
\end{align}
%%%%%%%%%%%%%%%%%%%%%%%%%%%%%%%%%%%%%%%%%%%%%%
\addcontentsline{toc}{section}{Appendix F: Bloch sphere}
\section*{Bloch sphere}
\label{app:Bloch-sphere}

\setcounter{equation}{0}
\renewcommand{\theequation}{F\arabic{equation}}

\begin{figure}[!ht]
\centering
\includegraphics[width=6cm]{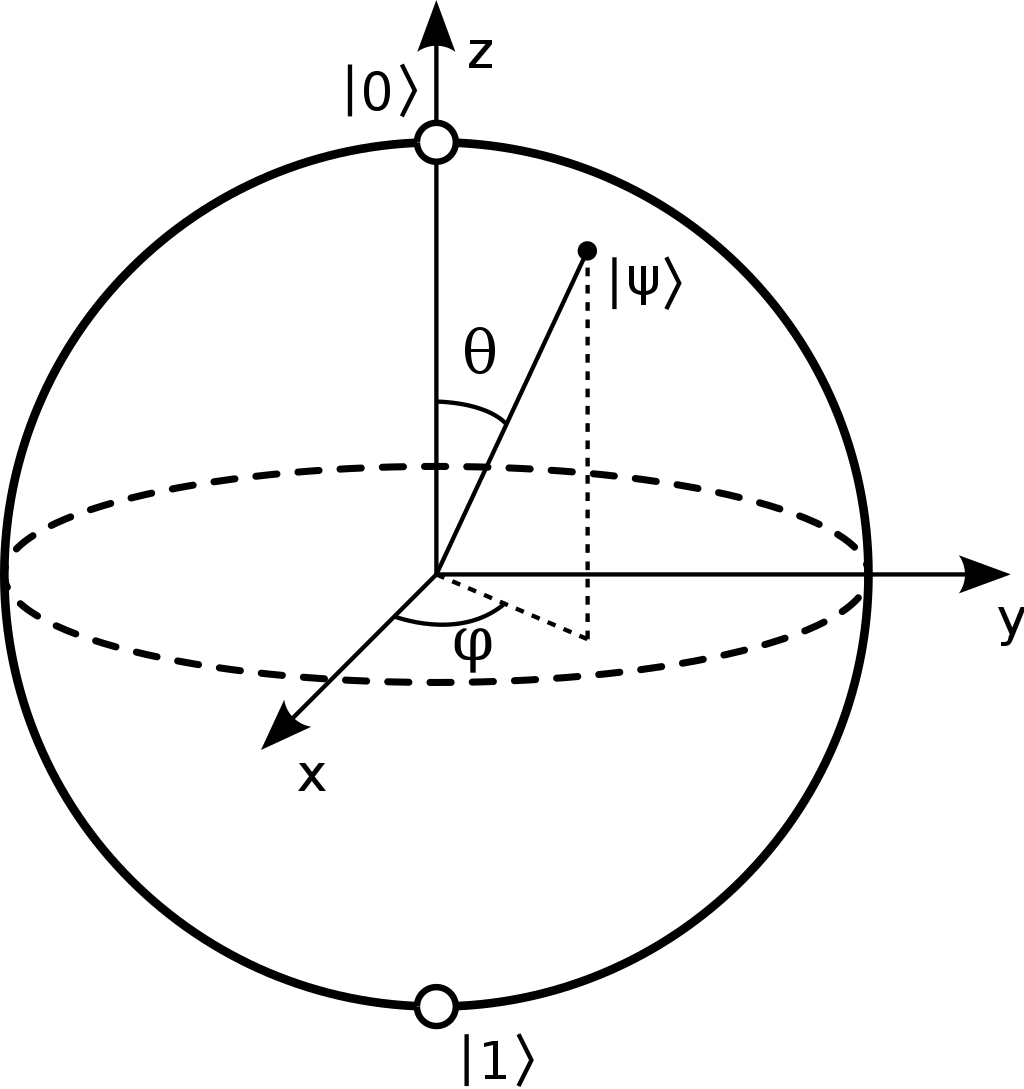}
\caption{Schematic representation of the Bloch sphere.}
\label{fig:Bloch-sphere}
\end{figure}

The Bloch sphere is a geometrical way to represent the pure states of a two-level quantum mechanical system (qubit).  To introduce the concept, let us note that the general state of a qubit may be written as 
\begin{align}
|\psi\rangle=\alpha|0\rangle+\beta|1\rangle;~~\alpha,\beta \in \mathbb{C}.
\end{align}
Normalization gives $|\alpha|^2+|\beta|^2=1$. Expressed in polar coordinates, we get
\begin{align}
|\psi\rangle=r_\alpha e^{i\phi_\alpha}|0\rangle+r_{\beta}e^{i\phi_\beta}|1\rangle
\label{eq:state1}
\end{align}
in terms of four real parameters $r_{\alpha,\beta}>0$ and $\phi_{\alpha,\beta}\in [0,2\pi)$.
The measurable quantities being the probabilities $|\alpha|^2$ and $|\beta|^2$ for the qubit to be in states $|0\rangle$ and $|1\rangle$, respectively, we may multiply the above state by $e^{-i\phi_\alpha}$ to obtain a state that is observationally indistinguishable from the state in Eq.~\eqref{eq:state1}; we continue to denote the new state by $|\psi\rangle$:
\begin{align}
 |\psi\rangle=r_\alpha |0\rangle+r_{\beta}e^{i\phi}|1\rangle;~~\phi \equiv (\phi_\beta-\phi_\alpha),
\end{align}
which involves three real parameters $r_{\alpha,\beta}$ and $\phi$. Switching back to Cartesian coordinates so that $r_\beta e^{i\phi}=x+iy$, with $x,y$ being real, normalization of $|\psi\rangle$ demands that
\begin{align}
r_\alpha^2+x^2+y^2=1,
\end{align}
which is the equation of a unit sphere in three dimensions with Cartesian
coordinates $(x,y,r_\alpha)$. Then, in terms of spherical polar coordinates $r_\alpha=\cos \theta'$, $x=\sin \theta' \cos \phi$, and $y=\sin \theta' \sin \phi$, with $\theta' \in [0,\pi]$ and $\phi \in [0,2\pi)$, we get
\begin{align}
|\psi\rangle=\cos \theta' ~|0\rangle+e^{\mathrm{i}\phi}\sin \theta' |1\rangle.
\end{align} 
We now have only two parameters $(\theta',\phi)$, which define points on the unit sphere. 
Consider a state $|\psi'\rangle$ corresponding to $|\psi\rangle$ but in the lower hemisphere, i.e., with coordinates $(\pi+\theta',\phi)$. We see that $|\psi'\rangle=-|\psi\rangle$, so that the two states are observationally indistinguishable.  This suggests that it suffices to only consider the upper hemisphere $ 0 \le \theta' \le \pi/2$, as corresponding points in the lower hemisphere differ only by a phase
factor of $-1$ and so are observationally indistinguishable.  Consequently,  we invoke a mapping 
\begin{align}
\theta = 2 \theta';~~\theta \in [0,\pi],
\end{align}
which allows mapping of points on the upper hemisphere onto points on a sphere with polar coordinates $(1,\theta,\phi)$, yielding the so-called Bloch sphere.  A point on a Bloch sphere represents the qubit 
\begin{align}
|\psi\rangle = \cos \frac{\theta}{2}~|0\rangle + \sin \frac{\theta}{2}~ e^{\mathrm{i}\phi}~|1\rangle.
\label{eq:state2}
\end{align}
At the North pole of the Block sphere, we have $\theta=0$ and $\phi$ undetermined, and so the state is $|\psi\rangle=|0\rangle$. At the South pole, we have $\theta=\pi$ and $\phi$ undetermined, and so the state is $|\psi\rangle=e^{\mathrm{i}\phi}|1\rangle$. But since the phase factor $e^{\mathrm{i}\phi}$ is inconsequential in as far as distinguishing this state from the state $|1\rangle$ is concerned, the South pole of the Bloch sphere is taken to represent the state $|1\rangle$. At the equator, we have $\theta=\pi/2$, so that the state is $|\psi\rangle=\frac{1}{\sqrt{2}}|0\rangle+e^{\mathrm{i}\phi}\frac{1}{\sqrt{2}}|1\rangle$.  As we vary $\phi$ to generate all points on the equatorial plane of the Bloch sphere,  the states so generated are indistinguishable from the state %which is a linear superposition of the $|0\rangle$ state and the $|1\rangle$ state: 
$|\psi\rangle=\frac{1}{\sqrt{2}}|0\rangle+\frac{1}{\sqrt{2}}|1\rangle$.  Consequently, the equator is taken to represent the state $\frac{1}{\sqrt{2}}|0\rangle+\frac{1}{\sqrt{2}}|1\rangle$. The Bloch sphere is shown schematically in Fig. \ref{fig:Bloch-sphere}. Corresponding to the pure state in Eq.~\eqref{eq:state2}, the density matrix reads as
\begin{align}
    \rho=\begin{pmatrix}\cos^2\frac{\theta}{2}& \cos\frac{\theta}{2}\sin\frac{\theta}{2}e^{-i\phi}\\
    \cos\frac{\theta}{2}\sin\frac{\theta}{2}e^{i\phi} & \sin^2\frac{\theta}{2}\end{pmatrix}.
    \label{eq:app-rho-0}
\end{align}

We now discuss how mixed states of the qubit are represented by points within the Bloch sphere, while pure states are represented by points on the surface of the sphere. To this end, we note that the density matrix of a qubit can always be expressed in terms of the $2\times 2$ identity matrix $I$ and the three Hermitian, traceless Pauli matrices $(\sigma_x,\sigma_y,\sigma_z)$, as 
\begin{align}
    \rho=\frac{1}{2}(I+a_x\sigma_x+a_y\sigma_y+a_z\sigma_z)=\frac{1}{2}(I+\vec{a}\cdot\vec{\sigma})=\begin{pmatrix}1+a_z & a_x -i a_y \\
    a_x+ia_y & 1-a_z\end{pmatrix},
    \label{eq:app-rho-1}
\end{align}
where $a_x,a_y,a_z$ are real quantities. It may be easily checked that the eigenvalues of $\rho$ are $(1/2)(1\pm |\vec{a}|)$. Since density matrices are defined to be positive semi-definite (i.e., having all non-negative eigenvalues), we then must have $|\vec{a}|\le 1$. Note that $\mathrm{Tr}(\rho)=1$; pure states have $\mathrm{Tr}(\rho^2)=1$, while mixed states have $\mathrm{Tr}(\rho^2)<1$. Noting that 
\begin{align}
    \mathrm{Tr}(\rho^2)=\frac{1}{2}(1+|\vec{a}|^2),
\end{align}
we have $|\vec{a}|=1$ for pure states. On the other hand, mixed states have $|\vec{a}|<1$. 

For pure states, comparing Eq.~\eqref{eq:app-rho-0} with Eq.~\eqref{eq:app-rho-1}, we find that $a_x=\sin 2\theta~\cos \phi,~a_y=\sin 2\theta~\sin \phi,~a_z=\cos 2\theta$, and thus, for a pure state, $|\vec{a}|$ represents a point on the surface of the Bloch sphere. For a mixed state, the condition $|\vec{a}|<1$ corresponds to points in the interior of the Bloch sphere. We thus arrive at the main conclusion: the surface of the Bloch sphere represents pure states of a qubit, whereas the interior corresponds to mixed states.

%%%%%%%%%%%%%%%%%%%%%%%%%%%%%%%%%%%%%%%%%%%%%%

%%%%%%%%%%%%%%%%%%%%%%%%%%%%%%%%%%%%%%%%%%%%%%
\section*{Acknowledgements}
SG gratefully acknowledges many useful discussions with Rohitashwa Chattopadhyay, Soumya Kanti Pal and Vinayak Pendse. This work is supported by the Department of Atomic Energy, Government of India, under Project Identification Number RTI-4012. SG would like to thank Ajay Salve and Kapil Ghadiali for computational support. 

\vspace{0.5cm}

\bibliographystyle{unsrt}
\bibliography{review}

\end{document}